\documentclass{aa}
\usepackage{amsmath,graphicx,amssymb}
\usepackage[varg]{txfonts}
\usepackage{natbib}
\bibpunct{(}{)}{;}{a}{}{,}
\usepackage{hyperref}
\usepackage{placeins}
\usepackage{capt-of}
\usepackage{longtable}
\usepackage{supertabular, booktabs}
\usepackage{multirow}
\usepackage{color}
\usepackage{siunitx}
\newcommand{\vsini}{\ensuremath{{\upsilon}\sin i}}
\newcommand{\rv}{\ensuremath{{V_\mathrm{r}}}}
\newcommand{\kms}{\,km\,s$^{-1}$}

\newcommand{\logg}{$\log g$}
\newcommand{\bz}{$\langle B_z \rangle$}
\newcommand{\nz}{$\langle N_z \rangle$}

\newcommand{\teff}{$T_{\rm eff}$}

\newcommand{\zav}[1]{\left(#1\right)}
\newcommand{\hzav}[1]{\left[#1\right]}
\newcommand{\szav}[1]{\left\{#1\right\}}
\newcommand{\cpd}{d$^{-1}$}
\renewcommand{\arraystretch}{1.2}

\begin{document}

\title{A comprehensive reanalysis of \object{HD\,99458}}
\subtitle{Not a magnetic Ap star but an oscillating Algol with ongoing mass transfer}

\author{
E.~Semenko\inst{1}\corrauth{eugene@narit.or.th}
\and Z.~Mikul\'a\v{s}ek\inst{2}\email{mikulas@physics.muni.cz}
\and D.~Mkrtichian\inst{1}\email{davidmkrt@narit.or.th}
\and P.~Walczak\inst{3}\email{walczak@astro.uni.wroc.pl}
\and M.~Gabdeev\inst{4}\email{maksim.gabdeev@upjs.sk}
\and J.~Jan{\'i}k\inst{2}\email{honza@physics.muni.cz}
\and O. Kochukhov\inst{5}\email{oleg.kochukhov@physics.uu.se}
\and J. Kol{\'a}\v{r}\inst{2,6}\email{460809@mail.muni.cz}
\and E.~Paunzen\inst{2}\email{epaunzen@physics.muni.cz}
\and M.~E.~Shultz\inst{7}\email{matt.shultz@gmail.com}
\and M.~Skarka\inst{8,2}\email{marek.skarka@asu.cas.cz}
\and V.N. Aitov\inst{9}\email{vit9517@sao.ru}
\and A.S. Beskakotov\inst{9}\email{beskakotov@sao.ru}
\and R.~Brahm\inst{10,11,12}\email{rafael.brahm@uai.cl}
\and V.V. Dyachenko\inst{9}\email{dyachenko@sao.ru}
\and L.~Eyer\inst{13}\email{Laurent.Eyer@unige.ch}
\and P.~Kab{\'a}th\inst{8}\email{petr.kabath@asu.cas.cz}
\and O. Kvaratskhelia\inst{14}\email{info@astronomia.ge}
\and T. Kvernadze\inst{14}\email{info@astronomia.ge}
\and J.~Lipt{\'a}k\inst{8}\email{jozef.liptak@asu.cas.cz}
\and A.V. Moiseeva\inst{9}\email{agreana@mail.ru}
\and V.~Schaffenroth\inst{15}\email{schaffenroth@tls-tautenburg.de}
\and L.~Vanzi\inst{16,17}\email{lvanzi@uc.cl}
\and J.~\v{Z}\'{a}k\inst{8}\email{jirizak1@seznam.cz}
}
\institute{
National Astronomical Research Institute of Thailand (Public Organization), Chiang Mai, 50180 Thailand
\and Department of Theoretical Physics and Astrophysics, Masaryk University,
Kotl\'a\v{r}sk\'a 2, 611\,37 Brno, Czech Republic
\and Astronomical Institute, University of Wroc{\l}aw, PL-51-622 Wroc{\l}aw, Poland
\and Pavol Jozef \v{S}af\'{a}rik University in Ko\v{s}ice, Faculty of Science, Ko\v{s}ice, Slovakia
\and Department of Physics and Astronomy, Uppsala University, Box 524, 75120, Uppsala, Sweden
\and Institute of Astronomy, Faculty of Physics, Astronomy and Informatics,
Nicolaus Copernicus University in Toru\'{n}, Piwnice k. Torunia, PL-87148 Łysomice,
Poland
\and Independent researcher, Kingston, Ontario K7L 3N6, Canada
\and Astronomical Institute of the Czech Academy of Sciences, Fri\v{c}ova 298, CZ-25165 Ond\v{r}ejov, Czech Republic
\and Special Astrophysical Observatory, Nizhnij Arkhyz, 369167, Russia
\and Facultad de Ingenier\'ia y Ciencias, Universidad Adolfo Ib{\'a}{\~n}ez, Av. Diagonal las Torres 2640, Pe{\~n}alol{\'e}n, Santiago, Chile
\and Millennium Institute for Astrophysics, Chile
\and Data Observatory Foundation, Chile
\and Department of Astronomy, University of Geneva, Chemin Pegasi 51, 1290 Versoix, Switzerland
\and E.~Kharadze Georgian National Astrophysical Observatory, 0301, Abastumani, Georgia
\and Th{\"u}ringer Landessternwarte, 07778 Tautenburg, Germany
\and Centre for Astro Engineering, Pontificia Universidad Cat{\'o}lica de Chile, Santiago, Chile
\and Department of Electrical Engineering, Pontificia
Universidad Cat{\'o}lica de Chile, Santiago, Chile
}

\date{Received date /
Accepted date }

\abstract
{Since 2019, \object{HD\,99458} has been known as a unique example of a $\delta$\,Sct variable with chemical peculiarities of Ap type in an eclipsing binary.}
{We aim to review the peculiar status of the star with new observational materials.}
{The current study is the result of meticulous analysis of space-based photometry from Kepler K2 and TESS, multi-site spectroscopic and spectropolarimetric observations, as well as speckle-interferometric data. Asteroseismic modelling of stellar pulsation is an essential part of the research.}
{We find \object{HD\,99458} to be an eclipsing Algol-type system with active mass transfer from the cool companion filling its Roche lobe to the pulsating primary. We identified signatures of a disc and accretion with varying rates in spectra and light curves. The system's orbital period increases with the rate $\dot{P}_\mathrm{orb}$ of 5.045\,s\,yr$^{-1}$, indicating potential multiplicity. We failed to detect traces of the primary's magnetic field or excessive abundances, both of which are necessary to classify \object{HD\,99458} as an Ap star. We have constrained the fundamental parameters of the pulsating primary and analysed its pulsation using asteroseismology. In space-based photometry, we identified pulsation modes with angular degree $\ell\leq3$. Two-dimensional Fourier analysis of spectroscopic time series reveals pulsations in the form of standing waves with the azimuthal orders $|m|=5\pm1$. Combined analysis of photometry and spectroscopy confirms the status of \object{HD\,99458} as a triaxial gravitationally distorted pulsator.}
{Our comprehensive study affirms the new classification for \object{HD\,99458}.}

\maketitle

\titlerunning{A comprehensive reanalysis of \object{HD\,99458}}
\authorrunning{E. Semenko et al.}

\section{Introduction} \label{introduction}

Amid the surge of interest in the search for new exoplanets in space-based photometry obtained by the Kepler K2 mission \citep{2010Sci...327..977B, 2014PASP..126..398H}, \citet{2016A&A...594A.100B} identified \object{HD\,99458} as a candidate exoplanetary system after recognising periodic eclipses with a depth of 1.494\%. In the follow-up study,  \citet{2019MNRAS.487.4230S} identified the secondary as a low-mass
red dwarf star rather than an exoplanet. The primary component was found to be an A-type pulsating $\delta$ Scuti (hereafter, $\delta$\,Sct) star with extensive spots and anomalous chemical composition. The reported excesses of Fe, Si, and Ti exceeded 1\,dex compared to the Sun, allowing the authors to classify the primary component of \object{HD\,99458} as an Ap star.

The Ap stars form a subgroup of the family of chemically peculiar (CP) stars, mostly main-sequence objects with abnormal spectra indicating a peculiar chemical composition. The fraction of CP stars among objects of a similar effective temperature is approximately 10\%. \citet{1974ARA&A..12..257P} subdivided CP stars into four subgroups, where A stars with peculiar spectra occur in the groups CP1 and CP2. The populations of these groups differ in the character of variability, chemical composition, magnetism, and binarity. Strong, globally organised magnetic fields are the hallmark of Ap variables of the $\alpha^{2}$ Canum Venaticorum \citep[ACV; ][]{2018A&A...619A..98H} type, or CP2 stars. Hereafter, we use Ap and CP2 interchangeably to refer to this class. According to the information presented in the Gaia Data Release 3 (DR3) catalogue, \object{HD\,99458} is classified as a variable of ACV|CP|MCP|ROAM|ROAP|SXARI type with a score of 0.27 \citep{2023A&A...674A..13E, 2023A&A...674A..22G, 2023A&A...674A..14R}.

In contrast to their non-magnetic siblings, the CP2 stars seem to avoid close binary systems. The fraction of the known systems harbouring magnetic components shrinks rapidly with a decreasing orbital period \citep[e.g.][]{2002A&A...394..151C, 2017A&A...601A..14M}. To the best of our knowledge, only three such systems with periods shorter than 3 days are known: HD\,15144 \citep{1981PASP...93..756B, 2008MNRAS.384.1588C}, HD\,200405 \citep{2002A&A...394..151C}, and \object{HD\,99458}. Of the three, only the object of this study undoubtedly has a companion, as its radial velocity varies with a high amplitude and photometry shows eclipses.

The $\delta$\,Sct variables are common among A-type stars. Many of them appear to be members of close binary systems. The $\delta$\,Sct variables are multi-periodic pulsators of luminosity classes V to III with masses between approximately 1.5$M_{\sun}$ and 2.5$M_{\sun}$ \citep{2009AIPC.1170..403H}. The observed light changes are caused by multiple radial and non-radial low-order pressure ($p$) modes, which are excited through the $\kappa$ mechanism \citep{2009AIPC.1170..403H}. In evolved $\delta$\,Sct stars, so-called mixed modes are often observed. These pulsation modes exhibit gravity ($g$) mode characteristics in the interior and $p$-mode characteristics near the stellar surface \citep{2016MNRAS.460.1970B}.

The coexistence of the $\delta$\,Sct-type pulsation with the strong magnetic fields typical of Ap stars is an important observational effect with a wide potential application \citep[e.g.][]{2017IAUS..329..146B, 2024bss..confE..74N}. On the one hand, asteroseismology is the only reliable method for probing the internal structure of stars. On the other hand, incorporating magnetic effects into the theory of stellar oscillations is challenging \citep{2018CoSka..48...73W}. Observationally, identifying $\delta$\,Sct pulsations in magnetic CP2 stars is also challenging. Spectral lines are simultaneously distorted by pulsation, Zeeman splitting, and chemical spots, complicating spectroscopic analyses \citep[see, e.g.][]{2026A&A...709A..43G}. Consequently, only a small number of CP2 stars are currently reported to exhibit $\delta$\,Sct pulsations, most of which have been identified from photometry alone. As an example, \citet{2025A&A...704A.134T} reported only 13 stars. Many of these pulsators fall outside the classical instability strip, suggesting either inaccuracies in physical parameters or incorrect identification of pulsators as CP2 due to hidden binarity. Stellar pulsation with eclipses, as observed in \object{HD\,99458}, provides a unique opportunity for a precise characterisation of the pulsator and its companion.

\citet{2019MNRAS.487.4230S} considered the presence of a cool inflated companion with an accretion disc as an alternative interpretation of the out-of-eclipse variation of \object{HD\,99458}. However, this scenario was rejected in favour of a fully detached binary with the CP2 primary due to weak observational support. The $\delta$\,Sct pulsators in eclipsing Algol-type semi-detached systems (oEA stars) exhibit characteristics similar to those of classical $\delta$~Sct stars, yet their evolutionary status differs markedly due to mass transfer between components \citep{2002ASPC..259...96M}. In systems with large orbital inclinations and no eclipses, these pulsators can be difficult to distinguish from classical $\delta$~Sct stars photometrically. They may instead be classified by the radial velocities and the pronounced variations in their spectral line profiles, which arise from the non-stationary nature of mass transfer and the presence of hot regions that formed by the interaction of the gas stream with the stellar atmosphere. The ongoing mass transfer, regulated by magnetic activity cycles of the Roche-lobe–filling late-type component, exerts a significant influence on the pulsational spectrum \citep{2020A&A...644A.121L,2022Galax..10...97M}.

Motivated by the growing interest in chemically peculiar stars with $\delta$\,Sct-type pulsations and by the lack of magnetic and asteroseismic analyses of \object{HD\,99458} in the literature, we carried out a dedicated study. The main goals are to verify the CP2 status of the pulsating primary, detect its magnetic field, and revise the physical picture of the system.

The paper is structured as follows. Sect.~\ref{sec:data} describes the data acquisition and reduction. Sect.~\ref{sec:char} presents the revised properties of the system and its components. We place the individual findings in context and discuss them in Sect.~\ref{sec:finale}. Supporting material is presented in the Appendices.

\section{Data acquisition and reduction} \label{sec:data}
The discovery of multi-mode stellar oscillations with a dominant frequency of 19.2\,\cpd, typical of $\delta$\,Sct-type variables, was reported by \citet{2019MNRAS.487.4230S} for \object{HD\,99458} as the result of the analysis of the Kepler K2 observations \citep{2014PASP..126..398H}. For the current study, we used multicolour ground-based observations of the star, which were conducted in early 2019 but not included in the original publication, and retrieved new high-precision data from the Transiting Exoplanet Survey Satellite (TESS; \citealt{2015JATIS...1a4003R}). The original Kepler K2 observations were also revised for consistency.

\subsection{Photometric observations from the ground} \label{photometry}

The dedicated observations of \object{HD\,99458} were carried out in five photometric filters (Johnson $B$, Str{\"o}mgren-Crawford $vby$ and $\Delta a$ $g_2$), switched cyclically, during two consecutive nights in January 2019 at the Mt. Suhora Observatory in Poland with the 0.6-m Zeiss telescope equipped with an Apogee Alta U47 detector. The short exposure time, which varied by filter from 2\,s ($y$ and $B$) to 20\,s ($v$), provided a high overall data cadence.

\begin{figure*}
\sidecaption
\includegraphics[width=12cm]{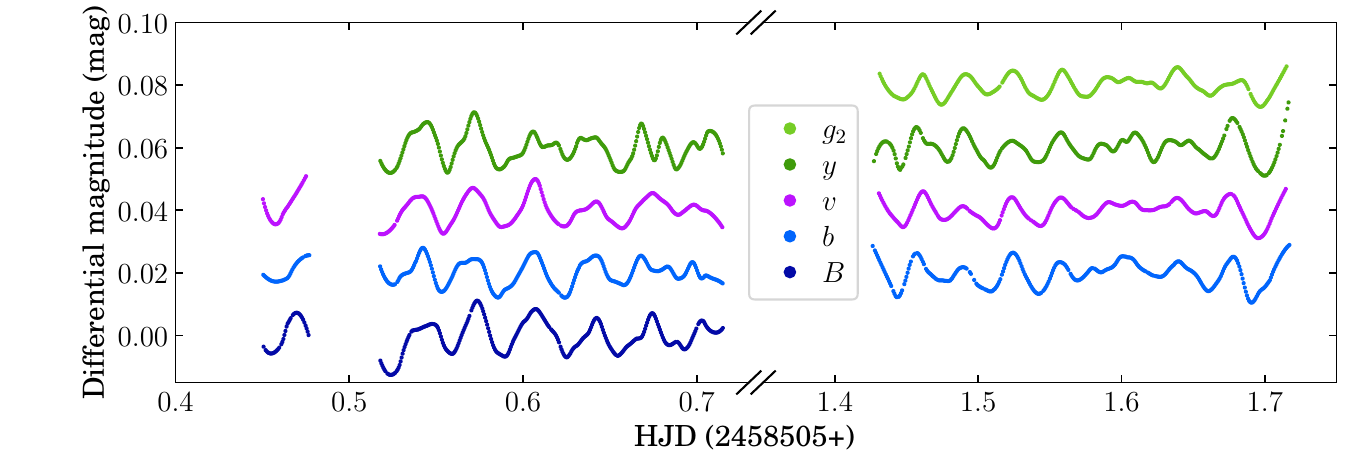}
\caption{Differential light curves of \object{HD\,99458} obtained in Johnson $B$, Str{\"o}mgren-Crawford $v$, $b$, $y$, and $\Delta a$ $g_2$ bands. An artificial vertical shift has been applied for demonstration purposes.}
\label{light_curves}
\end{figure*}

Data reduction and differential photometry were performed using the Image Reduction and Analysis Facility \citep[IRAF, ][]{1987PASP...99..191S} and \texttt{Munipack} \citep{2014ascl.soft02006H}. Both packages employ classic aperture photometry. As the target field is not very crowded, this approach seems to be appropriate. To account for varying seeing conditions, several apertures were tested. For each dataset, we chose the aperture that yielded the smallest scatter in the target star's differential light curves. This criterion was always satisfied by the presence of several comparison stars. The final light curves are shown in Fig.~\ref{light_curves}.

The excellent quality of the differential light curves, ensured by the relative brightness of \object{HD\,99458}, unambiguously demonstrates multi-periodic pulsations of the $\delta$\,Sct type. The data were further analysed using standard Fourier techniques and the Phase-Dispersion Method (PDM) implemented in the programme package \texttt{Peranso} \citep{2016AN....337..239P}. At this stage, we found a broad set of frequencies between 24 and 32\,\cpd, clearly incompatible with the 19--22\,\cpd\ range reported earlier. This apparent discrepancy originates from the Nyquist limit for the K2 data. The frequencies reported by \citet{2019MNRAS.487.4230S} are mirrored (reflected) at the Nyquist frequency and must therefore be interpreted in the super-Nyquist regime \citep{2014MNRAS.445..946C}.

\subsection{Space-based observations and their phenomenological modelling}

Our thorough analysis of the photometric behaviour of \object{HD\,99458} is based on a set of observations obtained from the Kepler K2 and TESS missions between 2014 and 2023. The Pre-search Data Conditioning Simple Aperture Photometry (PDCSAP) light curves were downloaded using \texttt{Lightkurve} \citep{2018ascl.soft12013L}. The K2 light curve consists of 3552 data points with a median uncertainty of 1.1 mmag. The three TESS sectors (45, 46, and 72), hereafter denoted S45, S46, and S72, yielded $17\,501$ photometric measurements with a median uncertainty of 0.85\,mmag. Table~\ref{tab:photolog} gives details about the individual light curves used in this study.

\begin{table}[t]
\begin{center}
\caption{Sources and characteristics of space-based photometry used in the study. }
\label{tab:photolog}
\begin{tabular}{l c c c c}
  \hline\hline
  Mission  & Band  &  Year    &   Data product   & Cadence (s) \\
  \hline
  Kepler K2 & $K_\mathrm{p}$ & 2014     &  K2           & 1800        \\
  TESS (S45) & $T$ & 2021               & TESS-SPOC     & 600         \\
  TESS (S46) & $T$ & 2021               & TESS-SPOC     & 600         \\
  TESS (S72) & $T$ & 2023               & TESS-SPOC     & 200         \\
  \hline\hline
\end{tabular}
\end{center}
\end{table}

\begin{figure*}[t]
\begin{center}
\sidecaption
\includegraphics[width=12cm]{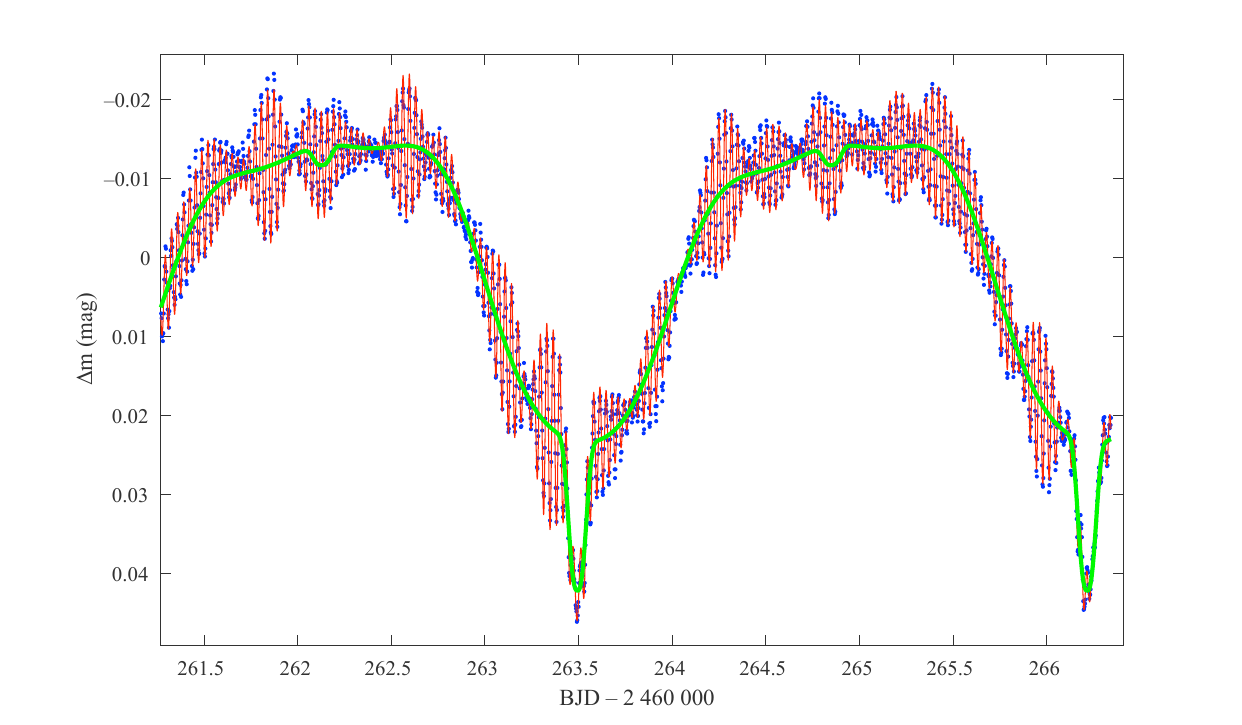}
\caption{Enlarged view of a five-day segment of the TESS light curve of \object{HD\,99458}. The individual measurements are shown as blue points. The phenomenological model consists of the multi-periodic $\delta$\,Sct pulsations shown as a red line and a combination of eclipses and a variable flux contribution of a bright spot on the surface of the primary component shown as a green line.}
\label{Detail_072}
\end{center}
\end{figure*}

One of us (Z.M.) constructed a three-component phenomenological model of the investigated object that accurately describes the observed changes to within a few tenths of mmag and removes all aperiodic instrumental deformations of the light curve. The model's complexity is illustrated in Fig.~\ref{Detail_072}, which shows a five-day segment of the TESS light curve from sector 72, together with its fit, which accounts for eclipses, multimode $\delta$\,Sct pulsations, and variability attributed to spots.

\subsection{Spectropolarimetry and \'{e}chelle spectroscopy of HD\,99458} \label{spectropolarimetry}

\subsubsection{Spectropolarimetry}\label{sec:specpol}

Four observations of \object{HD\,99458} were obtained with the 6-m telescope of the Special Astrophysical Observatory of the Russian Academy of Sciences (SAO RAS). Spectra of the target star and four standards were acquired using the Main Stellar Spectrograph (MSS) equipped with the circular polarisation analyser~(\cite{2004mast.conf..286C, 2014AstBu..69..339P}). Data were acquired and reduced following the standard procedure described by \citet{2022MNRAS.515..998S}. The extracted, calibrated, and continuum-normalised spectra cover the range 442.4--497.9\,nm with an average resolving power of $R=\lambda/\Delta\lambda=15\,000$. The signal-to-noise (S/N) ratio at 455\,nm is 150--170 for \object{HD\,99458} and 300--700 for the standard stars.

The longitudinal magnetic field $\langle B_\mathrm{z}\rangle$, which represents the line-of-sight component averaged over the visible stellar disc, was measured using two methods, detailed in \citet{2022MNRAS.515..998S}. The results are summarised in Table~\ref{bzfield}, where the indices `COG' and `reg' denote the centre-of-gravity and regression method, respectively.

\begin{table}[t]
\begin{center}
\caption{Longitudinal magnetic field $\langle B_\mathrm{z}\rangle$ of \object{HD\,99458} and the standard stars measured using SAO spectropolarimetry.}
\label{bzfield}
\begin{tabular}{l c c c}
  \hline\hline
  Star   &  BJD $-$     &   $\langle B^\mathrm{COG}_\mathrm{z}\rangle$   & $\langle B^\mathrm{reg}_\mathrm{z}\rangle$ \\
  & 2\,458\,000 & (G) & (G) \\
  \hline
  \object{HD\,99458}  &  532.50040  &  $-20\pm115$  & $-45\pm60$   \\
             &  533.53655  &  $190\pm130$  & $70\pm60$    \\
             &  921.46063  &  $-35\pm60$ &  $15\pm35$  \\
             &  922.29396  & $-45\pm90$ & $27\pm35$ \\
  \hline
  \multicolumn{4}{c}{Magnetic standards} \\
  53\,Cam    &  532.55482  &  $2\,630\pm100$ & $2\,260\pm60$  \\
             &  533.55547  &  $3\,370\pm80$  & $3\,060\pm40$  \\
             &  921.40202  &  $1\,300\pm140$  &  $915\pm50$  \\
  $\alpha^2$ CVn & 922.34122 & $-780\pm45$ & $-635\pm20$ \\
  \hline
  \multicolumn{4}{c}{Non-magnetic standards} \\
  $o$ UMa    &  532.56329  &  $-20\pm5$    & $-20\pm5$    \\
             &  533.56394  &  $-5\pm5$     & $0\pm5$      \\
             &  921.40920  &  $15\pm5$ & $-20\pm5$ \\
  HD\,33256  &  922.18476  &  $15\pm15$ & $5\pm5$ \\
  \hline\hline
\end{tabular}
\end{center}
\end{table}

Between June 11th and June 15th 2020, five Stokes $V$ observations of \object{HD\,99458} were acquired with the Echelle Spectropolarimetric Device for the Observation of Stars (ESPaDOnS; \citealt{1997MNRAS.291..658D}), a high-resolution ($R = 65\,000$) \'{e}chelle spectropolarimeter at the Canada-France-Hawaii Telescope (CFHT). The data cover the range 370--1050\,nm across 40 spectral orders. Each spectropolarimetric sequence consists of four sub-exposures obtained with different orientations of the polarisation optics, yielding four unpolarised (Stokes $I$) spectra, one circularly polarised (Stokes $V$) spectrum, and two null $N$ spectra. The $N$ spectra are acquired by combining the beams so that the intrinsic polarisation of the source cancels out, thereby providing both an independent measurement of the noise and a diagnostic to ensure normal instrument operation. The characteristics of ESPaDOnS and data reduction steps were described in detail by \cite{2016MNRAS.456....2W}.

The ESPaDOnS observation log is given in Table~\ref{esptab}. Each observation had a sub-exposure time of 430\,s (total exposure time of 1720\,s), yielding a mean peak S/N per spectral pixel of 412, with minimum and maximum S/N values of 366 and 441. Order-by-order normalisation was performed by fitting polynomial splines to continuum pixels in each spectral order.

\begin{table}
\centering
\caption[]{Log of ESPaDOnS observations and results of the magnetic analysis.}
\label{esptab}
\begin{tabular}{l c c c}
\hline\hline
BJD $-$   & \rv    & \bz & \nz \\
2\,459\,000 & (\kms)   & (G) & (G) \\
\hline
11.74414 & 2.2     &    $19.4\pm11.5$ &   $1.0\pm11.5$ \\
12.74320 & 39.3    &   $-11.9\pm14.4$ &   $5.1\pm14.4$ \\
13.74345 & 51.8    &   $-21.8\pm11.3$ & $-13.0\pm11.3$ \\
14.74375 & $-3.2$  &     $9.2\pm11.6$ &  $12.9\pm11.6$ \\
15.74368 & 59.0    &    $18.2\pm13.4$ &  $-0.1\pm13.4$ \\
\hline\hline
\end{tabular}
\end{table}

\begin{figure}
   \centering
   \includegraphics[width=0.4\textwidth]{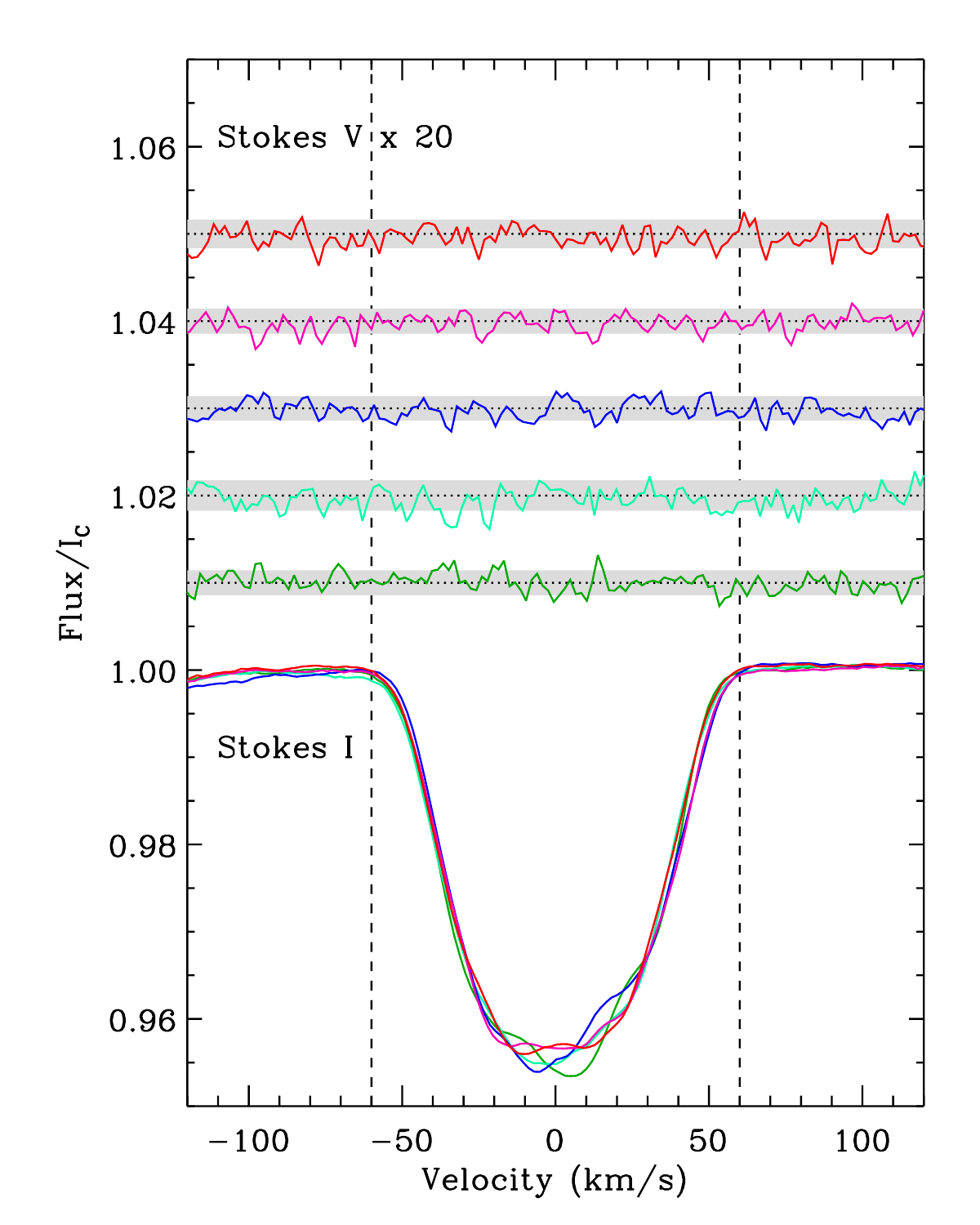}
      \caption[]{LSD Stokes $I$ (bottom) and Stokes $V$ (top) profiles of \object{HD\,99458}, where the latter profiles have been vertically shifted and amplified by a factor of 20. The LSD profiles are shown in the stellar rest frame. Grey-shaded regions indicate the mean uncertainty in Stokes $V$. The vertical dashed lines mark the integration range used to calculate the false alarm probability and \bz.}
         \label{lsd}
   \end{figure}

In order to measure the magnetic field with the best possible S/N, least-squares deconvolution \citep[LSD;][]{1997MNRAS.291..658D} mean line profiles were extracted using the \texttt{iLSD} package \citep{2010A&A...524A...5K}. LSD profiles are extracted by combining a line mask (delta functions at the laboratory wavelengths of the lines expected in the stellar spectrum, with corresponding depths and Land\'{e} factors) with the observed spectrum. The line mask was obtained from the Vienna Atomic Line Database \citep[VALD3;][]{piskunov1995, ryabchikova1997, kupka1999, kupka2000,2015PhyS...90e4005R} using an `extract stellar' request for $T_{\rm eff} = 7\,600$~K, $\log{g}=3.6$, and solar metallicity. Following the procedure described by \cite{2018MNRAS.475.5144S}, the line mask was then cleaned by removing all hydrogen lines, lines blended with the pressure-broadened wings of hydrogen lines, and lines blended with interstellar or telluric lines, in order to ensure that the LSD Stokes $I$ and $V$ profiles are extracted from a set of lines with relatively similar profiles. The line mask was then `tweaked' by adjusting the line depths to match the depths of the observed lines. Of the 8209 lines in the initial mask, 5685 lines remained for the analysis. The resulting LSD profiles are shown in Fig.\ \ref{lsd}, where the profiles have been shifted to the rest frame of the primary. Line-profile variations due to pulsations  are clearly visible in the Stokes $I$ profile.

The strength of the magnetic field was measured using \bz\, via the centre-of-gravity method \citep[e.g.][]{mat1989,wade2000}. These measurements are given in Table~\ref{esptab}, along with \nz~(i.e. the centre-of-gravity measurement applied to the $N$ spectrum). The mean uncertainty of the measurements is 12.4\,G.

\subsubsection{\'{E}chelle spectroscopy} \label{add_spectroscopy}

The original paper by \citet{2019MNRAS.487.4230S} reported \object{HD\,99458} as a $\delta$\,Sct-type pulsator and simultaneously recognised it as a CP2 star according to the classification introduced by \citet{1974ARA&A..12..257P} based on measured individual abundances. Both types of stars are characterised by distinct patterns of line-profile variability (LPV), which rarely occur in the same object. To study the LPV and to quantitatively re-evaluate the photospheric composition of \object{HD\,99458}, we utilised \'{e}chelle spectra obtained in 2019--2026 using the instruments listed in Table~\ref{tab:devices}.

\begin{table*}[bht]
\begin{center}
\caption{List of the \'{e}chelle spectrographs used in this study.}
\label{tab:devices}
\begin{tabular}{l l c c c}
\hline\hline
\multirow[c]{2}{*}{Device}    &  \multirow[c]{2}{*}{Telescope}  & Wavelength range   &  Resolving power           &  \multirow[c]{2}{*}{Reference}  \\
          &             & ($\mathrm{\AA}$)   &  $R=\lambda/\Delta\lambda$ &     \\
\hline
HERMES    & 1.2-m Mercator (Spain)      &  3\,765--9\,000   & 85\,000  & {\citet{2011A&A...526A..69R}} \\
MRES      & 2.4-m TNT (Thailand)        &  4\,200--10\,050  & 15\,000 (pre-2025)  &   \\
          &                             &  3\,710--10\,050  & 18\,000 (post-2025) &   \\
ESPaDOnS  & 3.6-m CFHT (Hawaii, USA)    &  3\,700--10\,050  & 65\,000             & {\citet{1997MNRAS.291..658D}} \\
PLATOSpec & 1.5-m ESO telescope (Chile) & 3\,947--7\,200    & 70\,000  & {\citet{2026MNRAS.545f1972K}} \\
Whoppshel & 1.5-m ASA AZ1500 (Georgia)  & 4\,000--7\,500    & 33\,000  & {\citet{2024CoBAO..71..340C}} \\
\hline\hline
\end{tabular}
\end{center}
\end{table*}

A total of nine high-resolution spectra with exposure times ranging from 600 to 1400\,s were obtained with the High Efficiency and Resolution Mercator Echelle Spectrograph (HERMES) in March 2019. On two occasions, observations were performed in time-resolved mode with a fixed exposure time $T_\mathrm{exp}=300$\,s, yielding 25 spectra in total. A ThArNe lamp was used as a wavelength reference close to the time of the science exposure. The data were reduced with the instrument-specific pipeline \citep{2011A&A...526A..69R}. The extracted spectra were normalised to the continuum using low-order polynomial fits.

We obtained 394 spectra over 15 observing runs, including seven intra-night time series, with the Medium-Resolution Echelle Spectrograph (MRES) on the 2.4-m Thai National Telescope at Doi Inthanon from 2020 to 2026. Data reduction was performed using the \texttt{PyYAP} pipeline\footnote{Available at \url{https://github.com/ich-heisse-eugene/PyYAP}} written in Python. The wavelength-calibrated spectra were normalised to the continuum level using custom scripts.

We carried out seven runs of high-resolution time-resolved spectroscopy in June--July 2025 using the Whoppshel instrument on the 1.5-m telescope at the Georgian National Astrophysical Observatory in Abastumani, Georgia. These observations produced a total of 90 spectra. The raw data were reduced following the same procedure as for MRES.

We obtained one night of time-resolved spectroscopy of \object{HD\,99458} in 2025 with the high-resolution \'{e}chelle spectrograph PLATOSpec at the 1.52-m telescope at La Silla in Chile, formerly operated by the European Southern Observatory (ESO). Data reduction was performed using standard procedures in IRAF.

Table~\ref{tab:obslog} presents an observational log, grouped by instrument. The columns list the barycentric Julian date (BJD), exposure time $T_\mathrm{exp}$, signal-to-noise ratio S/N, measured radial velocity \rv, and projected rotational velocity \vsini.

The radial and rotational velocities of \object{HD\,99458} were derived by fitting a rotation profile function $G(\lambda)$, as defined in \citep{2008oasp.book.....G}, to the LSD profiles extracted using the mask described in Sect.~\ref{sec:specpol}. The same procedure was also applied to the original spectra from \citet{2019MNRAS.487.4230S}. The new values of \rv\ are listed in Table~\ref{tab:oes}. Although our approach to measuring \rv\ is less sensitive to deformations of the line profiles than, e.g., Gaussian profile fitting, the results are still significantly affected by LPV (Fig.~\ref{lsd}). The mean rotational velocity $\langle\vsini\rangle$ of the star obtained across all observations is $49.9\pm2.0$\,\kms.

\subsection{Speckle interferometry}
Speckle interferometric observations of \object{HD\,99458} were obtained with the 6-m telescope of SAO RAS on 27 May 2024 \citep{maksimov2009emccd}. Three imaging series were recorded using filters with the central wavelength and bandpass combinations of 550/50, 700/50, and 900/80 nm. Each series contains 2\,000 frames with exposure times ranging from 20\,ms for the 550 and 700\,nm filters to 60\,ms for the 900\,nm filter.

\section{Revised characteristics of HD\,99458} \label{sec:char}

The volume and quality of the observational material collected in our study allow us to reappraise the status of \object{HD\,99458} as a pulsating, chemically peculiar A-type star in an eclipsing binary system. Furthermore, extensive asteroseismic modelling of the pulsating $\delta$\,Sct component is used to verify the consistency of the analysis.

\subsection{HD\,99458 as a system of stars}

In the following sections, we develop a physical model for the stellar system \object{HD\,99458}. We begin with a phenomenological description of the eclipses, as an incomplete or inaccurate accounting of them can compromise the interpretation of all subsequent results.

\subsubsection{Phenomenological model of eclipses} \label{sec:eclipmod}

The most stable and strictly periodic components of the observed light curve are the primary and secondary eclipses produced by the binary components, which orbit each other in relatively circular orbits. The shape of the eclipses and their depths remain constant throughout the observation period. Both eclipses can be described satisfactorily by a simple phenomenological model $F(\vartheta)$ proposed by \citet{Mik_ecl} and expressed as
\begin{align}
        F(\vartheta,A_{1,2},d_{\mathrm{e}},c_2,\mathit{\Gamma})=A_1\szav{1-\hzav{1-\exp\zav{1-\cosh\zav{\frac{\varphi_1}{d_{\mathrm{e}}}}}}^{\mathit{\Gamma}_1}} \nonumber\\ +\ A_2\szav{1-\hzav{1-\exp\zav{1-\cosh\zav{\frac{\varphi_2}{d_{\mathrm{e}}}}}}^{\mathit{\Gamma}_2}}. \label{eclipse}
\end{align}

\begin{figure}[t]
\begin{center}
\includegraphics[width=0.43\textwidth]{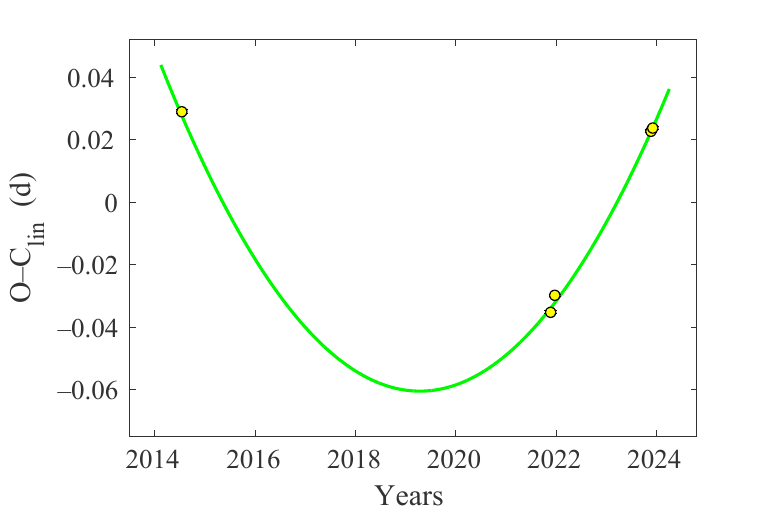}
\caption{Virtual O$-$C diagram for the centre of the primary eclipse with respect to the linear ephemeris given by the reference epoch $M_0$ and the period $P_0$.}
\label{O-C}
\end{center}
\end{figure}

In the model, the dimensionless phase function $\vartheta(t)$ is the sum of the integer epoch $E$ and the fractional phase $\varphi$ \citep[for details, see ][]{Mik_phasefun}, the parameter $d_{\mathrm{e}}$ is proportional to the duration of each eclipse, and the dimensionless parameters $\mathit{\Gamma}_{1,2}$ describe the profile of the eclipses. The arbitrary phases $\varphi_1$ and $\varphi_2$ are defined as
\begin{align}
      \varphi_1=\vartheta-\mathrm{round}(\vartheta);\ \ \varphi_2=(\vartheta\!-\!c_2)-\mathrm{round}\hzav{(\vartheta\!-\!c_2)}, \label{ficka}
\end{align}
where $c_2$ is the phase of the centre of the secondary eclipse, and the phase of the primary eclipse centre is put at zero. The depths $A_1$ and $A_2$ of the primary and secondary eclipses may vary over time.

The best-fit model describing eclipses in \object{HD\,99458} has the following parameters: $d_{\mathrm{e}}=0.014\,61(8),\ \mathit{\Gamma_1}=\mathit{\Gamma_2}=\mathit{\Gamma}=1.354(3),\ c_2=0.499\,6(3)$. As $c_2$ is nearly equal to 0.5, the orbit of the eclipsing binary is probably circular. The value of the parameter $\mathit{\Gamma}$, which is equal for both eclipses, suggests that the inclination angle is markedly different from 90 degrees, so the eclipses are very likely partial. These phenomenological conclusions are consistent with the physical solution of the eclipsing binary presented in \citet{2019MNRAS.487.4230S}.

\begin{table}
\centering
\caption[]{Amplitudes of eclipse minima.}
\label{ampec}
\begin{tabular}{c c c c}
\hline\hline
Filter & $\lambda_{\mathrm{eff}}$ (\AA) & $A_1$ (mag)   & $A_2$ (mag) \\
\hline
$K_\mathrm{p}$ & 5500  & 0.01848(17)    &    0.00191(16) \\
$T$ & 8900  & 0.01874(7)    &    0.00244(6) \\
\hline\hline
\end{tabular}
\tablefoot{The depth of the primary minimum $A_1$ shows no dependence on the effective wavelength $\lambda_{\mathrm{eff}}$, whereas the secondary minimum $A_2$ appears deeper in the $T$ band.}
\end{table}
\begin{table}
\centering
\setlength{\tabcolsep}{3pt}
\caption[]{Times of eclipse minima and corresponding O--C values derived from photometry.}
\label{octab}\tiny
\begin{tabular}{c c c  c c c}
\hline\hline
$E$ & O$-2400000 $   & O$-$C$_{\mathrm{lin}}$  & O$-$C$_\mathrm{{sqr}}$ & O$_{\mathrm{II}}$-C$_{\mathrm{II\,sqr}}$& Source\\
& BJD & (d) & (d)  & (d) &  \\
\hline
$-1013$ & 56\,852.8756    &    0.0289      &  0.0012(7)   & 0.0060(32)& K2\\
$-27$   & 59\,536.6707    &   $-0.0353$    &  $-0.0013(5)$ & 0.0004(19)& TESS\,S45\\
$-16$   & 59\,566.6178    &   $-0.0298$    &  0.0024(4)    & 0.0019(18)& TESS\,S46\\
241     & 60\,266.2157    &    0.0226      &  0.0002(4)    & $-0.0006(18)$ & TESS\,S72a\\
246     & 60\,279.8266    &    0.0237      &  $-0.0001(4)$ & $-0.0019(17)$ & TESS\,S72b\\
\hline\hline
\end{tabular}
\tablefoot{$E$ is the epoch number; O and O$_{\mathrm{II}}$ denote the observed times of the primary and secondary minima, respectively; C$_{\mathrm{lin}}$ and C$_{\mathrm{sqr}}$ are the calculated primary-minimum times from the linear and quadratic ephemerides, respectively, while C$_{\mathrm{II\,sqr}}$ is the calculated secondary-minimum time from the quadratic ephemeris. All calculated times are obtained from relations ($\ref{phasefun1}$) and ($\ref{phasefun2}$) with $\vartheta=E$ (or $\vartheta=E+0.5$ for the secondary minimum). Uncertainties of the residuals are given in parentheses.  The last column lists the photometric data sources.}
\end{table}

The analysis of the course of the phase function $\vartheta(t)$ and the virtual O$-$C diagram for \object{HD\,99458} (Fig.~\ref{O-C}), performed in the same manner as in \citet{Semen_24}, clearly shows the presence of a quadratic term in $\vartheta(t)$, indicating approximately linear growth of the period $P(t)$ during the observed interval. More explicitly,
\begin{eqnarray}
    &\displaystyle\vartheta_0=\frac{t-M_0}{P_0};\quad \vartheta=\vartheta_0-\textstyle{\frac{1}{2}}\dot{P}\,(\vartheta_0-\eta_2)\,(\vartheta_0-\eta_3),\label{phasefun1}\\
    & t_{\mathrm{C}}(\vartheta)=M_0+P_0\,\vartheta+\textstyle{\frac{1}{2}}\dot{P}\,P_0\,(\vartheta-\eta_2)\,(\vartheta-\eta_3),\label{phasefun2}\\
   & P(t)=\dot{\vartheta}^{-1}= P_0\szav{1+\dot{P}\,\hzav{\vartheta_0(t)-\textstyle{\frac{1}{2}}(\eta_2+\eta_3)}},\label{phasefun3}
\end{eqnarray}
where parameters $\eta_2=150.8$ and  $\eta_3=-903.8$ ensure the orthogonality of the relations (\ref{phasefun1}--\ref{phasefun2}), $P_0=2\fd721\,966\,9(3),\,M_0=2\,459\,610.199\,08(10),\ \dot{P}=1.599(6)\times 10^{-7}=5.045(18)\ \mathrm{s\,yr}^{-1}$.

The extremely smooth shape of the O$-$C diagram points to the possible extrinsic nature of the observed change in the orbital period. This could be due to either additional bodies in the system or a precession of the orbital plane on a timescale of dozens of years.

We paid special attention to the possible temporal development of the parameters describing the primary and secondary eclipses, in particular to the changes in the depth and the phase of the secondary minimum compared to the primary minimum. The analysis confirmed that the depth of the primary minimum $A_1$ (see equation (\ref{eclipse})) remains constant within the modelling uncertainties of the observed light curve to an accuracy of a few per cent. Likewise, the position of the secondary minimum relative to that of the primary does not change, as finally indicated by the fifth column of Table~\ref{octab}. The eccentricity of the binary orbit apparently tends to zero.

\subsubsection{Spectroscopic orbit} \label{sec:system}

Fitting the new \rv\ dataset with the orbital solution presented by \cite{2019MNRAS.487.4230S} revealed a systematic phase offset between the observations and the model. This discrepancy may stem from an increasing orbital period ($P_\mathrm{orb}$), as discussed in the previous section, or from the evolving argument of periastron ($\omega$). However, the latter, known as apsidal motion, appears unlikely when applied to \object{HD\,99458} given its circular orbit.

\begin{figure}[t]
\begin{center}
\includegraphics[width=0.76\columnwidth]{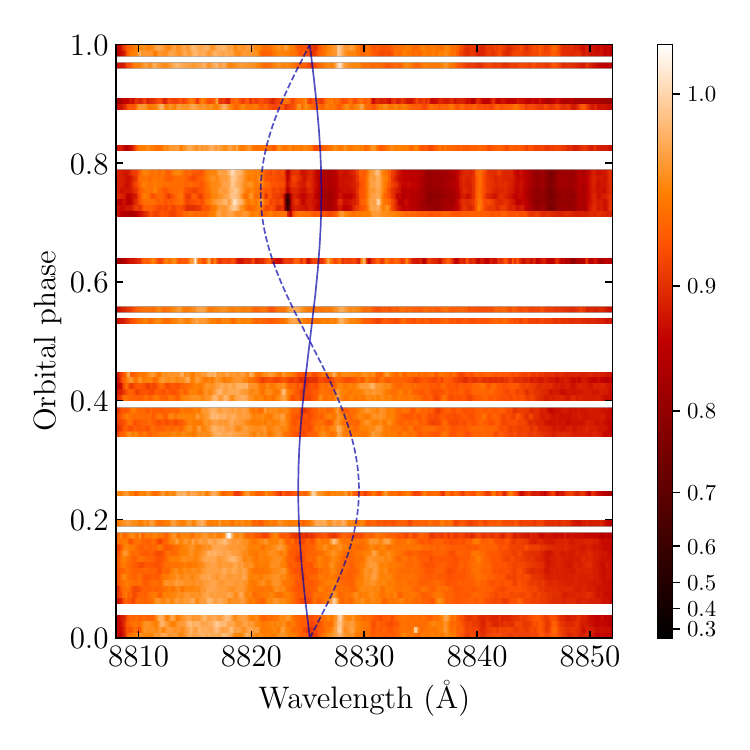}
\caption{Dynamic spectra showing the \ion{Fe}{i} 8824.219\,\AA\ line in the two components of \object{HD\,99458}. Radial velocity variations are shown by the solid line for the primary component and the dashed line for the secondary.}
\label{fig:IRFe}
\end{center}
\end{figure}

\begin{figure}
  \centering
  \includegraphics[width=0.84\columnwidth]{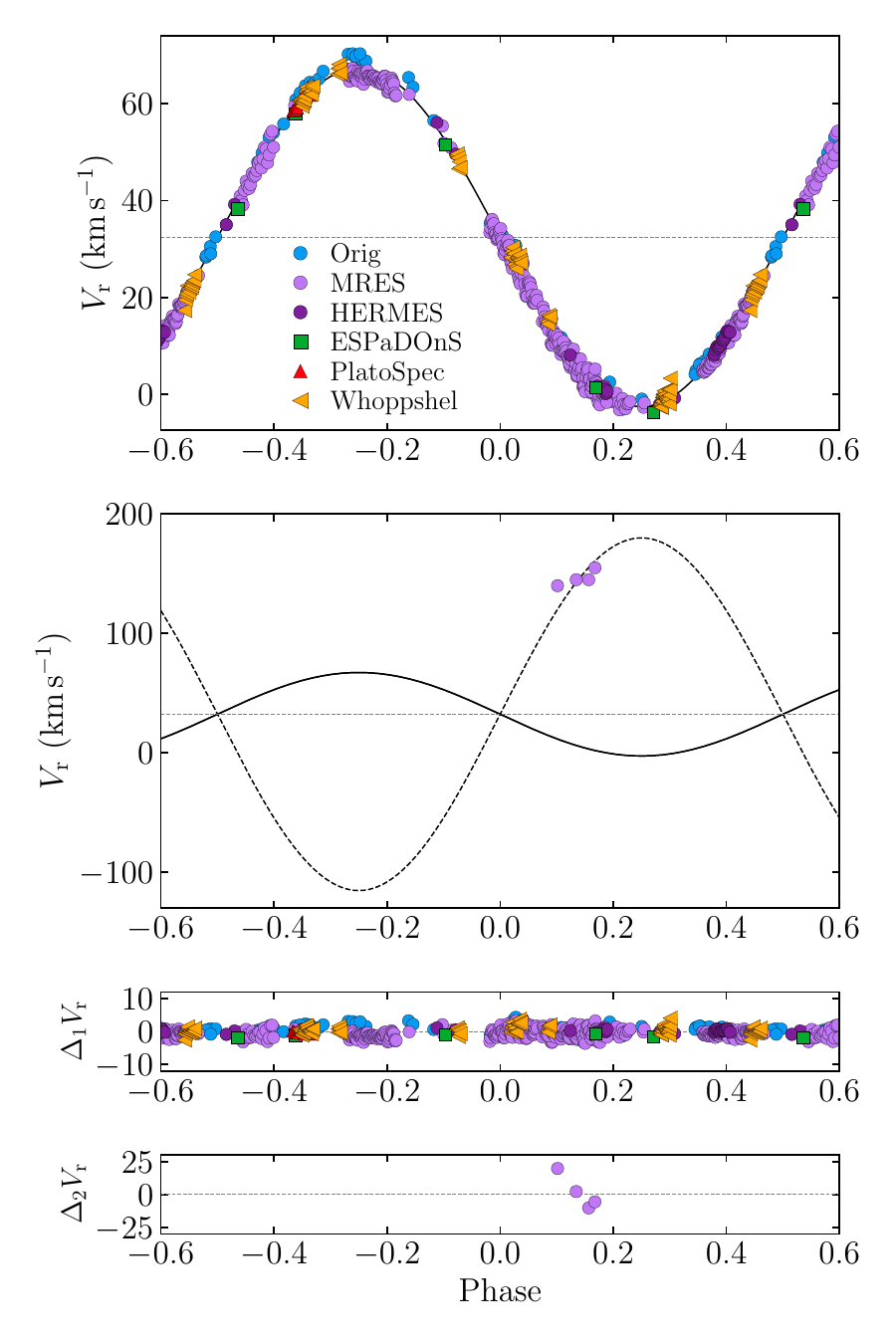}
      \caption[]{From top to bottom: Measured radial velocities of the primary (panel 1) and secondary (panel 2) components, phased with an orbital period of $2\fd722\,172$. The best-fitting orbital solution is shown as the solid and dashed lines for the primary and secondary components, respectively. The horizontal dashed lines in both panels indicate the systemic velocity $\gamma$. Panels 3 and 4 show the corresponding residuals. The measurements labelled as `Orig.' are listed in Table~\ref{tab:oes} and were derived from the spectra used by \citet{2019MNRAS.487.4230S}.}
      \label{fig:HD99458_orb}
  \end{figure}

To revise the orbital solution, we used the extended \rv\ dataset from Table~\ref{tab:obslog} using the \texttt{rvfit} code \citep{2015ascl.soft05020I, 2015PASP..127..567I}. While the spectrum of the secondary is generally difficult to detect, we were able to identify a single \ion{Fe}{i} 8824.219\,\AA\ line in which the contributions of both components are clearly discernible (Fig.~\ref{fig:IRFe}). By measuring the five most visible profiles, we obtained the secondary's radial velocity, which varied over the course of one night from 140 to 155\,\kms\ with an uncertainty of 10--20\,\kms, thereby identifying the system as an SB2 binary. The best-fitting parameters of the system are presented in Table~\ref{tab:orbipar}, and the corresponding \rv\ fits for each component are shown in Fig.~\ref{fig:HD99458_orb}.

\begin{table}
	\begin{center}
		\caption{Revised orbital elements for \object{HD\,99458}.}\label{tab:orbipar}
		\begin{tabular}{lr}
			\hline
			\hline
			Parameter & Value \\
			\hline
			$T_\mathrm{p}$  &  $2\,456\,814.492\,89\pm0.000\,68$ \\
			$P$ (days) & $2.722\,172\pm0.000\,001$ \\
			$e$ & 0.0 (fixed)  \\
            $\gamma$ (km\,s$^{-1}$) & $32.38\pm0.02$  \\
            $K_\mathrm{1}$ (km\,s$^{-1}$)    & $34.94\pm0.03$  \\
            $K_\mathrm{2}$ (km\,s$^{-1}$)    & $147.6\pm4.2$  \\
			\hline
	        $M_{1}\sin^{3} i$ ($M_{\sun}$) & $1.39\pm0.10$ \\
            $M_{2}\sin^{3} i$ ($M_{\sun}$) & $0.33\pm0.01$ \\
            $q$ ($M_{2}/M_{1}$)  & $0.24\pm0.01$ \\
            $a\sin i$ ($R_{\sun}$)  & $9.82\pm0.22$ \\
            \hline
            RMS$_1$ \rv\ (km\,s$^{-1}$) & 1.4 \\
            RMS$_2$ \rv\ (km\,s$^{-1}$) & 11.5 \\
			\hline \hline
		\end{tabular}
	\end{center}
\end{table}

The orbital period we derived from spectroscopy is 10.9\,s longer than that published by \citet{2019MNRAS.487.4230S}. However, the observed difference is only about one-third of that expected between the two average observational epochs, given the rate $\dot{P}$ measured via eclipse timing. This apparent discrepancy, while broadly consistent with a variable orbital period, may result from the limited time span and quality of the data used in the 2019 study or from the pulsation-induced scatter in the measured radial velocities. We return to this problem in Sect.~\ref{sec:finale}.

\subsubsection{Physical model of eclipses}

\begin{figure}
  \centering
  \includegraphics[width=0.98\columnwidth]{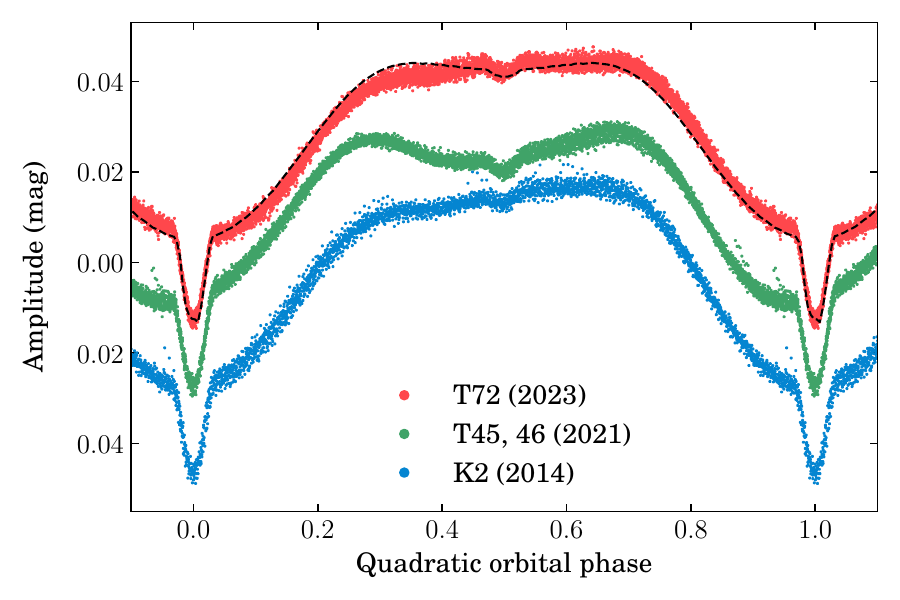}
      \caption[]{Eclipse light curves in K2 and TESS data shifted vertically for clarity. The black dashed line shows the \texttt{PHOEBE} model fit without spots.}
  \label{fig:phoebe}
\end{figure}

We utilised the TESS light curves, cleaned of the pulsation using our phenomenological model of photometric variability, to build a physical model of the binary system \object{HD\,99458}. For this, we employed two codes: the PHysics Of Eclipsing BinariEs (\texttt{PHOEBE}; \citealt{2005ApJ...628..426P}) and the Eclipsing binary Learning and Interactive System (\texttt{ELISa}; \citealt{2021A&A...652A.156C}). \texttt{PHOEBE} was used solely to fit eclipses, ignoring out-of-eclipse variability (Fig.~\ref{fig:phoebe}). For the model fitting, we utilised the differential corrections method. The modelling in \texttt{ELISa} included an MCMC-based uncertainty analysis. The optimisation proceeded through successive releases of the free parameters. This approach was chosen to reduce degeneracies between orbital and stellar-surface parameters during sampling. The mass priors were tied to independent constraints. The mass of the primary component was anchored to the asteroseismic value of \(M_1\approx1.9\,M_{\sun}\)  (Sect.~\ref{plus_model}). The secondary-mass bounds were derived from the \(K_1\) value and the mass function, propagated across a plausible inclination interval. Under these constraints, we estimated inclination, masses and effective temperatures. The final models included configurations with both two fixed spots and two to three spots with free latitude. The spots were located on the primary component. Each spot was parametrised by longitude, latitude, angular radius, and the temperature factor. Table~\ref{tab:detached_mcmc_models} summarises the modelling with both codes used. The posterior medians and 68\% credible intervals are provided only for the five principal detached solutions obtained using the \texttt{ELISa} code.

In comparison to the results presented by \citet{2019MNRAS.487.4230S}, our final model is characterised by a slightly smaller mass (1.8$M_\sun$ vs 2.15$M_\sun$) and a substantially smaller radius (2.3--2.8$R_\sun$ vs 3.47$R_\sun$) of the primary. The most important difference is that the radius of the secondary star is larger, revised from 0.59 to 2.6--2.8\,$R_\sun$. Introducing an inflated secondary reduces the angle $i$ by about 10$\degr$ from the originally reported $73.2\degr$. These changes make the secondary component fill its Roche lobe and provide the best overall agreement with the observations. The spots on the stellar surface that we used in the modelling with the \texttt{ELISa} code do not significantly improve the fit.

\begin{table*}[ht]
\setlength{\tabcolsep}{4pt}
\centering
\small
\caption{Physical solutions for the TESS light curve.}
\label{tab:detached_mcmc_models}
\begin{tabular}{@{}lccccccccp{4.0cm}@{}}
\hline\hline
Model &  $\mathcal{R}^2$ & $M_1$ ($M_\sun$) & $M_2$ ($M_\sun$) & $R_1$ ($R_\sun$) & $R_2$ ($R_\sun$) & $T_{\mathrm{eff,1}}$ (K) & $T_{\mathrm{eff,2}}$ (K) & $i$ ($\degr$) & Spot geometry \\
\hline
PHOEBE  &   &  1.8 & 0.4 & 2.34 & 2.80 & 7600 K (fixed)  & $3800\pm100$ & $64.4\pm0.1$ & --- \\
Spotless  & 0.981 & $1.81^{+0.12}_{-0.07}$ & $0.44^{+0.02}_{-0.02}$  & 2.73 & 2.55 & $7350^{+440}_{-420}$ & $4060^{+310}_{-230}$ & $63.4^{+1.9}_{-1.9}$ & --- \\
1 spot    & 0.977 & $1.85^{+0.12}_{-0.09}$ & $0.44^{+0.02}_{-0.03}$ & 2.89 & 2.69 & $7600^{+300}_{-300}$ & $4080^{+270}_{-240}$ & $62.6^{+1.8}_{-2.3}$ & $\lambda=201\degr$, $\phi=89\degr$, $r=9.4\degr$, $T_{\mathrm{sp}}=\SI{6840}{K}$ \\
2 spots,  fixed $\phi$ & 0.983 & $1.85^{+0.12}_{-0.08}$ & $0.44^{+0.02}_{-0.02}$ & 2.67 & 2.69 & $7750^{+180}_{-330}$ & $3970^{+210}_{-170}$ & $63.4$ (fixed) & $\phi=90\degr$ fixed; A: $\lambda=162\degr$, $r=7.9\degr$, $T_{\mathrm{sp}}=\SI{7130}{K}$; B: $\lambda=200\degr$, $r=7.7\degr$, $T_{\mathrm{sp}}=\SI{6670}{K}$ \\
2 spots   & 0.982 & $1.88^{+0.08}_{-0.11}$ & $0.45^{+0.01}_{-0.02}$ & 2.83 & 2.63 & $7780^{+180}_{-430}$ & $4080^{+290}_{-250}$ & $63.2^{+1.3}_{-1.7}$ & A: $\lambda=193\degr$, $\phi=92\degr$, $r=10.1\degr$, $T_{\mathrm{sp}}=\SI{7160}{K}$; B: $\lambda=203\degr$, $\phi=94\degr$, $r=8.6\degr$, $T_{\mathrm{sp}}=\SI{7240}{K}$ \\
3 spots & 0.986 & $1.86^{+0.11}_{-0.10}$ & $0.45^{+0.01}_{-0.02}$ & 2.83 & 2.65 & $7690^{+240}_{-360}$ & $4080^{+300}_{-250}$ & $62.8^{+1.9}_{-1.7}$ & A: $\lambda=191\degr$, $\phi=93\degr$, $r=8.5\degr$, $T_{\mathrm{sp}}=\SI{7380}{K}$; B: $\lambda=200\degr$, $\phi=94\degr$, $r=7.7\degr$, $T_{\mathrm{sp}}=\SI{7000}{K}$; C: $\lambda=212\degr$, $\phi=93\degr$, $r=8.6\degr$, $T_{\mathrm{sp}}=\SI{7230}{K}$ \\
\hline\hline
\end{tabular}
\tablefoot{The columns list the Bayesian $\mathcal{R}^2$ metric, the masses $M_1$ and $M_2$, the radii $R_1$ and $R_2$, and the effective temperatures $T_{\mathrm{eff,1}}$ and $T_{\mathrm{eff,2}}$ of the primary and secondary components, respectively; the inclination angle $i$, and the spot parameters: the radius $r$, longitude $\lambda$, latitude $\phi$, and the temperature $T_{\mathrm{sp}}$.}
\end{table*}

\subsubsection{Potential third body}

A gravitational perturbation from a third body could explain the observed variation in the orbital period of \object{HD\,99458}. We used speckle interferometry to search for a potential close-in companion.

\begin{figure}
   \centering
   \includegraphics[width=\columnwidth]{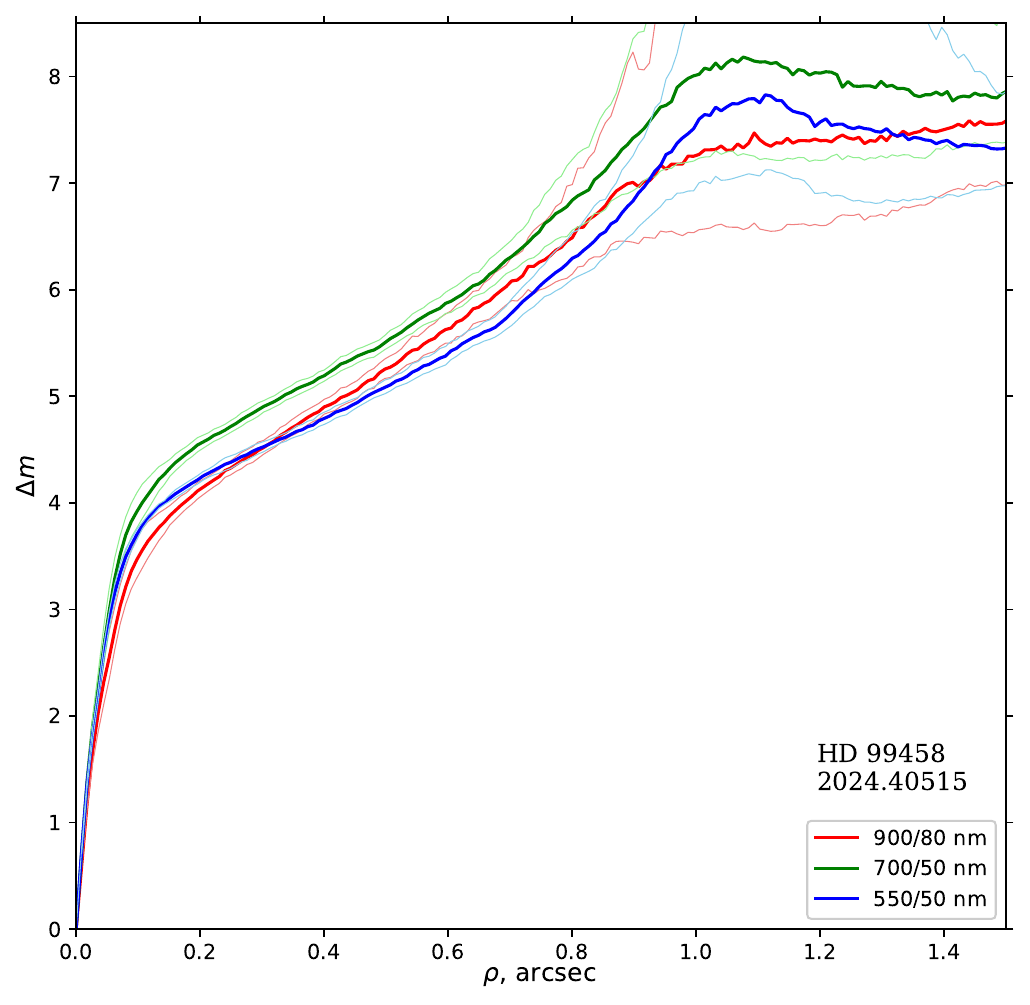}
      \caption[]{Speckle-interferometric limits on the presence of additional components in \object{HD\,99458}. The coloured lines show the angle-averaged limits on $\Delta m$ at various distances. The semi-transparent lines mark the 1-$\sigma$ uncertainty limits.}
         \label{fig:spcl}
   \end{figure}

Fig.~\ref{fig:spcl} presents the speckle-interferometric detection limits as a function of angular separation $\rho$ and magnitude difference $\Delta m$ for potential companions. In the analysis of the autocorrelation functions and power spectra of the speckle-interferometric data series, we find no clear evidence of additional components. Yet we cannot completely rule out a hypothetical third body, as the power spectra near the resolving limit of the 6-m telescope in the 700 and 900\,nm filters appear asymmetric. This effect may result from an incomplete compensation of atmospheric dispersion at zenith distances $z>60\degr$. In this range of $z$, the atmospheric dispersion compensator of our instrument becomes less effective. The observations of \object{HD\,99458} through the 550\,nm filter are unaffected, as they were conducted under more favourable conditions. As the total recording time of a single series is less than 10 minutes, corresponding to a change in the zenith distance of only about $2\degr$, the available data do not allow us to reliably distinguish between the residual effects of atmospheric dispersion and the presence of an unresolved component. Consequently, we cannot exclude a companion with $\Delta m > 2$\,mag at $\lambda=550$\,nm and $\Delta m < 2$\,mag in the 700 and 900\,nm filters. Additional observations are necessary to obtain a more reliable result.

\subsection{Reassessment of the CP status} \label{sec:abund}
To be classified as a CP star, \object{HD\,99458} must possess a strong globally organised magnetic field and demonstrate chemical anomalies. The presence of stable spots caused by the inhomogeneous distribution of chemical elements is another requirement. In our study, we tested the star against each of these criteria.

\subsubsection{Magnetic field}
Strong, large-scale magnetic fields are a defining characteristic of chemically peculiar A-type stars. However, \citet{2012MNRAS.422.2072F} showed that in exceptional cases a magnetic field can be detected in very young stars with a normal composition or even with anomalies of $\lambda$ Boo type. Our dedicated spectropolarimetric observations of \object{HD\,99458} made with the two spectropolarimeters showed no measurable longitudinal magnetic field.

The MSS observations place an upper limit of approximately 150\,G on the longitudinal field \bz\ of \object{HD\,99458}, verified by observations of two magnetic (53\,Cam and $\alpha^2$\,CVn) and two non-magnetic ($o$ UMa and HD\,33256) standard stars. The only spurious signal was detected in the data for $o$ UMa, although it is vanishingly small and does not affect our conclusion.

The analysis of the ESPaDOnS observations yielded a non-detection with even higher confidence due to a substantially larger number of measured lines contributing to the LSD analysis. In this case, the presence of a magnetic field was evaluated using false alarm probabilities \citep[FAP; ][]{1997MNRAS.291..658D}, via calculation of the $\chi^2$ statistic of Stokes $V$ and $N$ inside and outside of the stellar lines to the null model of zero polarisation. An FAP above $10^{-3}$ indicates a non-detection. All five ESPaDOnS observations are above the FAP threshold.

\begin{figure}
   \centering
   \includegraphics[width=0.85\columnwidth]{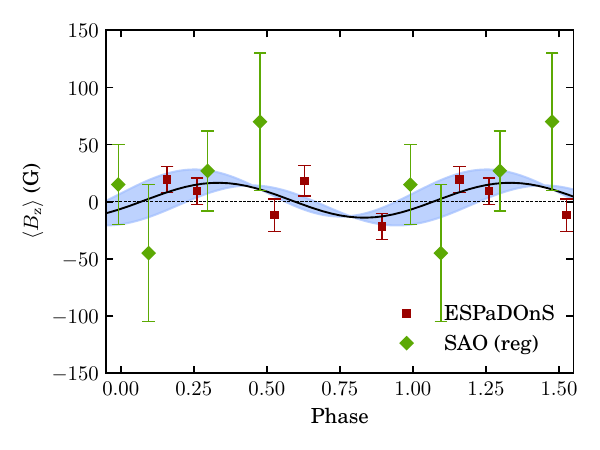}
      \caption[]{ESPaDOnS and SAO \bz\ measurements phased with nominal rotation period. The black curve shows the harmonic fit to the measurements, and the shaded region indicates the associated uncertainties.}
         \label{HD99458_bz}
   \end{figure}

Fig.~\ref{HD99458_bz} shows \bz\ folded with the period $P_\mathrm{0}$ (Sect.~\ref{sec:eclipmod}), assuming the primary's rotation is synchronised with the orbital motion. A first-order harmonic fit to the ESPaDOnS measurements yields an amplitude of $16 \pm 11$~G, consistent with the uncertainty in \bz, and the reduced $\chi^2$ is identical to that of the null model (i.e., no variation aside from noise).

\begin{figure}
   \begin{center}
   \includegraphics[width=1\columnwidth]{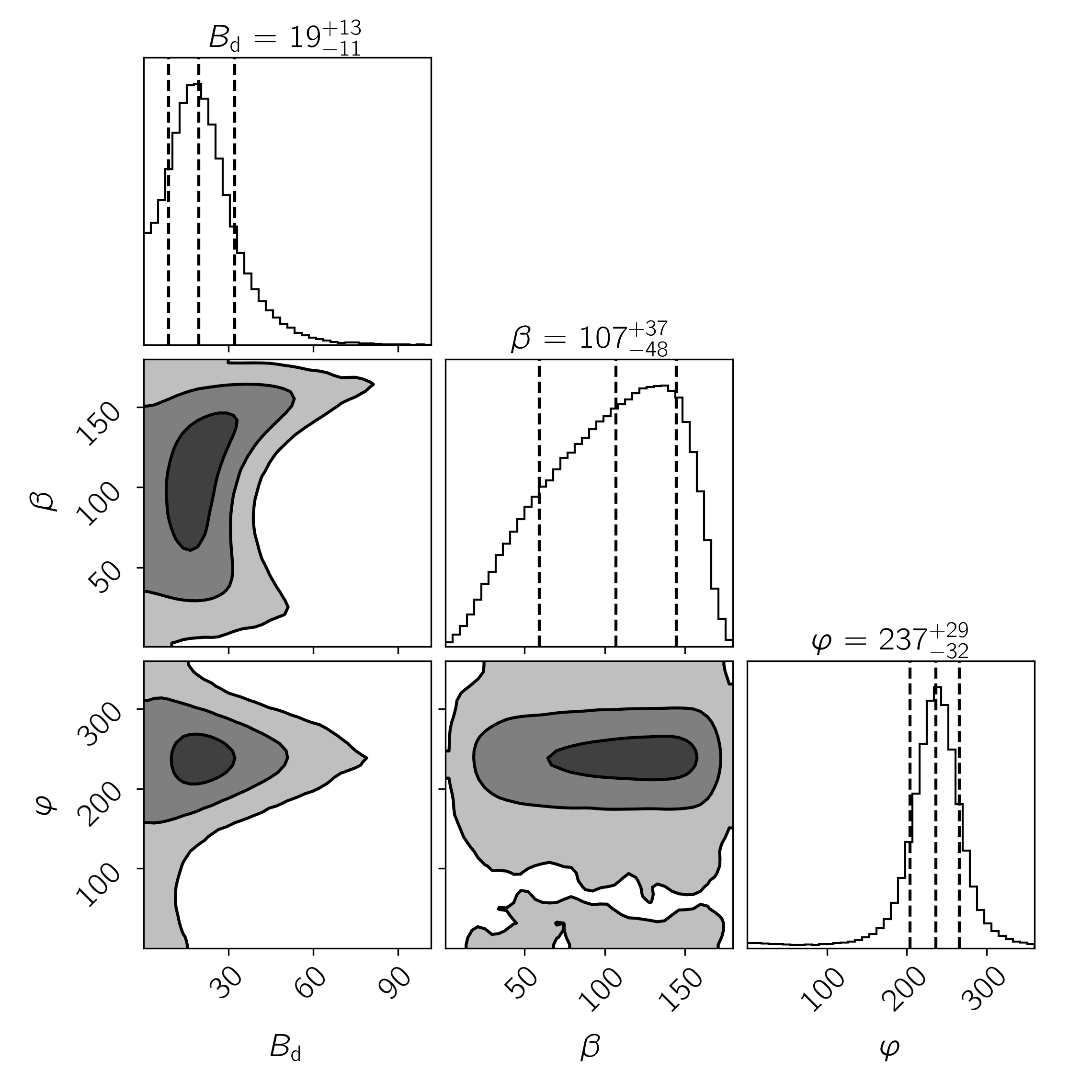}
      \caption[]{Marginalised posterior probability distributions for a dipolar fit to the Stokes $V$ LSD profiles of \object{HD\,99458}. Contours correspond to 1, 2, and 3-$\sigma$ confidence levels.}
         \label{fig:globalfield}
         \end{center}
\end{figure}

To obtain a more accurate estimate of the upper limit of the global dipolar field strength, we performed a Bayesian comparison of the ESPaDOnS LSD Stokes $V$ profiles with theoretical polarisation spectra calculated using the \texttt{InversLSD} code \citep{kochukhov:2014}. This software was applied to compute a library of predicted circular polarisation profiles for five rotational phases, assuming a coplanar synchronised configuration of the binary ($i_{\rm rot}=i_{\rm orb}$ and $P_{\rm rot}=P_{\rm orb}$). The Stokes $V$ spectra were treated in the weak-field approximation \citep[e.g.][]{kochukhov:2010}, assuming that the LSD profile behaves as a pseudo-line with $\lambda=500$~nm and an effective Land\'e factor of 1.2. The equivalent width and broadening of this line were adjusted by fitting the LSD Stokes $I$ spectra. With these assumptions, Markov chain Monte Carlo (MCMC) sampling was applied to explore the entire parameter space $B_{\rm d}$, $\beta$, $\phi$ with the help of \texttt{SoBAT} routines written in IDL \citep{anfinogentov:2021}. A uniform prior was used for $B_{\rm d}$ and a random orientation of the dipolar field axis was implemented with uniform priors in $\phi$ and $\cos{\beta}$. The MCMC walkers were run for $10^6$ steps after the initial $10^4$ step burn-in. The resulting posterior distribution of the three dipolar field parameters is illustrated in Fig.~\ref{fig:globalfield}. Formally, we find $B_{\rm d}=19\substack{+13 \\ -11}$\,G but the statistical significance of this detection is below 2-$\sigma$. The upper limits for $B_{\rm d}$ are 32, 52, and 81~G for the 1, 2, and 3-$\sigma$ confidence levels, respectively. Thus, any magnetic field possessed by the star must be in the `ultra-weak' regime occupied by stars such as Vega or Alhena \citep[e.g.][]{2020MNRAS.492.5794B}, as it is certainly below the 300\,G lower limit for magnetic Ap stars found by \cite{2007A&A...475.1053A} and \cite{2019MNRAS.483.3127S}.

\subsubsection{Abundance analysis}\label{sec:chemcomp}

\object{HD\,99458} was recognised by \citet{2019MNRAS.487.4230S} as a chemically peculiar object. The anomalous elemental abundances are the hallmark of this class of stars.

We used the IDL-based version of Spectroscopy Made Easy (SME) \citep{1996A&AS..118..595V, 2012ascl.soft02013V, 2017A&A...597A..16P} to fit the HERMES spectra of \object{HD\,99458}. From the set of atmospheric models distributed with SME, we utilised the grid of \texttt{LLmodels} \citep{2004A&A...428..993S} as the most suitable for A-type stars. The atomic line parameters were extracted from VALD3 in accordance with the preliminary effective temperature \teff\ and surface gravity \logg\ of the star.

First, we verified the stellar parameters $T_\mathrm{eff}=7600$\,K and $\log g=3.6$ reported by \citet{2019MNRAS.487.4230S}. Spectrum synthesis of the first three Balmer lines and a 100\,nm region centred at 520\,nm showed mostly satisfactory agreement with the observations. Thus, we decided to use the originally published values for \teff\ and \logg\ throughout this study. The only visible exception, where the fit did not match the observed profiles, is H$_{\alpha}$. This line demonstrates strong variability across spectra obtained at different epochs. While such variability is not expected from the main A-type star, it appears consistent with emission from a cool and active companion. Apart from H$_{\alpha}$ and the near-IR \ion{Fe}{i} line used in Sect.~\ref{sec:system}, we have identified a few spectral features, described in Sect.~\ref{sec:specvar}, which are potentially attributable to the companion star. However, none of these findings allows us to infer the physical parameters or chemical composition of the secondary from the spectra. Moreover, the lines of the hot primary component completely dominate in the visible spectrum, so that we can effectively treat the star as single-lined in the abundance analysis.

After the preliminary phase, we evaluated the microturbulent velocity $\xi_\mathrm{micro}=2.9\pm0.8$\,\kms\ and the rotational velocity \vsini$=47.9\pm0.5$\,\kms\ by fitting the \ion{Fe}{i} and \ion{Fe}{ii} lines. The latter parameter is in full agreement with the value found from the LSD profiles (Table~\ref{tab:obslog}).

\begin{figure}[t]
  \centering\includegraphics[width=\linewidth]{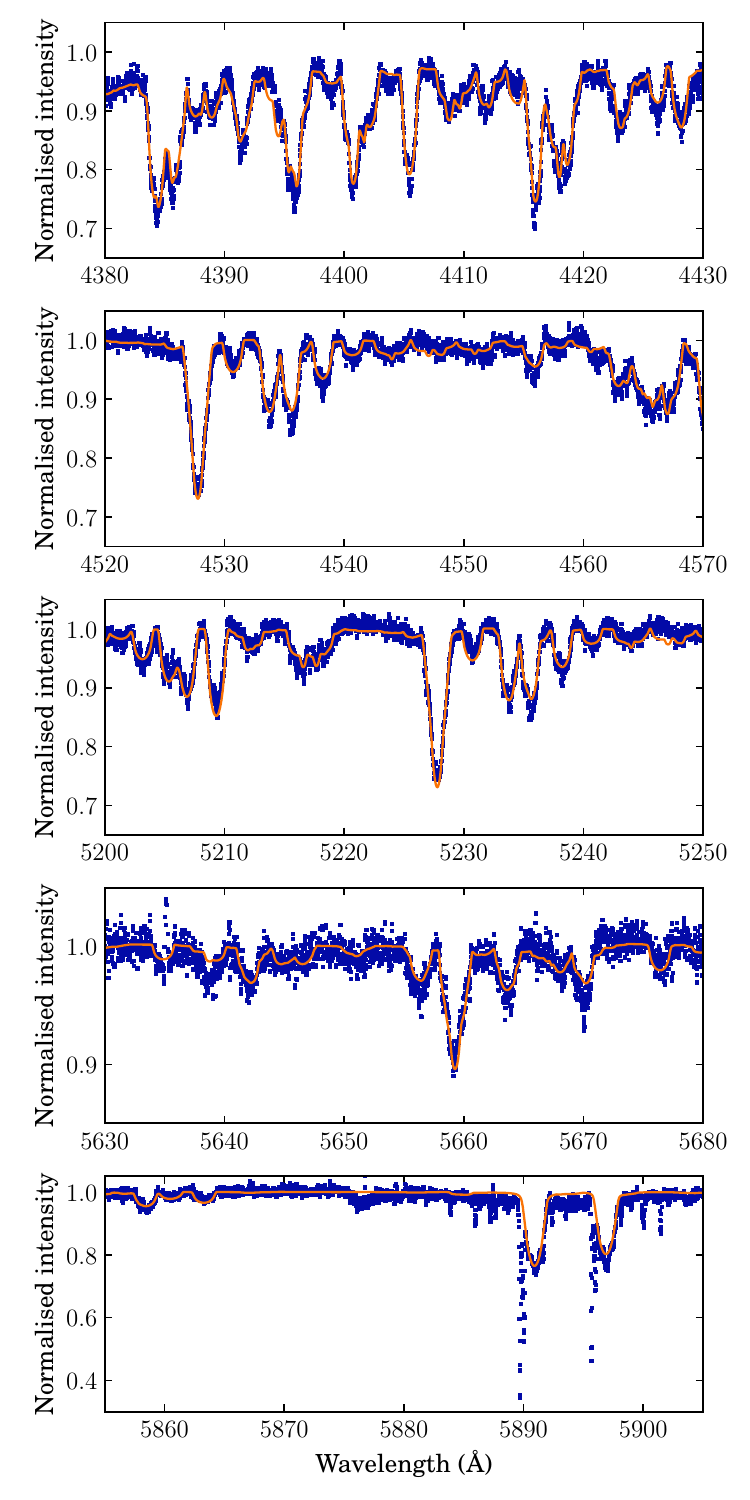}
  \caption{Fit of a synthetic spectrum (orange) to the HERMES observations (blue).}
  \label{fit_spectrum}
\end{figure}

\begin{figure}
  \centering\includegraphics[width=\columnwidth]{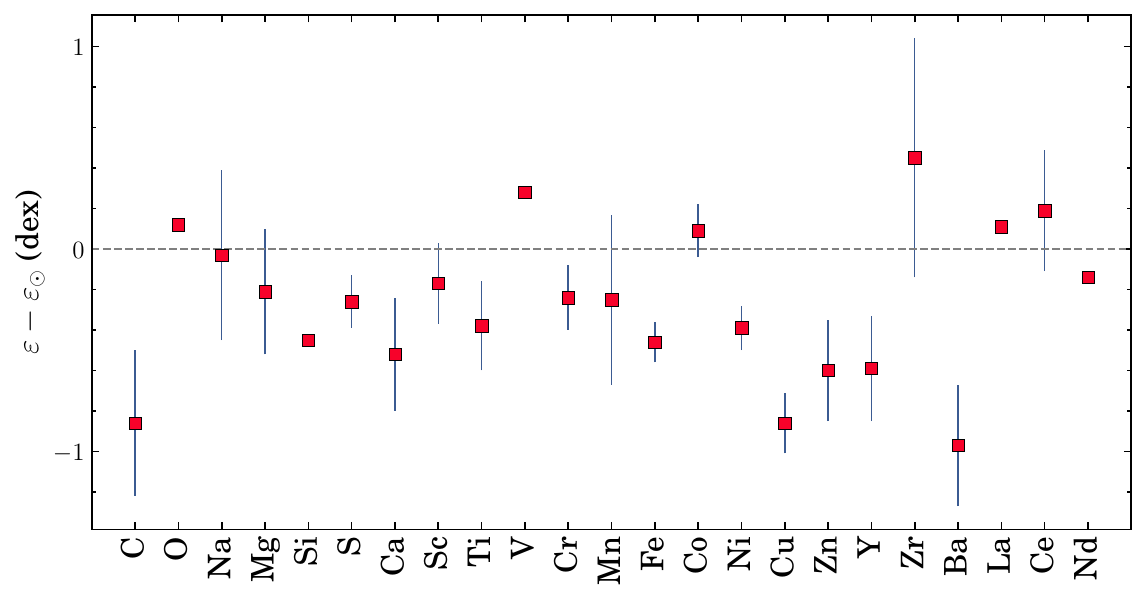}
  \caption{Chemical composition of \object{HD\,99458} compared to the Sun.}
  \label{plot_abundances}
\end{figure}

Finally, by fitting large portions of the HERMES spectra, we evaluated the atmospheric chemical composition of \object{HD\,99458}. The goodness of fit is demonstrated in Fig.~\ref{fit_spectrum}. Table~\ref{abundances} lists the individual abundances of 23 species and their corresponding solar values, and Fig.~\ref{plot_abundances} illustrates the relative abundance pattern. Our results show that most elements in \object{HD\,99458} are mildly depleted, in contrast to the 1\,dex enrichment in Fe, Si, and Ti reported by \citet{2019MNRAS.487.4230S}.

\begin{table}[t]
\begin{center}
\caption{Atmospheric abundances ($\log N/N_\mathrm{tot}$) in \object{HD\,99458} together with the corresponding solar values (in parentheses).}
\label{abundances}
\begin{tabular}{l|l}
  \hline\hline
	Element and abundance & Element and abundance \\
	\hline
	C: $-$4.33$\pm$0.36  ($-$3.47)  & Fe: $-$4.98$\pm$0.10 ($-$4.52) \\
    O: $-$3.14  ($-$3.26)           & Co: $-$7.00$\pm$0.13 ($-$7.09) \\
   Na: $-$5.78$\pm$0.42 ($-$5.81)   & Ni: $-$6.18$\pm$0.11 ($-$5.79) \\
   Mg: $-$4.69$\pm$0.31 ($-$4.48)   & Cu: $-$8.71$\pm$0.15 ($-$7.85) \\
   Si: $-$4.89 ($-$4.44)            & Zn: $-$8.07$\pm$0.25 ($-$7.47) \\
    S: $-$5.17$\pm$0.13 ($-$4.91)   & Y: $-$10.41$\pm$0.26 ($-$9.82) \\
   Ca: $-$6.18$\pm$0.28 ($-$5.66)   & Zr: $-$8.99$\pm$0.59 ($-$9.44) \\
   Sc: $-$9.06$\pm$0.20 ($-$8.89)   & Ba: $-$10.73$\pm$0.30 ($-$9.76) \\
   Ti: $-$7.44$\pm$0.22 ($-$7.06)   & La: $-$10.81 ($-$10.92) \\
    V: $-$7.85 ($-$8.13)            & Ce: $-$10.26$\pm$0.30 ($-$10.45) \\
   Cr: $-$6.65$\pm$0.16 ($-$6.41)   & Nd: $-$10.75 ($-$10.61) \\
   Mn: $-$6.86$\pm$0.42 ($-$6.61)   &  \\
  \hline\hline
\end{tabular}
\tablefoot{The uncertainties are given only for elements with a sufficient number of identified lines. The solar composition is compiled from \citet{2021A&A...653A.141A} and \citet{2022A&A...661A.140M}.}
\end{center}
\end{table}

\subsubsection{Line profile variability} \label{sec:specvar}

Given the absence of any chemical anomalies or a measurable magnetic field characteristic of CP2 stars, we can conclude that the earlier classification of \object{HD\,99458} is not supported by our analysis. However, a careful inspection of the high-resolution spectra has revealed an LPV, which can be attributed to at least two distinct mechanisms. One of these typically occurs in hot CP stars \citep[e.g.][]{2005MNRAS.357..251T}, produces variable hydrogen lines, and, in the case of \object{HD\,99458}, may indicate a magnetosphere in either component.

\begin{figure}
  \centering
  \includegraphics[width=\columnwidth]{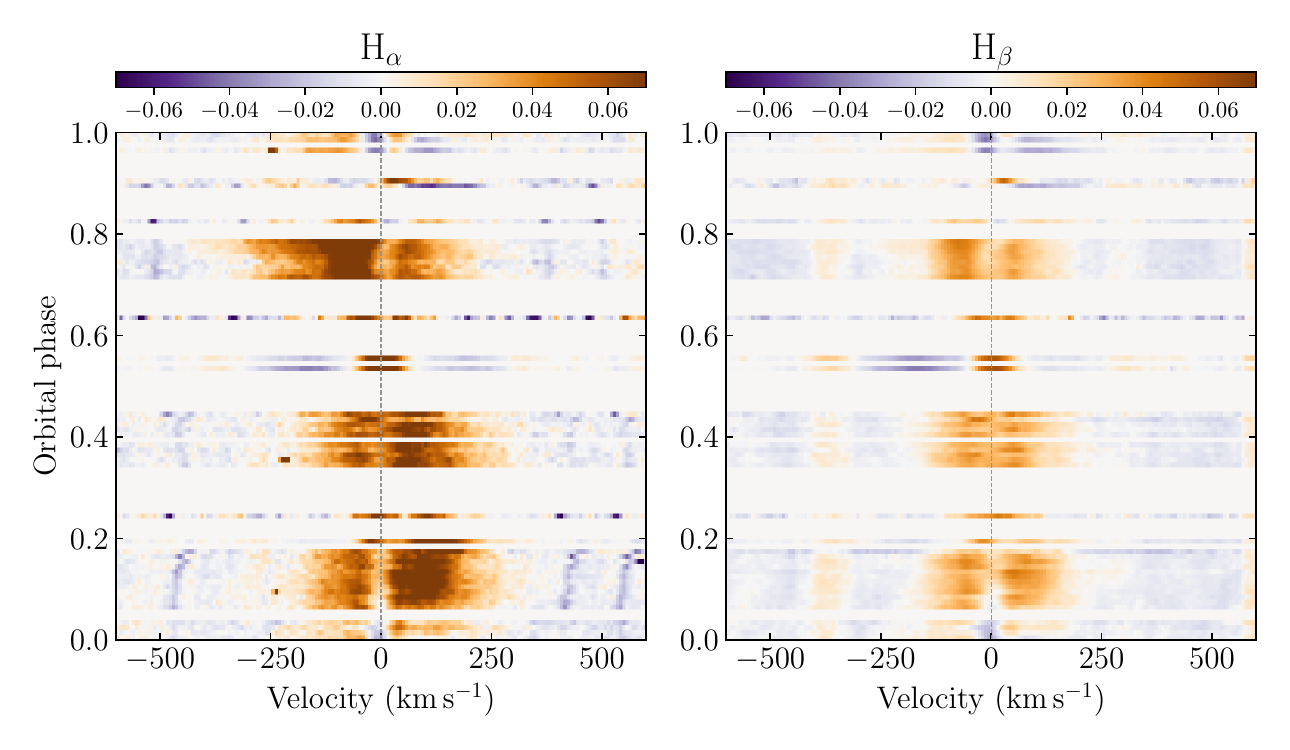}
      \caption[]{Residual $H_{\alpha}$ (left) and $H_{\beta}$ (right) profiles demonstrating LPVs as a function of orbital phase. The residuals were obtained by subtracting the synthetic spectrum from the observations. The velocity scale is given in the reference frame of the primary component.}
         \label{fig:hd99458_halpha}
  \end{figure}

The $H_{\alpha}$ line exhibits significant line-profile variability in all high-S/N spectra obtained in this study. Moreover, somewhat weaker yet phase-coherent changes in $H_{\beta}$ also persist throughout the orbital cycle. The lower resolution of MRES suppresses the fine structure of these profiles, but the spectra from this instrument provide superior phase coverage in our observations and were therefore selected to illustrate the effect of the hydrogen-line LPVs in Fig.~\ref{fig:hd99458_halpha}. We subtracted the synthetic spectrum obtained using \texttt{SME} in Sect.~\ref{sec:chemcomp} from the observations to examine the hydrogen-line profiles in greater detail.

In the residual spectra, both the $H_{\alpha}$ and $H_{\beta}$ lines exhibit emission that reaches maximum strength at the approximate orbital phases of 0.1--0.2 and 0.7--0.8. These emission peaks are located at radial velocities of approximately +110 and $-65$\,\kms. At orbital phases close to 0, 0.5, and 1, the emission profiles become nearly symmetric and double-peaked. In the PLATOSpec data, obtained near an orbital phase of 0.64, we observed intranight evolution of the absorption, with the absorption minimum centred at $-220$\,\kms.

\begin{figure}
  \centering
  \includegraphics[width=0.9\columnwidth]{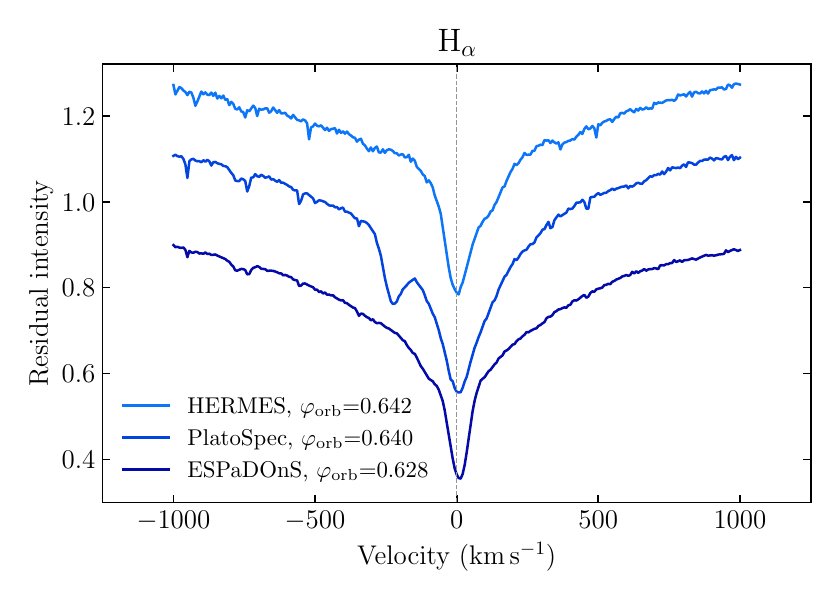}
      \caption[]{Three spectral regions centred on $H_{\alpha}$ obtained at similar orbital phases. The corresponding dates are BJD\,2\,458\,547.5964 (HERMES), BJD\,2\,459\,015.7437 (ESPaDOnS), and BJD\,2\,460\,850.526 (PLATOSpec).}
         \label{fig:H_same_phi}
  \end{figure}

The $H_{\alpha}$ profiles obtained at similar orbital phases but at widely separated epochs can differ substantially, as demonstrated in Fig.~\ref{fig:H_same_phi}. Given the large gaps between observations (HERMES in 2019, ESPaDOnS in 2020, and PLATOSpec in 2025), these differences suggest that the mechanism driving this hydrogen-line variability is not stable on multi-year timescales, and is therefore not typical of CP stars.

\begin{figure*}
  \centering
  \includegraphics[width=0.8\textwidth]{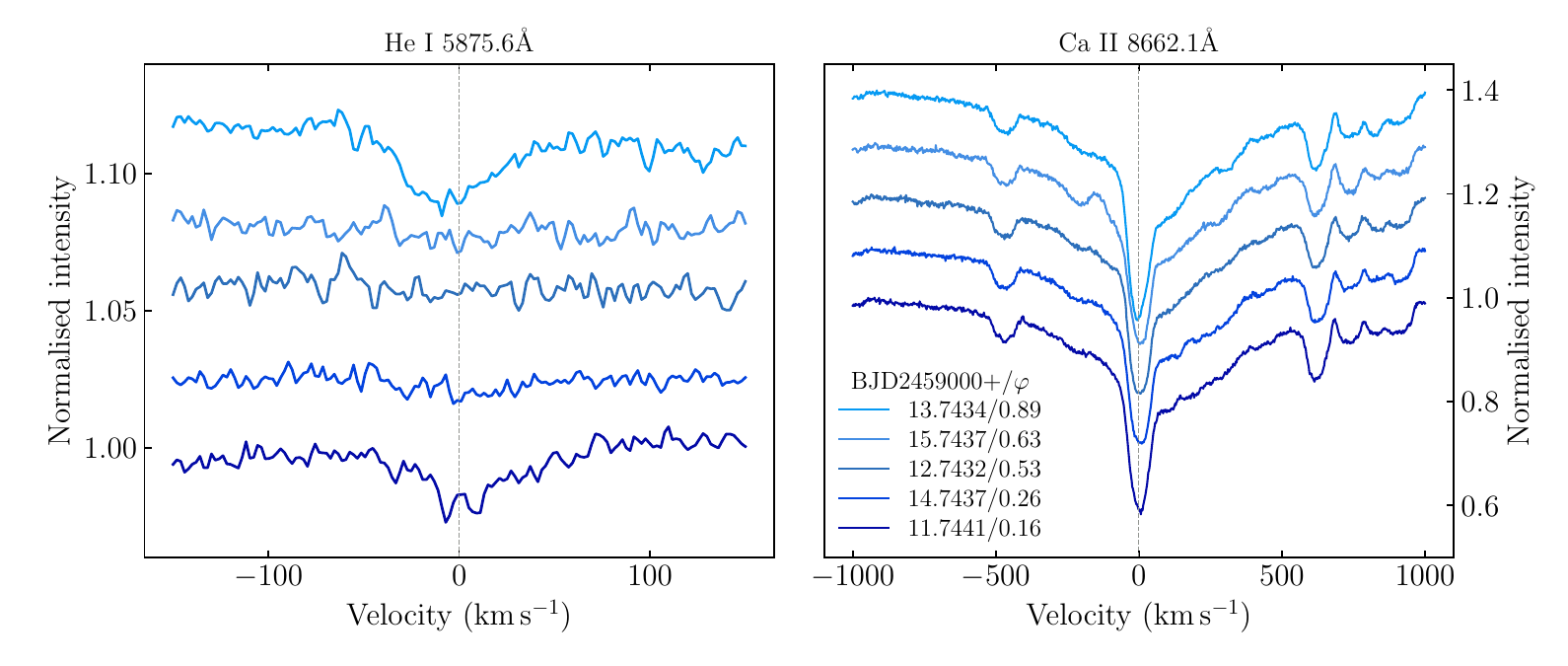}
      \caption[]{Two spectral regions centred on \ion{He}{i} 5875\,\AA\ (left) and \ion{Ca}{ii} 8662\,\AA\ (right), showing LPVs with orbital phase $\varphi$. The velocity scale is corrected for orbital motion.}
    \label{fig:helium}
  \end{figure*}

Our spectra contain a broad spectral feature of variable intensity that appears in the vicinity of the \ion{He}{i} 5875\,\AA\ line in the high-S/N spectra obtained within $\pm0.2\,\varphi_\mathrm{orb}$ of the primary eclipse (left panel of Fig.~\ref{fig:helium}). The measured maximum equivalent width $W_{\lambda}$ of this feature is 170\,m\AA. With \teff=7600\,K, the primary star is too cool to produce the photospheric \ion{He}{i} 5875\,\AA\ line. Despite this, we associate this feature with helium and examine its origin in a broader context in Sect.~\ref{sec:finale}.

\begin{figure*}
  \centering
  \includegraphics[width=0.9\textwidth]{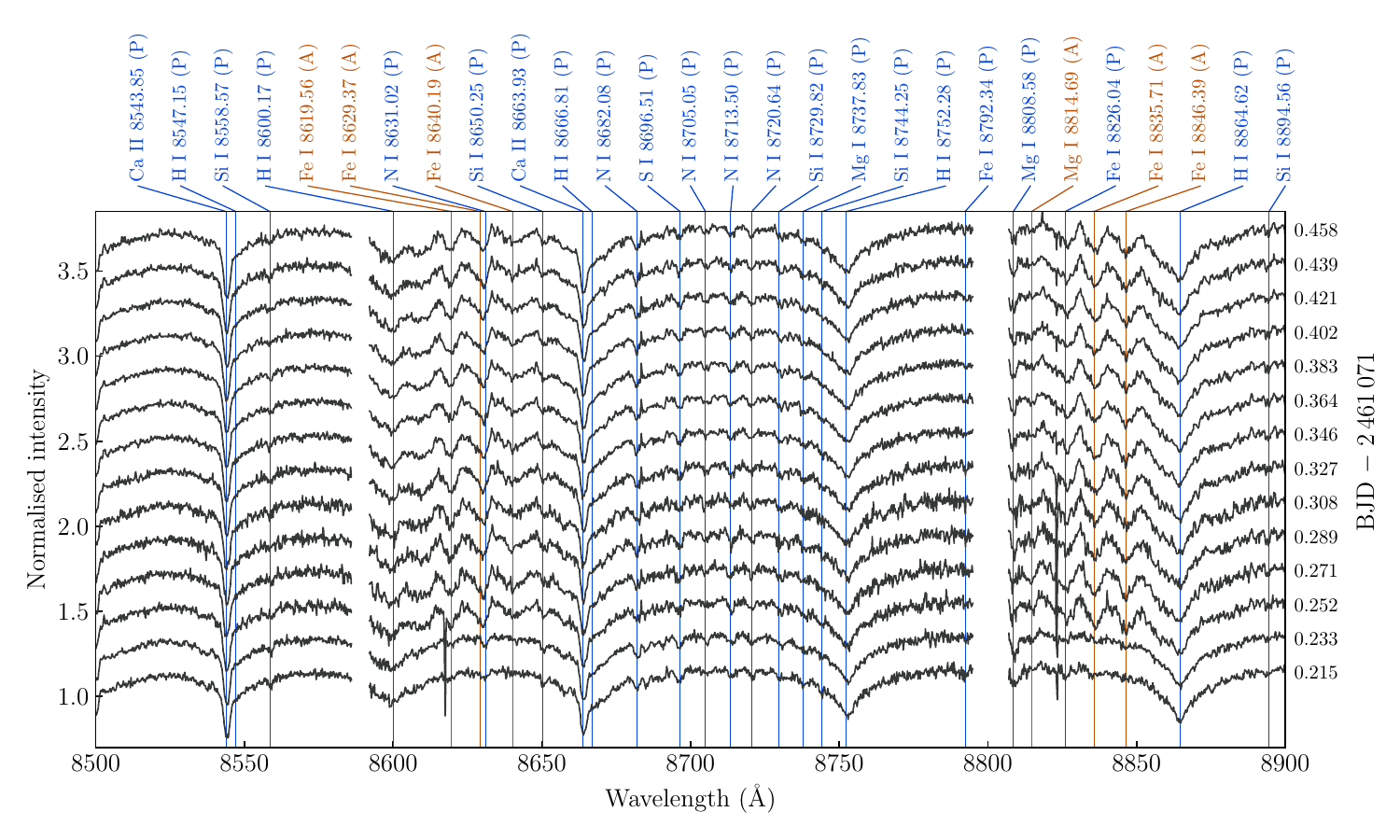}
      \caption[]{Selected series of spectra demonstrating rapid spectral variability resulting in additional absorption features. The absorptions and their tentative identification are labelled with `(A)'. The lines of the primary component are denoted in the same manner with `(P)'.}
    \label{fig:rapidIR}
  \end{figure*}

Finally, the spectral region between the Paschen jump and 9000\,\AA\ contains multiple lines with substantially varying profiles. The most prominent feature here is the \ion{Ca}{ii} 8662\,\AA\ line (right panel of Fig.~\ref{fig:helium}). Given the cool companion's contribution to this region, the variation is likely partly related to binarity. The secondary component may also be responsible for the phenomenon we observed in the data at BJD\,2\,461\,071. In these observations, numerous absorptions developed rapidly over a timescale of $\sim$30 minutes. These absorptions appear as broadened lines with irregular and often asymmetric profiles. Fig.~\ref{fig:rapidIR} illustrates this effect. To identify the lines, we compared the observations with two synthetic spectra computed for \teff$=7600$\,K (primary) and $\approx4500$\,K (secondary). The comparison suggests that the observed absorptions are most likely abnormally enhanced high-excitation-potential ($\chi>2$\,eV) \ion{Fe}{i} lines belonging to the secondary star, shifted by $\simeq+200$\,\kms\ from the rest position. All identified features have similarly broadened asymmetric profiles with a typical $W_{\lambda}\approx1$\,\AA. The results of the tentative identification are labelled as `(A)' in Fig.~\ref{fig:rapidIR}. For comparison, we also show several reference lines of the primary component. A possible explanation of this phenomenon is discussed in Sect.~\ref{sec:finale}.

\subsubsection{Model of spot variability} \label{Spotmodel} 

\begin{figure}[t]
\begin{center}
\includegraphics[width=0.48\textwidth]{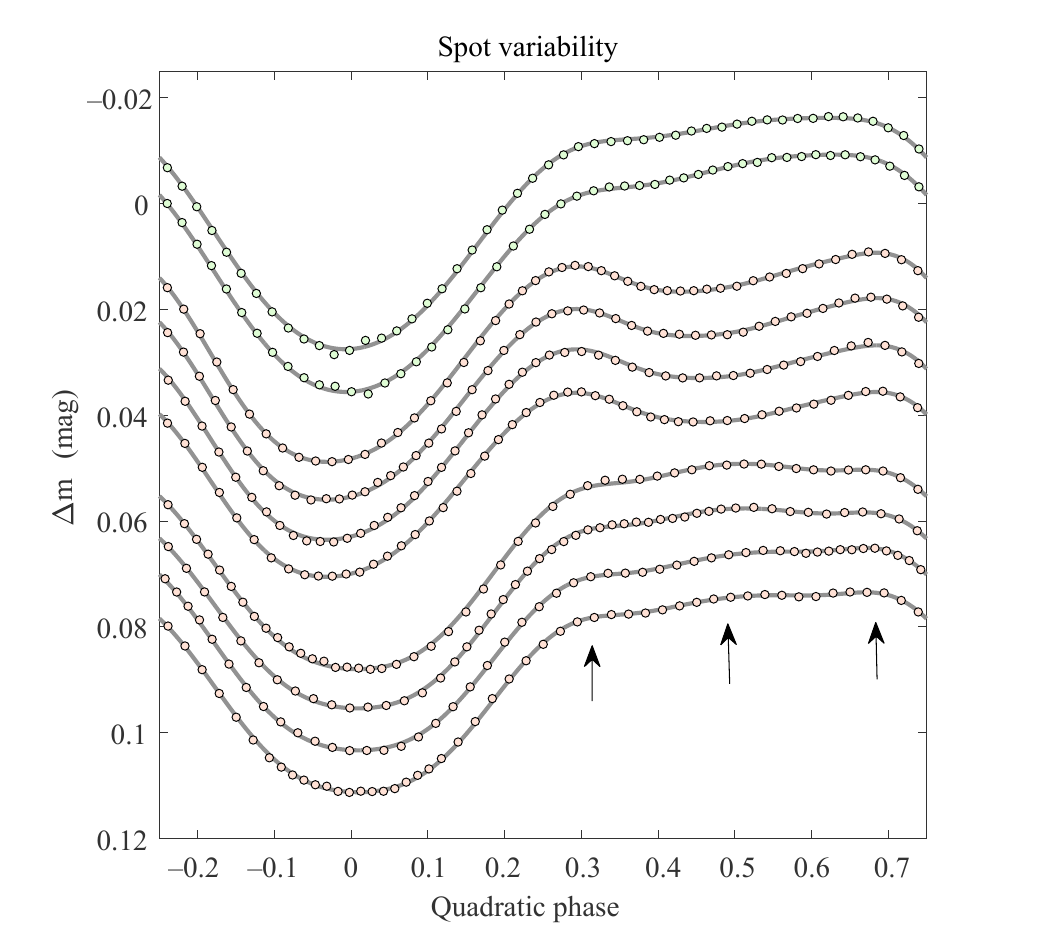}

\caption{Long-term evolution of the \object{HD\,99458} light curve after removing eclipses and $\delta$\,Sct pulsations. The phase curves derived from K2 (2 curves), TESS S45 and S46 (2 curves each), and S72 (4 curves) can be interpreted as evidence of three hot surface spots located at phases 0.3, 0.5, and 0.7, each with a variable amplitude. The phenomenological model is defined by the relation (\ref{spotky}).}
\label{Spotvar}
\end{center}
\end{figure}

The main and most controversial part of the phenomenological model of photometric variability is the residual after subtraction of eclipses (Sect. \ref{sec:eclipmod}) and pulsations (Sect. \ref{sec:lcdSct}). It manifests as a periodic, asymmetric modulation of the system brightness, characterised by a phase curve that evolves with time. Formally, this signal can be approximated by a fourth-order harmonic polynomial with time-varying parameters.
An alternative, more physically intuitive description represents this variability as a linear combination of three symmetric bright (warmer) surface spots with fluctuating intensities located near quadratic phases 0.3, 0.5, and 0.7 (Fig.~\ref{Spotvar}). The corresponding phenomenological model is
\begin{eqnarray}
        \displaystyle F_{\mathrm{s}}(\vartheta,B,d,\varphi_{0})=\sum_{i=1}^{3} B_i(t)\szav {\exp\hzav{1-\cosh\zav{\frac{\varphi_i}{d_i}}}-2.289\,d_i},\nonumber\\
     \mathrm{where}\ \ \ \varphi_i=\zav{\vartheta-\varphi_{0i}}-\mathrm{round}\zav{\vartheta-\varphi_{0i}}, \label{spotky}
\end{eqnarray}
where $\varphi _{0 i}$ and $d_i$ denote the instantaneous position and half-width of the $i$-th spot, respectively, and $\vartheta$ is the quadratic orbital phase defined from the eclipses. The instantaneous parameter values derived for the individual data groups are given in Table~\ref{spottab}. From this table, one can conclude that the spot amplitudes $B_i(t)$ evolve on a timescale of about 10 days.

If our hypothesis about the spots is correct, their variability would be inconsistent with long-term stability of chemical spots observed in chemically peculiar stars \citep{2019MNRAS.487.4230S}. Instead, such formations could represent sites of elevated temperature where, for example, accreting streams strike the stellar surface. Although the development of the spot component of the light curve appears substantial in Fig.~\ref{Spotvar}, a detailed analysis of the data in Table~\ref{spottab} demonstrates only marginal changes in the parameters of the three identified spots. If the spots are associated with accretion onto the stellar surface, the corresponding process in \object{HD\,99458} may therefore be relatively stationary.

\begin{table*}
\small
\centering
\caption[]{Phenomenological parameters of the three bright spots.}
\label{spottab}
\begin{tabular}{c | c c c | c c c | c c c}
\hline\hline
 S     &    $\varphi_{01}$ &  $d_1$  &  $B_1$ & $\varphi_{02}$&  $d_2$   &  $B_2$    &  $\varphi_{03}$ &  $d_3$ &  $B_3$   \\
       &                   &        &  (mag) &               &      & (mag)     &
       &          & (mag) \\
 \hline
K2        &    0.3178(20)      &  0.1432(13)  & $-0.03937$(8)   &  0.4977(15)   &  0.0849(18)  & $-0.01105$(8)  &    0.6755(17)  &    0.1376(15) & $-0.04461$(8)\\
S45, S46&    0.2928(4)      &  0.1369(4)   & $-0.03665$(4)   &   0.5153(9)   &  0.0806(4)   & $-0.01834$(4)  &  0.6981(7)     &    0.1118(6) & $-0.03750$(5)  \\
S72 &    0.3144(8)       & 0.1216(6)    &  $-0.03351$(4)  &   0.4956(7)   &  0.0812(4)   & $-0.01889$(4)  &    0.6856(6)   &    0.1283(5)   & $-0.03772$(4)  \\
\hline
median     &  0.314 & 0.137 & $-0.036$ & 0.498 & 0.081 & $-0.018$ & 0.686 & 0.128 &  $-0.038$ \\
\hline\hline
\end{tabular}
\tablefoot{Data sources S: Kepler K2, TESS S45, S46, and S72. The parameters $\varphi_{0j}$, $d_j$, and $B_j$, where $j=1,2,3$, denote the central phase, half-width, and amplitude of the $j$th spot, respectively. Sizes of individual spots vary on timescales of 10--20~days.}
\end{table*}

Detailed analysis of the phase light curves of magnetic chemically peculiar stars obtained from photometric satellite observations identifies relatively narrow brightness dips that persist in position and shape over many decades. These photometric dips are observed in the light curves of many mCP stars \citep[e.g.][]{Semen_24}. The most likely explanation for this phenomenon is the presence of semi-transparent clouds of charged particles frozen in the corotating magnetospheres of the stars, as proposed by \citet{2020pase.conf...46M} and developed by \citet{2022A&A...659A..37K} and \citet{2024bss..confE..11M}. We also tested \object{HD\,99458} for the presence of these dips.

\begin{figure}[t]
\begin{center}
\includegraphics[width=\columnwidth]{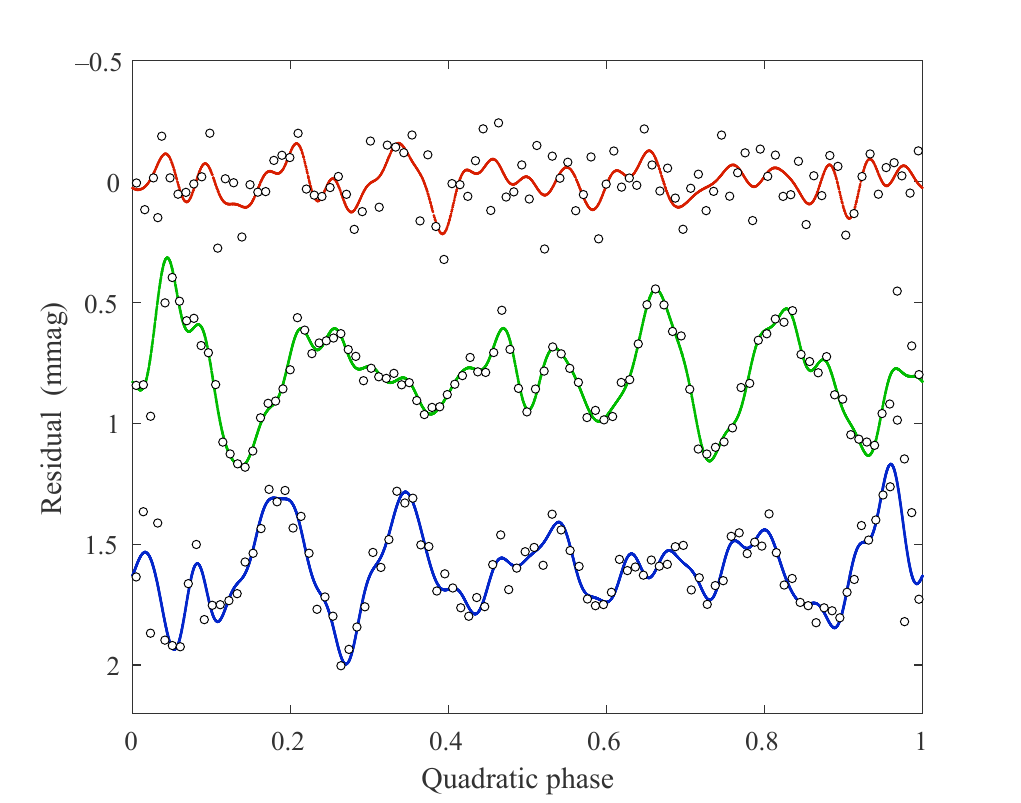}
\caption{Light curves shown relative to the quadratic ephemeris and corrected for occultations, trends, spot variability, and pulsations still exhibit residual variability with amplitudes up to 1\,mmag. Red, green, and blue lines correspond to the K2, TESS S45+S46, and S72 observations, respectively.}
\label{dips}
\end{center}
\end{figure}

Fig.~\ref {dips} shows the residual light curves after subtracting eclipses, long-term trends, spot variability, and pulsations for three data groups, each with a gap of several years. Since these curves do not appear to correlate, they are likely artefacts arising from the removal of the aforementioned variability mechanisms. This again confirms that the studied object is not a magnetic CP star, as other, primarily spectroscopic investigations have conclusively shown.

\subsection{Observational manifestation of stellar pulsation}

\subsubsection{$\delta$\,Sct pulsations in photometry and radial velocities}\label{sec:lcdSct}

The shape of the \object{HD\,99458} light curves is determined mainly by strictly periodic brightness variations caused by eclipses, as discussed in Sect. \ref{sec:eclipmod}, and by the spots considered in the previous section. As \citet{2019MNRAS.487.4230S} first noted, the observed light curve of \object{HD\,99458} is also modulated significantly by multiperiodic, $\delta$\,Sct-type pulsations. In our study, this is convincingly demonstrated by the TESS S72 light-curve detail shown in Fig.~\ref{Detail_072}.
\begin{figure}
    \centering
    \includegraphics[width=\columnwidth]{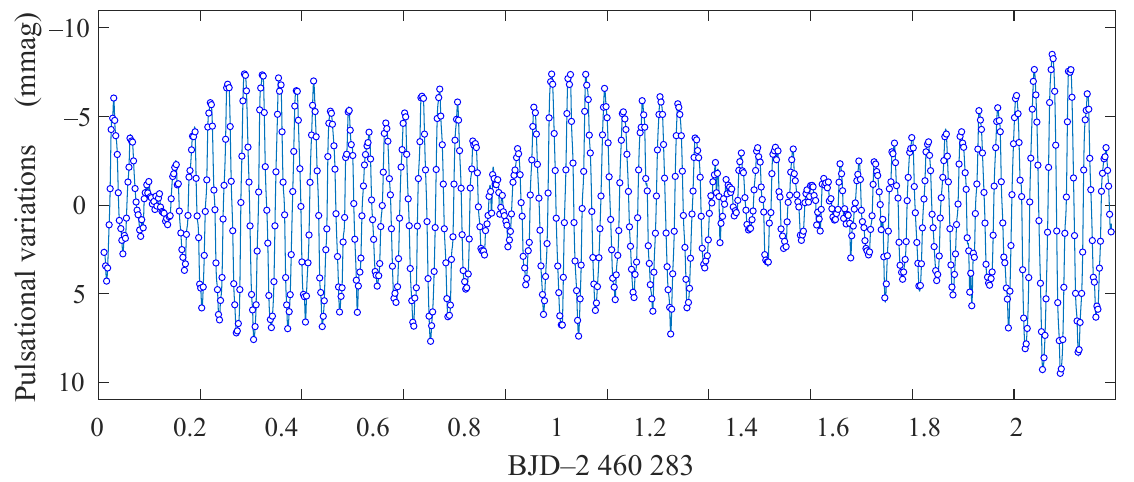}
    \caption{Disentangled pulsational variability during a two-day interval.}
    \label{fig:pulsolo}
\end{figure}

To study the pulsation component of stellar variability, we divided the data into three time-restricted groups: observations from K2, then from TESS sectors 45 and 46, and finally from sector 72, which we investigated separately. We assumed that light changes are periodic with the quadratic ephemeris, approximated by a smooth light curve, while we approximated the pulsations as a sum of sinusoidal changes with different frequencies and amplitudes (Fig.~\ref{fig:pulsolo}). By analysing the frequency spectrum of the variability, we identified several dozen frequencies in each group. In Table~\ref{tab:frequencies}, we summarise the reliably identified frequencies and their characteristics.

Multiperiodic pulsations are also visible in the measured radial velocities. After subtracting the orbital variation (Sect.~\ref{sec:system}), we performed a standard Fourier analysis of the residuals using the \texttt{Period04} programme \citep{2005CoAst.146...53L}. By prewhitening the time series obtained with MRES, PLATOSpec, and Whoppshel, we detected the nine pulsation frequencies listed in Table~\ref{tab:specfreq}. Seven of these frequencies are potentially cross-identified in Table~\ref{tab:frequencies}.

\begin{table}
\begin{center}
\caption{Pulsation frequencies derived from the radial velocity time series.}\label{tab:specfreq}
\begin{tabular}{l c c c}
\hline\hline
ID & $f$   &  $A$   & CID \\
   & ($\pm0.05$\,\cpd) & ($\pm0.03$\,\kms)  &  \\
\hline
$f_\mathrm{s1}$   & 28.98  & 0.80  & $f_{16}$  \\
$f_\mathrm{s2}$   & 26.77  & 0.56  & $f_{46}$  \\
$f_\mathrm{s3}$   & 32.95  & 0.55  & $f_{94}$  \\
$f_\mathrm{s4}$   & 33.05  & 0.50  & $f_{56}$ or $f_{106}$  \\
$f_\mathrm{s5}$   & 26.32  & 0.48  & $f_{65}$  \\
$f_\mathrm{s6}$   & 31.60  & 0.39  & $f_{29}$  \\
$f_\mathrm{s7}$   & 32.70  & 0.37  & $f_{49}$  \\
$f_\mathrm{s8}$   & 31.24  & 0.29  &           \\
$f_\mathrm{s9}$   & 23.83  & 0.27  &           \\
\hline\hline
\end{tabular}
\tablefoot{Columns give mode ID, frequency $f$ (in \cpd), amplitude $A$ (in \kms), and possible cross-identification (CID) with frequencies listed in Table~\ref{tab:frequencies}.}
\end{center}
\end{table}

\subsubsection{Pulsation-induced LPV}
First high-S/N spectra of \object{HD\,99458} obtained with the MSS and HERMES in 2019 revealed line-profile variability characteristic of pulsating $\delta$\,Sct stars. Since then, we collected nine time-resolved datasets, including seven nights with MRES and one night each with HERMES and PLATOSpec, as listed in Table~\ref{tab:log_series}. The observations sample a range of orbital phases: three datasets are close to primary eclipse, one is near secondary eclipse, and two are nearly symmetric between secondary eclipse and quadratures.

From the beginning of our study, we aimed to test the influence of the close secondary body on stellar pulsations. Recent progress in interpreting tidally perturbed pulsations achieved in \citet{2024MNRAS.528.3378Z}, \citet{2024ApJ...975..121J}, and most recently \citet{2025ApJ...979...80F} has reinforced the relevance of our work. Given the good coverage of orbital phases and the high temporal resolution of the data collected in our study, we can directly assess the spectroscopic manifestation of tidally perturbed pulsations across different configurations.

\begin{table}
\small
\centering
\caption{Datasets used in seismic LPV analysis, sorted by average orbital phase $\varphi_\mathrm{orb}$.}
\label{tab:log_series}
\begin{tabular}{lcccc}
\hline\hline
ID & BJD (first--last)  & $\varphi_\mathrm{orb}$ (first--last)
 & $N$  & Instrument \\
\hline
DS1 & 2\,461\,069.197--.426  & 0.964--0.048  & 62 & MRES   \\
DS2 & 2\,461\,102.116--.427  & 0.056--0.170  & 84 & MRES   \\
DS3 & 2\,461\,072.215--.462  & 0.072--0.163  & 67 & MRES   \\
DS4 & 2\,458\,859.381--.462  & 0.185--0.214  & 12 & MRES   \\
DS5 & 2\,458\,549.545--.623  & 0.357--0.386  & 20 & HERMES \\
DS6 & 2\,461\,070.220--.467  & 0.340--0.431  & 67 & MRES   \\
DS7 & 2\,458\,860.313--.475  & 0.527--0.586  & 23 & MRES   \\
DS8 & 2\,460\,850.475--.576  & 0.621--0.658  & 22 & PLATOSpec \\
DS9 & 2\,461\,071.214--.469  & 0.705--0.799  & 69 & MRES   \\
\hline\hline
\end{tabular}
\tablefoot{$N$ denotes the number of spectra in each dataset.}
\end{table}

\begin{figure*}
  \centering
  \includegraphics[width=0.95\textwidth]{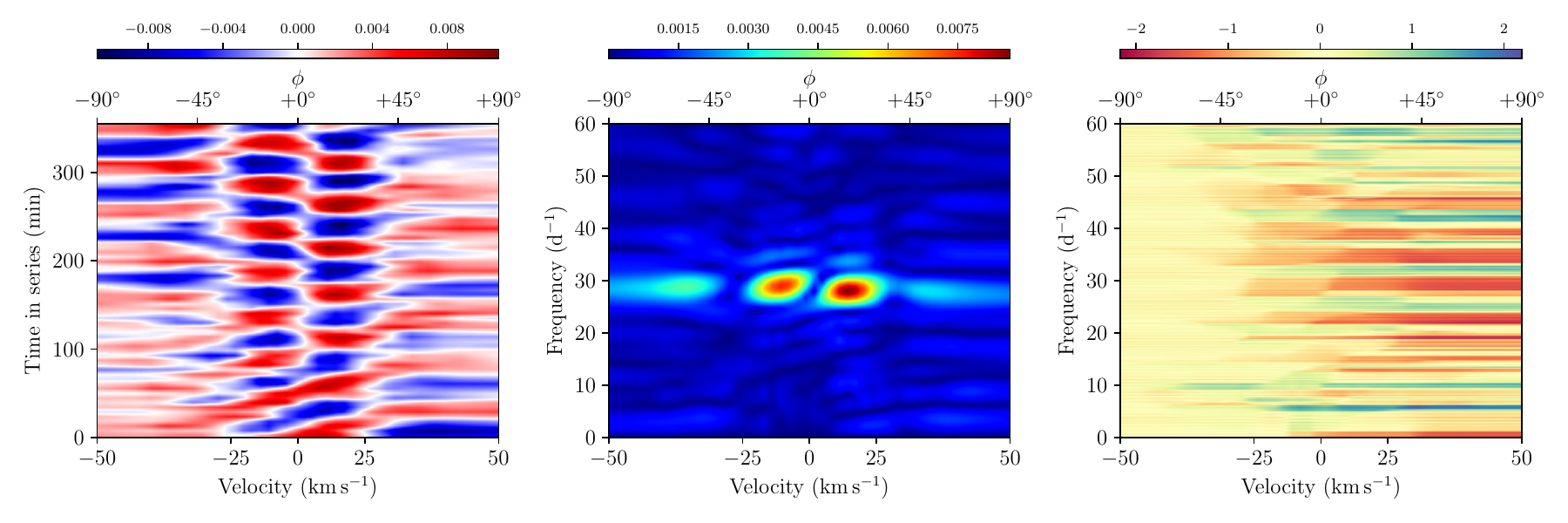}
      \caption[]{2D-DFT summary of LPV analysis of dataset DS6. The corresponding range of $\varphi_\mathrm{orb}$ is 0.340--0.431. The left panel shows LPV in residual spectra, the middle panel demonstrates the amplitude in the 2D Fourier transform of LPV, and the right panel shows the corresponding phase.}
  \label{fig:2DDFT_main}
\end{figure*}

For LPV analysis, we use a pixel-to-pixel technique based on the two-dimensional Discrete Fourier Transform (2D-DFT). The implementation follows the principles developed in \citet{1992ApJ...400L..71K}, \citet{1994PhDT........40K}, and \citet{1997A&A...317..723T}. Our variant of the 2D-DFT maps pulsations across the line profile and extracts the amplitude and phase of the selected modes, which are required for spectroscopic mode identification. The analysis is combined with the method developed by Mkrtichian et al. (in prep.) for reconstructing the surface nodal structure of tidally inclined pulsations in oEA stars. Fig.~\ref{fig:2DDFT_main} shows the three-panel summary diagram produced by our 2D-DFT code for dataset DS6.

The left panel of Fig.~\ref{fig:2DDFT_main} illustrates the average-subtracted line profiles. The vertical axis of the panel shows time in minutes that elapsed from the beginning of the series. The top axis in all panels gives the apparent longitude $\phi$, spanning from $-90\degr$ to $+90\degr$, with $\phi=0$ at the centre of the visible hemisphere. The lower axis of the panels shows the velocity scale, which varies across the line profile from $-$\vsini\ to \vsini. The middle panel demonstrates the amplitude of the Fourier transform of the LPV computed in the time domain. The corresponding phase is shown in the right panel.

In LPV data, pulsation modes produce characteristic patterns resembling waves travelling across the line profile. Propagation from negative to positive longitude $\phi$ corresponds to a prograde mode, whereas propagation in the opposite direction indicates a retrograde mode. In the Fourier domain (middle panels), the pulsation appears as a signal at the frequency $f_\mathrm{2D}$. For non-radial modes, the amplitude distribution of the LPV exhibits alternating maxima and minima associated with the nodes and antinodes of the pulsation pattern, the number and spacing of which are related to the azimuthal order $m$. The spherical degree $\ell$ can be estimated from the phase variation across the line profile following schemes described in \citet{1997A&AS..121..343S} and \citet{1997A&A...317..723T}.

\begin{figure}
  \centering
  \includegraphics[width=0.9\columnwidth]{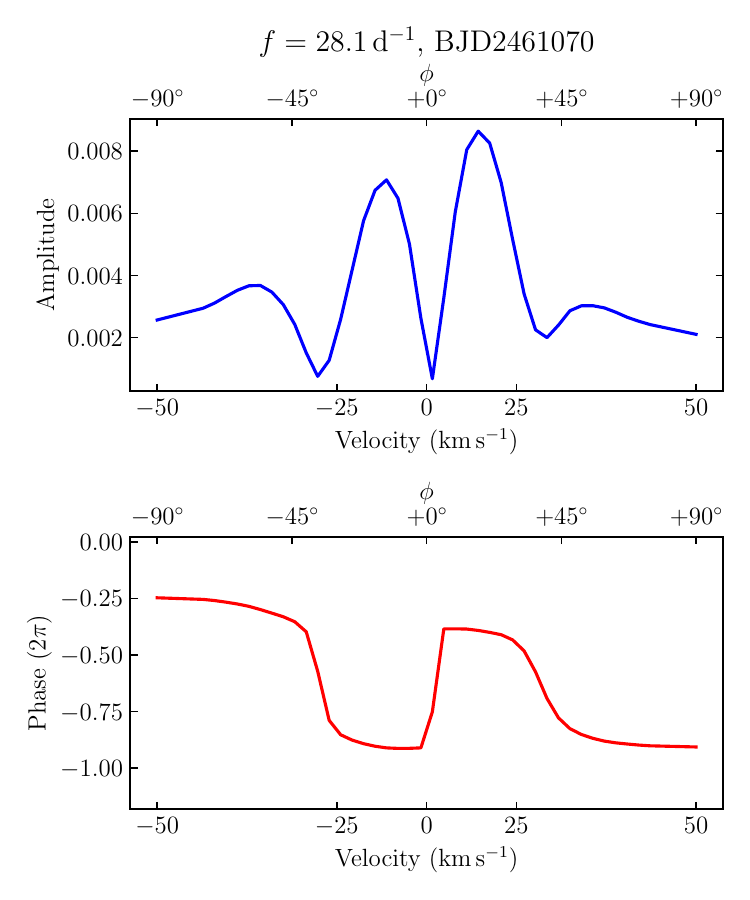}
      \caption[]{Amplitude and phase variations of two modes near $f=28.1$\,\cpd\ observed in dataset DS6.}
  \label{fig:amp-phase}
\end{figure}

The observed LPV pattern of \object{HD\,99458} is considerably more complex than the ideal case described above, with traces of multiple modes, including both prograde and retrograde components. Eclipses, changes in orbital configuration, and limited temporal resolution further complicate the interpretation. However, the most striking feature of the LPV in our observations is the presence of standing waves, clearly visible in the middle panel of Fig.~\ref{fig:2DDFT_main}. Standing waves are a distinctive feature of gravitationally perturbed pulsators \citep{2025ApJ...979...80F}. The 2D-DFT diagrams for the remaining datasets (Sect.~\ref{ap:dft}) show how the pulsation pattern appears at different orbital phases. In each dataset, there is at least one pair of bright antinodes with a phase difference of $\pi$ between them, as shown in Fig.~\ref{fig:amp-phase}. More regularly, however, the Fourier phase exhibits a staircase-like distribution. By localising the antinodes on the surface of the pulsating primary, we sought to identify the corresponding modes.

To identify the spectroscopic modes, we first measured the apparent frequencies $f_\mathrm{2D}$, amplitudes, and apparent longitudes $\phi$ of the antinodes listed in Table~\ref{tab:2DDFT_result}. Due to differences in temporal resolution and coverage across the datasets, the accuracy of the extracted values varies. The typical frequency uncertainty is of order 1\,\cpd\ and depends strongly on the spot's angular position. The position of the antinode also defines the accuracy of the longitude measurements. Antinodes located closer to the stellar limb are more strongly affected by projection effects. The quoted uncertainty $\pm5\degr$ in $\phi$ therefore reflects only the accuracy with which the antinode's position can be inferred from the diagram, and does not account for the underlying physics.

\begin{table}
\renewcommand{\arraystretch}{1.05}
\centering
\caption{Apparent longitudes, frequencies, and amplitudes of the antinodes measured from the 2D-DFT diagrams.}
\label{tab:2DDFT_result}
\begin{tabular}{ccc | ccc}
\hline\hline
$\phi$  & $f_\mathrm{2D}$  & $A$  & $\phi$  & $f_\mathrm{2D}$  & $A$  \\
($\pm5\degr$)  & ($\pm1.0$\,\cpd)  & ($\times10^{-3}$) & ($\pm5\degr$)  & ($\pm1.0$\,\cpd)  & ($\times10^{-3}$)  \\
\hline
\multicolumn{3}{c|}{DS1 ($-13.0$--17.3)} &  \multicolumn{3}{c}{DS5 (128.5--139.0)} \\
$-41$  &  28.1 & 4.6                    &  $-39$  & 26.1  & 1.9 \\
$-34$  &  34.1 & 1.9                    &  $-6$   & 27.9  & 2.5 \\
$-14$  &  33.0 & 2.5                    &  $+11$  & 35.7  & 1.7 \\
$-11$  &  27.1 & 4.8                    &  $+26$  & 26.6  & 3.9 \\
$+18$  &  27.2 & 3.3                    &  $+55$  & 28.5  & 2.5 \\
$+56$  &  28.3 & 3.3                    &  \multicolumn{3}{c}{DS6 (122.4--155.2)} \\
\multicolumn{3}{c|}{DS2 (20.2--61.2)}    &  $-58$  & 28.9  & 3.9 \\
$-22$  &  28.6 & 7.5                    &  $-14$  & 29.0  & 7.2 \\
$+8$   &  29.5 & 7.6                    &  $+19$  & 28.1  & 8.2 \\
$+13$  &  33.7 & 3.5                    &  $+56$  & 28.0  & 3.1 \\
$+41$  &  27.0 & 5.5                    &  \multicolumn{3}{c}{DS7 (189.7--211.0)} \\
\multicolumn{3}{c|}{DS3 (25.9--58.7)}    &  $-39$  & 32.3  & 1.5 \\
$-55$  & 27.3  & 4.4                    &  $-2$   & 30.2  & 2.3 \\
$-44$  & 31.3  & 2.8                    &  $+30$  & 28.0  & 2.2 \\
$-15$  & 27.6  & 6.3                    &  \multicolumn{3}{c}{DS8 (223.6--236.9)} \\
$+4$   & 33.4  & 3.7                    &  $-25$  & 30.1  & 3.9 \\
$+24$  & 29.1  & 3.2                    &  $+11$  & 29.1  & 4.8 \\
$+44$  & 30.1  & 3.3                    &  $+35$  & 28.3  & 4.1 \\
\multicolumn{3}{c|}{DS4 (66.6--77.0)}    &  \multicolumn{3}{c}{DS9 (253.8--287.6)} \\
$-26$  & 29.7  & 1.2                    &  $-17$ & 26.4  & 5.6 \\
$+16$  & 26.9  & 1.5                    &  $-17$ & 31.6  & 5.2 \\
$+59$  & 31.3  & 0.8                    &  $+3$  & 32.7  & 6.0 \\
       &       &                        &  $+9$  & 32.2  & 6.3 \\
       &       &                        &  $+17$ & 26.9  & 5.9 \\
\hline\hline
\end{tabular}
\tablefoot{Range of phase angles spanning the time series is given in parentheses for each dataset.}
\end{table}

\begin{figure}
  \centering
  \includegraphics[width=\columnwidth]{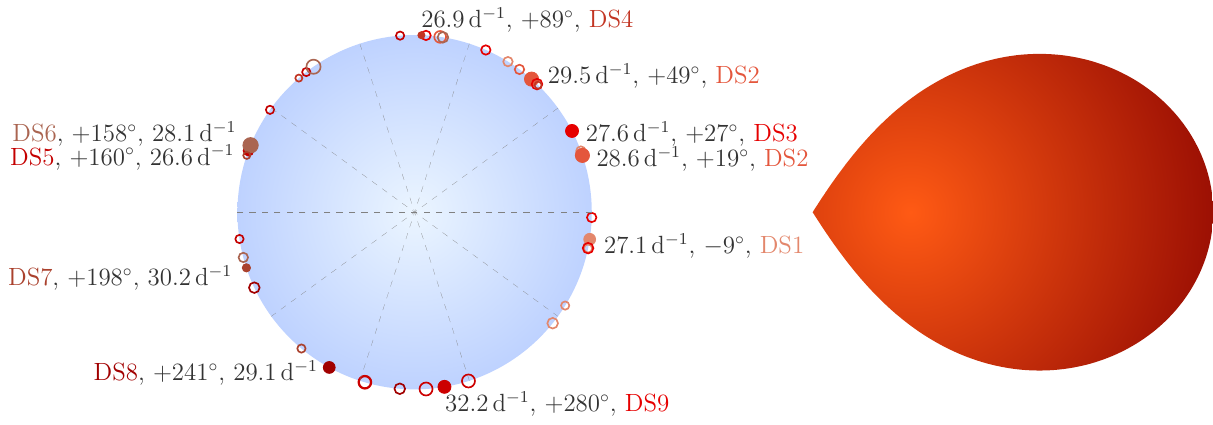}
      \caption[]{Pole-on projection showing the distribution of surface antinodes on the pulsating component, derived from 2D-DFT diagrams. Datasets are colour-coded. The filled, labelled circles indicate the highest-amplitude peak in each dataset. Labels contain the dataset ID, frequency, and physical longitude of the antinode. Dashed lines identify nodal positions for the azimuthal order $|m|=5$. The secondary component is shown for reference.}
  \label{fig:pulsamap}
\end{figure}

In the second step, the antinode positions were mapped onto the stellar surface by correcting the apparent longitude $\phi$ for the orbital phase of the system. This procedure revealed zones where the antinodes preferentially occur (Fig.~\ref{fig:pulsamap}).

Finally, we averaged the positions of the individual antinode groups and compared their distribution with sectoral patterns containing different numbers of sectors. The best agreement was obtained for patterns corresponding to azimuthal orders $|m|=5\pm1$, implying $\ell \geq |m| \geq 4$. The inferred pulsation pattern itself appears stable and synchronised with the orbital motion. However, we cannot exclude that some peaks originated from overlapping low-degree modes, as the frequencies in Table~\ref{tab:2DDFT_result} form two distinct clusters below and above $f_\mathrm{2D}\approx30$\,\cpd. To resolve this ambiguity, detailed mode modelling is required, which is beyond the scope of this paper.

\subsection{Seismic modelling} \label{plus_model}
The detection of $\delta$\,Sct-type pulsations makes \object{HD\,99458} a suitable target for asteroseismic modelling. Determining the star's evolutionary status places an important constraint.

The luminosity of \object{HD\,99458} can be estimated from the apparent brightness $V = 8.16(9)$\,mag \citep{2000A&A...355L..27H} and the parallax $\pi = 3.827(41)$\,mas \citep[Gaia DR3;][]{2016A&A...595A...1G,2021A&A...649A...1G}. The 3D reddening maps of \cite{2018MNRAS.478..651G} give the colour excess $E(B-V) = 0.033(17)$\,mag at the star's parallax-based distance of $261.3\pm2.8$\,pc. Adopting the total-to-selective extinction ratio for the Johnson $V$ band, $R_{\mathrm{V}} = 3.089$ \citep{2004ASPC..309...33F}, we derive an extinction of $A_{\mathrm{V}} = 0.103(52)$\,mag. The bolometric correction for the Johnson $V$ band, taken from \cite{1996ApJ...469..355F}, is $BC_{\mathrm{V}} = 0.0321(21)$, which yields a luminosity of $\log{L/L_{\sun}} = 1.489(31)$. The corresponding position in the Hertzsprung–Russell diagram (Fig.~\ref{fig:HR}) places \object{HD\,99458} in the category of evolved stars.

\begin{figure}
   \centering
	    	\includegraphics[width=1\columnwidth]{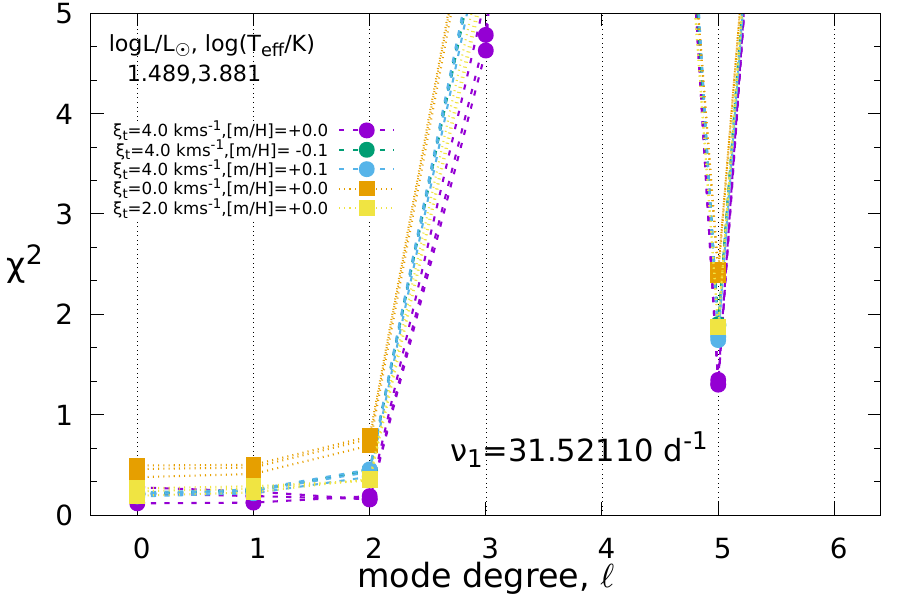}
	    \caption{$\chi^2$ as a function of the mode degree, $\ell$, for the dominant frequency $f_{1} = 31.52110$\,\cpd.}
	    \label{fig:mode_ident}
\end{figure}

Available Str\"omgren $bvy$ photometry allowed us to perform preliminary mode identification. We applied the method developed by \citet{JDD2003}, which provides constraints on the spherical degree $\ell$ of the pulsation modes. In particular, we constrained the mode degree of the dominant frequency, $f_{1} = 31.52110$\,\cpd\ from Table~\ref{tab:frequencies}. The results are presented in Fig.~\ref{fig:mode_ident}, where the $\chi^2$ values are shown as a function of $\ell$. Here, $\chi^2$ quantifies the goodness of fit between observed and calculated amplitudes and phases.

The calculated amplitudes and phases depend slightly on the adopted atmospheric models. In this work, we used the Vienna model atmospheres \citep{2002A&A...392..619H} and adopted the non-linear limb-darkening law of \citet{2000A&A...363.1081C}. We also examined the effects of different metallicities, $[\mathrm{m}/\mathrm{H}]$, and microturbulent velocities, $\xi_\mathrm{t}$. The effective temperature, adopted from the spectroscopic analysis, and the luminosity were fixed at the centre of the observational error box.

Our results indicate that in the observed pulsation spectrum, modes with $\ell = 0$, $1$, and $2$ are all plausible. Although the photometric mode identification is not fully conclusive and cannot replace modelling, it does not rule out the $\ell = 2$ hypothesis for the dominant mode, which is independently suggested by the seismic models discussed in the next section.

\subsubsection{Multiplets}

\begin{figure*}
   \sidecaption
	\includegraphics[width=12cm]{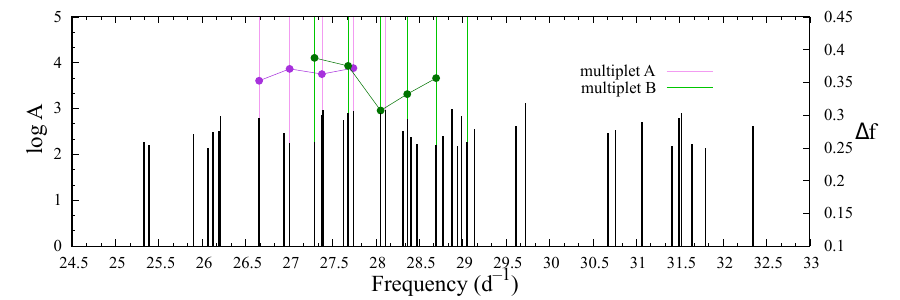}
	\caption{Frequency spectrum of \object{HD\,99458}. Left-hand y-axis shows the amplitudes of the detected modes. Frequencies belonging to multiplet~A are indicated by violet lines, while those associated with multiplet~B are marked in green. Right-hand y-axis shows the frequency separations between consecutive multiplet components, indicated by circles. Only frequencies derived from the K2 observations are shown.}
	\label{fig:pm:df_f}
\end{figure*}

In the frequency spectrum of \object{HD\,99458}, we identified two regular structures that can be interpreted as rotationally split modes of the same spherical degree and radial order, but with different azimuthal orders. The frequencies belonging to multiplet~A and multiplet~B are indicated in Fig.~\ref{fig:pm:df_f}. The frequency separations between consecutive components are also shown on the right-hand y-axis. Both structures exhibit comparable frequency spacings and span a similar frequency range. The individual components of the multiplets are listed in Table~\ref{tab:pm:fs}. For multiplet analysis, we adopted the $f_{\rm best}$ frequencies listed in Table~\ref{tab:frequencies}. The K2 spectrum shown in Fig.~\ref{fig:pm:df_f} is used only to illustrate the frequency pattern.

Such regular patterns enable mode identification even when observations are limited to a single photometric band. Multiplet~A consists of five components and can be interpreted as the complete set of rotationally split quadrupole modes ($\ell = 2$). Multiplet~B comprises six frequencies, generally with lower amplitudes, and may correspond to octupole modes ($\ell = 3$). However, this multiplet is less regular than multiplet~A, and one component appears to be missing, as $\ell = 3$ modes are expected to split into seven components.

\begin{table}
	\begin{center}
		\caption{Frequencies, amplitudes, and frequency separations of the
		multiplets detected in \object{HD\,99458}.}
		\label{tab:pm:fs}
		\begin{tabular}{c c c c c c}
			\hline\hline
			ID &
			$f$ &
			$\sigma_f$ &
			$A$ &
			$\Delta f$ &
			$\sigma_{\Delta f}$ \\
			&
			(\cpd) &
			(\cpd) &
			(mmag) &
			(\cpd) &
			(\cpd) \\
			\hline

			\multicolumn{6}{c}{multiplet A} \\

			$f_{20}$ & 26.65500 & 0.00030 & 0.640 & 0.3489 & 0.00042 \\
			$f_{21}$ & 27.00390 & 0.00030 & 0.580 & 0.3828 & 0.00036 \\
			$f_{4}$  & 27.38670 & 0.00020 & 1.550 & 0.3549 & 0.00022 \\
			$f_{2}$  & 27.74160 & 0.00010 & 1.800 & 0.3710 & 0.00022 \\
			$f_{5}$  & 28.11260 & 0.00020 & 1.460 & ---    & --- \\

			\multicolumn{6}{c}{multiplet B} \\

			$f_{42}$ & 27.29660 & 0.00090 & 0.210 & 0.3840 & 0.00092 \\
			$f_{7}$  & 27.68060 & 0.00020 & 1.180 & 0.3749 & 0.00028 \\
			$f_{9}$  & 28.05550 & 0.00020 & 1.030 & 0.3070 & 0.00028 \\
			$f_{12}$ & 28.36250 & 0.00020 & 0.920 & 0.3311 & 0.00073 \\
			$f_{32}$ & 28.69360 & 0.00070 & 0.300 & 0.3579 & 0.00086 \\
			$f_{30}$ & 29.05150 & 0.00050 & 0.350 & ---    & --- \\

			\hline\hline
		\end{tabular}
        \tablefoot{The uncertainties of the frequency separations were calculated as
		$\sigma_{\Delta f} =
		(\sigma_{f_i}^2+\sigma_{f_{i+1}}^2)^{1/2}$.}
	\end{center}
\end{table}

\subsubsection{Evolutionary models}\label{Sect:pmem}

Evolutionary tracks were computed using the Modules for Experiments in Stellar Astrophysics (MESA) code \citep{Paxton2011, Paxton2013, Paxton2015, Paxton2018, Paxton2019}, version 12115. Unless stated otherwise, we adopted the OPAL opacity tables \citep{OPAL}, supplemented at low temperatures with data from \cite{2005ApJ...623..585F}. The solar chemical composition was taken from \cite{AGSS09}.

The equation of state (EOS) implemented in MESA is a combination of several formulations, including OPAL \citep{Rogers2002}, SCVH \citep{Saumon1995}, PTEH \citep{Pols1995}, HELM \citep{Timmes2000}, and PC \citep{Potekhin2010}. Nuclear reaction rates were adopted from the JINA REACLIB database \citep{Cyburt2010}, complemented by tabulated weak reaction rates from \cite{Fuller1985, Oda1994, Langanke2000}. Screening effects were included following the prescriptions of \citet{Salpeter1954}, \citet{Dewitt1973}, \citet{Alastuey1978}, and \citet{Itoh1979}, while thermal neutrino loss rates were taken from \citet{Itoh1996}. The evolutionary calculations include atomic diffusion as well as exponentially decaying convective core overshooting \citep{2000A&A...360..952H}.

In a subset of models, we also considered rotationally induced mixing processes and angular momentum transport, including the Solberg--H{\o}iland instability, secular shear instability, Eddington--Sweet circulation, Goldreich--Schubert--Fricke instability, and dynamical shear instability. A detailed description of these mechanisms can be found in \citet{2000ApJ...528..368H}. In these computations, we adopted the scaling factor $f_{\mathrm{c}} = 1/30$, which governs the efficiency of rotational mixing in the diffusion coefficient \citep{1989ApJ...338..424P, 1992A&A...253..173C, 2000ApJ...528..368H}, and $f_\mu = 0.05$, which describes the sensitivity of rotational mixing to the mean molecular weight gradient, $\nabla_\mu$ \citep{2000ApJ...528..368H}.

\subsubsection{Pulsational models}

Adiabatic and non-adiabatic pulsation models were computed using our in-house code, which solves the linearised stellar pulsation equations \citep{2024MNRAS.529.4176W}. The effects of rotation were included using a perturbative approach, by solving first-order perturbed eigenfunctions \citep[e.g.][]{1951ApJ...114..373L,1992ApJ...394..670D,1998A&A...334..911S,2006A&A...449..673S}. Radial differential rotation was also taken into account. The rotational velocity profiles were obtained from MESA, which treats angular momentum transport in a diffusion approximation \citep{1978ApJ...220..279E, 1989ApJ...338..424P, 2000ApJ...528..368H}. We considered modes with degrees $\ell = 0$, $1$, $2$, and $3$, and all corresponding azimuthal orders, $m \in [-\ell, \ell]$.

We aimed to reproduce the observed frequencies of \object{HD\,99458} using models located within the observational error box in the Hertzsprung--Russell diagram. We explored a broad parameter space, covering metallicities in the range $Z = 0.008$--0.02, stellar masses from $M = 1.8\,M_{\sun}$ to $2.2\,M_{\sun}$, and convective core overshooting parameters from $f_{\mathrm{ov}} = 0$ to 0.04. We also tested two values of the initial hydrogen abundance, $X = 0.70$ and $X = 0.725$, as well as two values of the mixing-length parameter, $\alpha_{\mathrm{mlt}} = 1.8$ and $1.0$.

In our modelling, we assumed that multiplet~A is associated with quadrupole modes ($\ell = 2$). We began by fitting the axisymmetric mode $f_{4} = 27.38670$\,\cpd. This frequency can be reproduced by modes of different radial orders, $n$, and therefore the position of the star in the Hertzsprung--Russell diagram was used to constrain the value of $n$. Initially, several solutions were possible, depending on the adopted model parameters. However, further analysis showed that only one solution was consistent with all constraints, corresponding to $n = 7$. Fitting the axisymmetric mode thus imposes a strong constraint on effective temperature (and hence the age) for a given set of parameters, including chemical composition ($Z$, $X$), stellar mass, rotational velocity, mixing-length parameter, and overshooting parameter.

Next, we attempted to reproduce the non-axisymmetric components of multiplet~A. The key parameter controlling frequency splitting is the rotational velocity, $\upsilon_\mathrm{rot}$. By adjusting this parameter and adopting appropriate \teff\ values, we were able to reproduce the observed frequencies of all components of multiplet~A with good accuracy. Fitting both the axisymmetric mode and its rotationally split components effectively constrains two parameters, \teff\ and $\upsilon_{\mathrm{rot}}$.

We then explored models computed with different values of $Z$, $X$, $M$, $\alpha_{\mathrm{mlt}}$, and $f_{\mathrm{ov}}$. By comparing observed and theoretical frequency spectra, we found that the high-amplitude frequency $f_{3} = 29.72200$\,\cpd\ can be uniquely identified as an ($\ell = 2$, $m = -1$, $p_{8}$) mode. In the next step, we determined the stellar mass required to reproduce this frequency. Although $f_{3}$ could also be matched by a ($\ell = 2$, $m = -1$, $p_{9}$) mode if $f_{4}$ were assumed to have $n = 8$, such models lie well outside the observational error box in the Hertzsprung--Russell diagram and were therefore rejected.

We then examined models that simultaneously reproduce the components of the quadrupole multiplet and frequency $f_{3}$, exploring a range of metallicities, initial hydrogen abundances, mixing-length parameters, and overshooting efficiencies. We found that, for a particular metallicity, the dominant frequency $f_{1} = 31.52110$\,\cpd\ can be matched by an ($\ell = 2$, $m = -1$, $p_{9}$) mode.

The final frequency we were able to reproduce accurately is $f_{23} = 26.12570$\,\cpd. Achieving this required adjusting the overshooting parameter. For an appropriate value of $f_{\mathrm{ov}}$, this frequency can be identified as a dipole mode ($\ell = 1$, $m = 0$, $p_{8}$).

We note that the initial hydrogen abundance and mixing-length parameter have only a minor impact on the remaining, unfitted frequencies, and therefore cannot be reliably constrained using this approach. This issue is discussed further below.

A representative seismic model that reproduces the quadrupole multiplet as well as the $f_{3}$, $f_{1}$, and $f_{23}$ frequencies is presented in Table~\ref{tab:parsei} (Model\,1). In this model, we adopt $X = 0.7$, $\alpha_{\mathrm{mlt}} = 1.8$, and OPAL opacities, while the remaining parameters are determined through frequency fitting. Rotationally induced mixing processes are not included in these calculations.

The resulting model requires a relatively large amount of convective core overshooting, $f_{\mathrm{ov}} = 0.03$. It is also slightly hotter than the blue edge of the observational error box in the Hertzsprung--Russell diagram, and its metallicity is subsolar.

The main discrepancy in this model concerns the rotational velocity. We derive a surface rotation of approximately $29$\,\kms, which is significantly lower than the observed projected rotational velocity, $\vsini = 49.9 \pm 2.0$\,\kms. This inconsistency can be alleviated by including rotationally induced mixing and angular-momentum transport. MESA provides several such mechanisms, as described in Sect.~\ref{Sect:pmem}.

Rotationally induced mixing processes modify both the chemical composition profile and  angular momentum transport within the star. In models that include rotational mixing, angular momentum is redistributed from the interior towards the surface, leading to significantly faster rotation in the outer layers. The resulting surface rotational velocity reaches approximately 50\,\kms, making these models potentially consistent with the spectroscopic rotational constraint. A quantitative comparison with the observed \vsini, including the inclination inferred from the orbital solution, is presented below. This result can therefore be regarded as indirect evidence for angular momentum transport mechanisms. The parameters of the seismic models computed with rotational mixing are listed in Table~\ref{tab:parsei}.

We examined two values of initial hydrogen abundance, corresponding to $X = 0.70$ for Model\,2 and $X = 0.725$ for Model\,3. Similarly, we tested two values of mixing-length parameter, with $\alpha_{\mathrm{mlt}} = 1.8$  for Model\,2 and $\alpha_{\mathrm{mlt}} = 1.0$ for Model\,4.

\begin{figure}
	\includegraphics[angle=0, width=1\columnwidth]{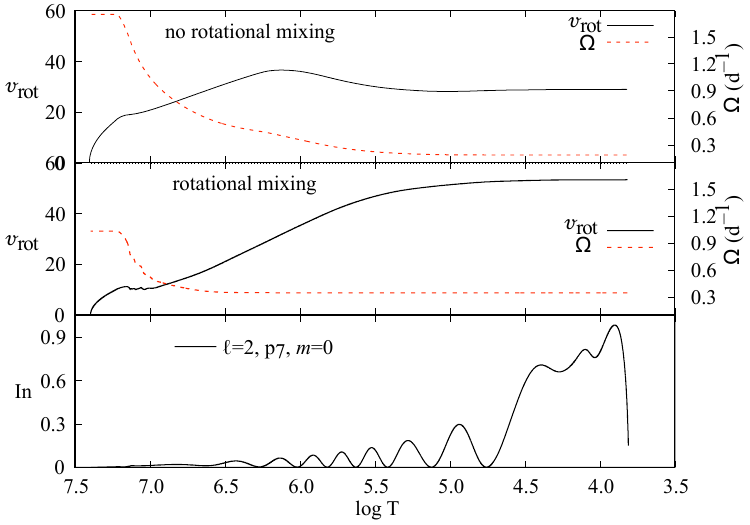}
	\caption{Upper panel: internal rotational profile of seismic Model\,1 listed in Table~\ref{tab:parsei}, given in terms of the linear rotational velocity (left-hand axis, in \kms) and the angular velocity (right-hand axis). Middle panel: as in the upper panel, but for Model\,2, which includes rotational mixing. Bottom panel: mode inertia profile (in arbitrary units) for the ($\ell = 2$, $m = 0$, $p_{7}$) mode computed for Model\,2.}
	\label{fig:rot_prof}
\end{figure}

Internal rotational profiles of models computed without and with rotational mixing are shown in Fig.~\ref{fig:rot_prof}. The left-hand axis represents the linear rotational velocity, $\upsilon_{\mathrm{rot}}$ (in \kms), while the right-hand axis gives the corresponding angular velocity, $\Omega$. In models without rotational mixing, the rotational velocity in the outer layers decreases to values of order $30$\,\kms. By contrast, models including rotational mixing exhibit a nearly monotonic increase in $\upsilon_{\mathrm{rot}}$ towards the surface, reaching approximately $50$\,\kms\ in the outermost layers.

It is noteworthy that Models\,1 and 2 differ substantially, although the principal change in the adopted physical treatment is the inclusion of rotationally induced mixing and angular-momentum transport. Re-optimising our seismic solution then leads to significant differences in the inferred stellar parameters, including metallicity, mass, overshooting efficiency, and other model properties.

The origin of this behaviour is related to the nature of high-order $p$ modes considered in our study, which propagate primarily in the outer stellar envelope. In the bottom panel of Fig.~\ref{fig:rot_prof}, we present mode inertia for the quadrupole axisymmetric mode ($\ell = 2$, $m = 0$, $p_{7}$), showing that most of the pulsation energy is concentrated in the outer layers above the region where $\log T \simeq 5.0$. As a result, rotational splitting of the multiplets is determined mainly by the rotation profile in these outer regions. Since these layers are most strongly affected by rotational mixing, including such processes has a substantial impact on the seismic modelling results.

\setlength{\tabcolsep}{5pt}
\begin{table*}
	\centering
    \setlength{\tabcolsep}{4pt}
    \caption{Seismic models of \object{HD\,99458}.}
	\label{tab:parsei}
	\begin{tabular}{ccccccccccccccc}
		\hline\hline
		Model & opacity & $Z$ & $X$ & $f_{\mathrm{ov}}$ & $\alpha_{\mathrm{mlt}}$ & $\upsilon_{\mathrm{rot,0}}$ & $\upsilon_{\mathrm{rot}}$ & $M$ & $\log{T_{\rm{eff}}}$ & $\log{\frac{L}{L_{\sun}}}$ & $R$ & $\log{g}$ & Age& $X_c$ \\
         &&&&&& (\kms) & (\kms) &(M$_{\sun}$) & & & (R$_{\sun}$) & & (Myr)&
        \\
		\hline
		\multicolumn{15}{c}{no rotational mixing}\\
        1 & OPAL & 0.0089 & 0.7 & 0.030 & 1.8 & 126.1 & 28.7 & 1.9016 & 3.8965 & 1.5153 & 3.0754 & 3.7400 & 963.447 & 0.173 \\
        \multicolumn{15}{c}{rotational mixing}\\
	    2 & OPAL & 0.0110 & 0.7 & 0.027 & 1.8 & 78.0 & 52.9 & 1.9467 & 3.8890 & 1.4895 & 3.0906 & 3.74599 & 934.616 & 0.178 \\
		3 & OPAL &0.0104 & 0.725 & 0.028 & 1.8 & 77.6 & 53.3 & 1.9861 & 3.8867 & 1.4862 & 3.1118 & 3.7487 & 999.334 & 0.194 \\
		4 & OPAL & 0.0109 & 0.7 & 0.022 & 1.0 & 77.3 & 52.9 & 1.9451 & 3.8882 & 1.4860 & 3.0898 & 3.7458 & 939.428 & 0.173 \\
		5 & OPLIB & 0.0117 & 0.7 & 0.030 & 1.8 & 76.9 & 52.7 & 1.9212 & 3.8801 & 1.4494 & 3.0748 & 3.7447 & 993.381 & 0.191 \\
		\hline\hline
	\end{tabular}
    \tablefoot{Columns list the model identifier (Model), adopted opacity tables (Opacity), metallicity ($Z$), initial hydrogen abundance ($X$), convective core overshooting parameter ($f_{\mathrm{ov}}$), mixing-length parameter ($\alpha_{\mathrm{mlt}}$), initial surface rotational velocity ($\upsilon_{\mathrm{surf,0}}$), current surface rotational velocity ($\upsilon_{\mathrm{surf}}$), stellar mass ($M$), effective temperature ($\log T_{\mathrm{eff}}$), luminosity ($\log L/L_{\sun}$), radius ($R$), surface gravity ($\log g$), age measured from the zero-age main sequence (Age), and central hydrogen abundance ($X_c$).}
\end{table*}

\begin{figure}
	\includegraphics[angle=0, width=0.95\columnwidth]{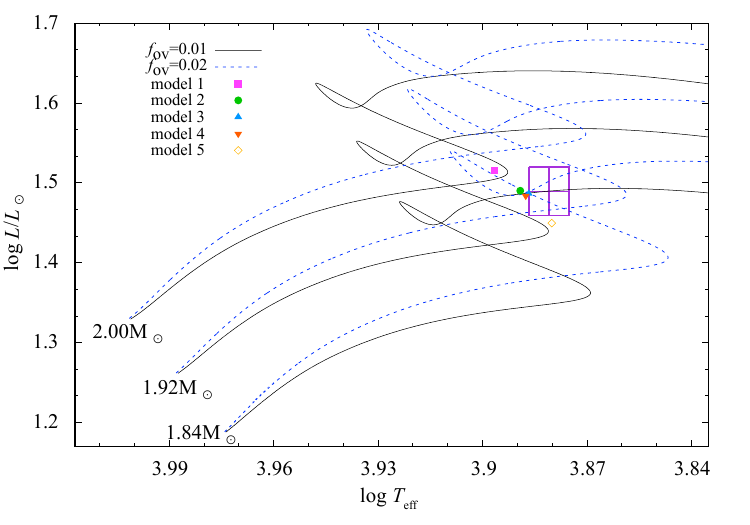}
	\caption{Hertzsprung--Russell diagram showing the position of \object{HD\,99458}. Evolutionary tracks were computed for $f_{\mathrm{ov}} = 0.02$, $\upsilon_{\mathrm{rot,0}} = 80$\,\kms, $Z = 0.01$, and $X = 0.7$. Locations of the seismic models discussed in the text are also indicated.}
	\label{fig:HR}
\end{figure}

\begin{figure*}
	\sidecaption
    \includegraphics[width=12cm]{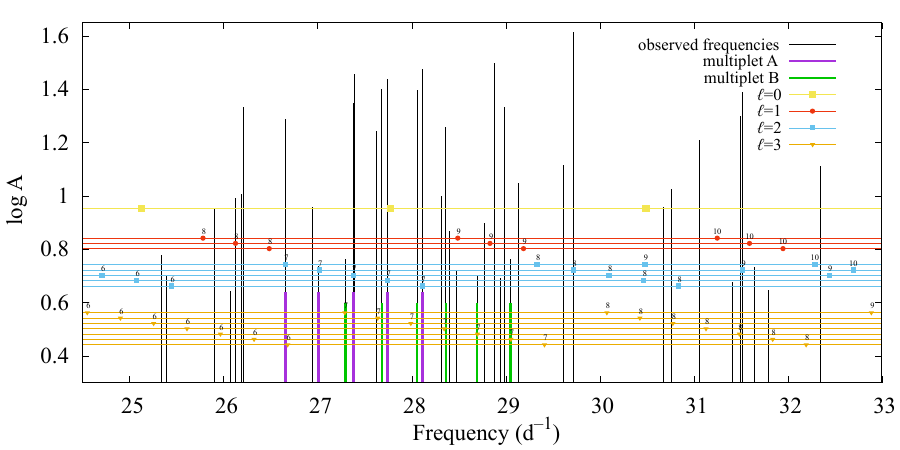}
	\caption{Comparison between the observed and theoretical frequency spectra for Model\,2. Components of multiplets~A and B are indicated by violet and green lines, respectively. For clarity, the amplitudes of the frequencies detected in the K2 photometry have been scaled by a factor of 0.5. The vertical scale for the theoretical modes is arbitrary.}
	\label{fig:nu_spectrum}
\end{figure*}

Theoretical modes predicted within the observed frequency range of our seismic models are found to be stable. The instability parameter, $\eta$ \citep{1978AJ.....83.1184S}, is typically of order $-0.4$, indicating difficulties in reproducing the excitation of the observed pulsation modes. However, these instability results should be interpreted with caution. In the present calculations, we adopted a frozen convective flux approximation, which is not ideally suited for $\delta$\,Sct pulsators. A more realistic treatment would require a time-dependent, non-local convection formalism \citep[e.g.][]{2016MNRAS.457.3163X,2019MNRAS.490.4040A}, although such modelling is beyond the scope of this work.

All models listed in Table~\ref{tab:parsei} reproduce the observed frequency spectrum of \object{HD\,99458} reasonably well. Comparison between observed and theoretical spectra for Model\,2 is presented in Fig.~\ref{fig:nu_spectrum}, where multiplets~A and B are highlighted. The theoretical spectrum includes modes with degrees $\ell = 0$--$3$ and all corresponding azimuthal orders; the vertical positions of the theoretical frequencies are arbitrary and serve only for visual clarity.

Multiplet~A is reproduced particularly well by our model. Good agreement is also obtained for several individual frequencies, including the high-amplitude mode $f_{1}$, as well as $f_{3}$ and $f_{23}$. Additional satisfactory matches are found for $f_{41} = 28.40390$\,\cpd, identified as an ($\ell = 1$, $m = -1$) mode, $f_{6} = 31.49280$\,\cpd, corresponding to ($\ell = 3$, $m = 1$), and $f_{18} = 30.75890$\,\cpd, identified as an ($\ell = 3$, $m = -1$) mode.

Interestingly, multiplet~B is also reproduced reasonably well, particularly for the $m = -3$, $0$, $1$, and $2$ components. The agreement is somewhat poorer for the $m = -2$ and $m = -1$ components. Many additional observed frequencies also have theoretical counterparts, although the agreement is not always exact. For example, the frequency $f_{8} = 28.87760$\,\cpd\ appears to correspond to an ($\ell = 1$, $m = 0$, $p_{9}$) dipole axisymmetric mode, but the predicted theoretical frequency is slightly lower than the observed value.

Propagating the frequency uncertainties to the individual splittings according to
$$
	\sigma_{\Delta f}=\left(\sigma_{f_i}^2+\sigma_{f_{i+1}}^2\right)^{1/2},
$$
we obtain consecutive splittings
$0.3489\pm0.00042$, $0.3828\pm0.00036$, $0.3549\pm0.00022$, and $0.3710\pm0.00022$\,\cpd\ for multiplet A, while for multiplet B they are
$0.3840\pm0.00092$, $0.3749\pm0.00028$, $0.3070\pm0.00028$, $0.3311\pm0.00073$, and $0.3579\pm0.00086$\,\cpd. Thus, if only the formal frequency uncertainties are considered, the departure from uniform splitting is highly significant. We note, however, that these uncertainties describe the precision of the frequencies determined from the individual datasets and do not account for the observed epoch-to-epoch frequency variations. Indeed, frequencies measured independently from the K2 data and the two TESS sectors show shifts that are typically one or two orders of magnitude larger than their formal uncertainties. For example, the four splittings of multiplet A are $0.3489$, $0.3791$, $0.3586$, and $0.3720$\,\cpd\ in Sector 45; $0.3475$, $0.3874$, $0.3544$, and $0.3714$\,\cpd\ in Sector 72; and $0.3526$, $0.3693$, $0.3634$, and $0.3714$\,\cpd\ in K2 data. Therefore, although the non-equidistant character of the multiplets appears to be robust, the very high formal significance inferred from the statistical frequency uncertainties alone should not be interpreted literally.

Perfectly equidistant multiplets are expected only in the first-order perturbative approximation for sufficiently slow rotation. At the precision reached by observations of \object{HD\,99458}, departures from this approximation may become important. Included radial differential rotation changes the magnitude of first-order splitting through a mode-dependent rotational kernel, but does not by itself produce non-equidistant components within a given multiplet. Additional deviations from a simple first-order Ledoux splitting may arise from second- and higher-order rotational effects, such as centrifugal distortion and higher-order Coriolis terms, which are not included in the present calculations. Near-degeneracy between modes of similar frequencies may further perturb the individual multiplet components. In addition, latitudinal differential rotation and perturbations related to the close binary configuration may contribute to the observed non-uniformity of the splittings.

Seismic rotation inferred from the multiplet structure is nevertheless consistent with the independent orbital and spectroscopic constraints. Using the spectroscopic constraint $M_1\sin^3 i=1.39\pm0.10\,M_\odot$ together with the masses of our seismic models, $M\simeq1.92\text{--}1.99\,M_\odot$, we obtain inclination angles of approximately $60^\circ\text{--}67^\circ$, with central values close to $63^\circ$. For the models including rotational mixing, the present-day surface rotational velocities are $\upsilon_{\mathrm{surf}}\simeq52.7\text{--}53.3$\,\kms, corresponding to $\vsini\simeq46.1\text{--}48.4$\,\kms. These values are broadly consistent with the spectroscopic estimate of $\vsini=49.9\pm2.0$\,\kms. The corresponding surface rotation periods are approximately $2\fd95$, that is slightly longer than the orbital period, $P_{\mathrm{orb}}=2\fd722172$.

We stress, however, that our seismic models include differential rotation, so the rotational splittings probe a kernel-weighted internal rotation rate rather than simply the surface angular velocity. Moreover, the exact surface rotation rate depends on the efficiency of angular momentum transport and has not been specifically calibrated to enforce synchronisation with the orbit. Thus, although the present models are not exactly synchronised at the surface, they reproduce multiplet A very well, provide a reasonable description of multiplet B, and simultaneously predict a projected surface rotational velocity consistent with the observed \vsini. We therefore consider the seismic interpretation and the spectroscopic rotational constraint to be mutually consistent.

We find a strong dependence of the overshooting parameter on the adopted mixing-length parameter. In Model\,4, computed with the lower value $\alpha_{\mathrm{mlt}} = 1.0$, the derived overshooting parameter $f_{\mathrm{ov}}$ is correspondingly smaller. The remaining stellar parameters, however, differ only marginally.

We also investigated the effect of opacity tables by computing models using OPLIB data \citep{OPLIB1, OPLIB2}. The best-fitting seismic model obtained with OPLIB opacities is presented in Table~\ref{tab:parsei} (Model\,5). Significant differences between models computed with OPAL and OPLIB opacities indicate a strong sensitivity of seismic solutions to the opacity profile. High-order $p$ modes primarily probe the outer stellar layers, where opacity profiles from different sources show the largest discrepancies. Consequently, $\delta$\,Sct pulsators with well-identified multiplets may provide a valuable diagnostic for testing stellar opacity calculations. A comprehensive analysis, however, would require a larger sample of well-constrained multiplets.

The OPLIB-based model occupies a different location in the Hertzsprung--Russell diagram compared to OPAL models, being cooler and less luminous. Nevertheless, it remains close to the observational error box.

Our seismic analysis of \object{HD\,99458} suggests moderate-to-efficient convective core overshooting, with $f_{\mathrm{ov}}$ in the range of approximately 0.02--0.03. The exact value depends primarily on the mixing-length parameter $\alpha_{\mathrm{mlt}}$, and to a lesser extent on opacity tables. The models also indicate that the stellar metallicity is likely subsolar, with $Z \sim 0.01$. By contrast, the hydrogen abundance remains only weakly constrained. Models with higher $X$ provide a slightly better match to the observational error box in the Hertzsprung--Russell diagram (see Fig.~\ref{fig:HR}).

In summary, our seismic modelling of \object{HD\,99458} demonstrates that the observed frequency spectrum, including the identified multiplets, can be reproduced consistently within a narrow range of stellar parameters. The models favour subsolar metallicity and moderate-to-efficient convective core overshooting. Comparison between OPAL- and OPLIB-based models highlights the strong sensitivity of high-order $p$ modes to the structure of the outer stellar layers, emphasising the diagnostic potential of $\delta$\,Sct pulsators for constraining opacity calculations. Although the hydrogen abundance remains weakly constrained, the agreement between the seismic models and observations supports the overall consistency of the seismic solutions. In addition, models including angular-momentum transport yield projected surface rotational velocities consistent with the spectroscopic \vsini\ constraint and reproduce the overall rotational splitting pattern of the identified multiplets.

\section{Discussion} \label{sec:finale}

Observational evidence collected in our study consistently supports the view that \object{HD\,99458} is a typical Algol-type binary system with mass transfer from the secondary component filling its Roche lobe (donor) to a pulsating $\delta$\,Sct star (gainer), specifically a member of the oEA group of variables \citep{2022Galax..10...97M}. This is precisely the interpretation that \citet{2019MNRAS.487.4230S} briefly mentioned in their original publication and later rejected in favour of the CP2 star hypothesis.

The evidence collected in our study suggests that variable mass transfer and the accretion disc are responsible for fluctuations in out-of-eclipse photometric flux. Cold spots used in our modelling with \texttt{ELISa} are unable to produce a substantial effect on the flux.

We consider remarkably different H$_{\alpha}$ patterns shown in Fig.~\ref{fig:hd99458_halpha} and observed in spectra separated by several years to be an observational manifestation of long-periodic activity. Presumably, the main driver of this slow variation is the donor's weak magnetic field. While this field remains undetected, its fluctuating activity provides a plausible explanation for the detection of an X-ray source at the position of \object{HD\,99458} in observations from eROSITA \citep{2021A&A...647A...1P, 2024A&A...682A..34M, 2026A&A...712A.171R}. The source 3eRASS J112636.1+010319 is located within 1.8\arcsec\ of \object{HD\,99458}. The X-ray flux $F_\mathrm{X}$ of this source is $4.10\times10^{-13}$ erg\,s$^{-1}$\,cm$^{-2}$. This yields an X-ray luminosity $L_\mathrm{X}=3.35\times10^{30}$ erg\,s$^{-1}$, which is comparable to the X-ray luminosity of Algol \citep{2025ApJ...995..152N}.

We attribute the periodic reappearance of the \ion{He}{i} 5875\,\AA\ line in spectra obtained around the orbital phases $\varphi_\mathrm{orb}=0.89$ and $\varphi_\mathrm{orb}=0.16$ to the presence of hot regions in the circumstellar environment. This feature naturally explains the excitation of the corresponding transition, which would be impossible in a star with the primary's effective temperature of 7600\,K.  Given the system geometry, the hot regions are most likely located where the gas stream hits the atmosphere of the pulsating star, clearly visible at orbital phases before the primary minimum, and in another zone on the back side of the gas stream interacting with gas flows circulating around the primary star. A similar periodic excitation of \ion{He}{i} line around the phases of primary minima was reported in the semi-detached Algol-type system \object{R~CMa} by \cite{2018A&A...615A.131L}.

Strong variable absorption in the H$_{\alpha}$ line, visible in the PLATOSpec data, and near-IR lines observed with MRES at BJD\,2\,461\,071 are part of the same mass-transfer process. The corresponding orbital phases are nearly symmetric around $\varphi_\mathrm{orb} = 0.75$, when the line connecting the stars is perpendicular to the line of sight. This geometry suggests that our spectra may have sampled the gas stream flowing from the donor to the gainer at different viewing angles.

\begin{figure}
  \centering
  \includegraphics[width=0.8\columnwidth]{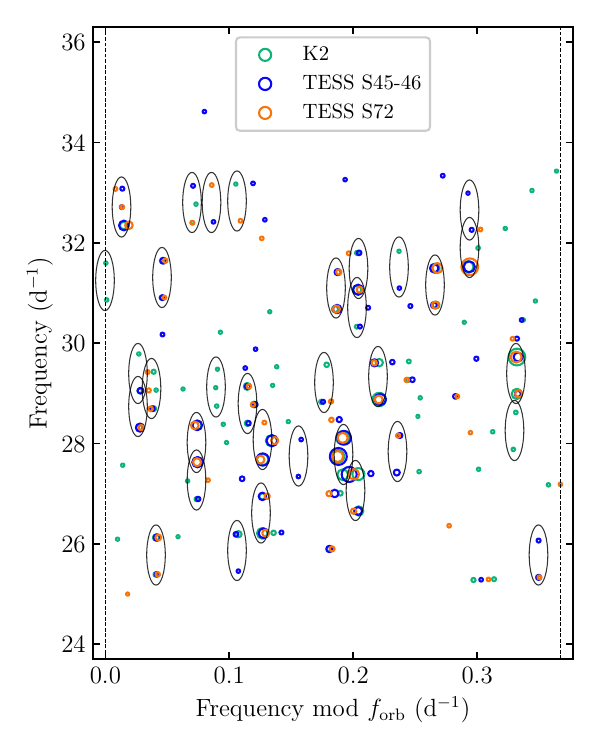}
      \caption[]{\'{E}chelle diagram showing pulsation peaks detected in Kepler K2 and TESS photometry (Table~\ref{tab:frequencies}) as a function of frequency modulo $f_\mathrm{orb}=0.36738$\,\cpd. Frequencies separated by $2f_\mathrm{orb}$ are indicated in pairs. Symbol size is proportional to the amplitude.}
  \label{fig:echelle}
\end{figure}

We have, for the first time, performed a detailed asteroseismic analysis of \object{HD\,99458}. We show that the rich pulsation spectrum extracted from the photometry can be partly explained by intrinsic pulsations of the primary. Comparison between observational and theoretical frequencies shown in Fig.~\ref{fig:nu_spectrum} confirms the identification of pulsation modes with degree $\ell$ up to 3. In addition, the star exhibits pulsation modes tidally perturbed by the secondary. The photometric \'{e}chelle diagram constructed with the orbital frequency $f_\mathrm{orb} = 0.36738$\,\cpd\ contains more than 30 pairs of frequencies separated by $2f_\mathrm{orb}$ (Fig.~\ref{fig:echelle}). This effect is well-known and has a natural explanation in the theory of triaxial pulsation in gravitationally distorted stars \citep{2025ApJ...979...80F}.

\begin{figure}
  \centering
  \includegraphics[width=0.95\columnwidth]{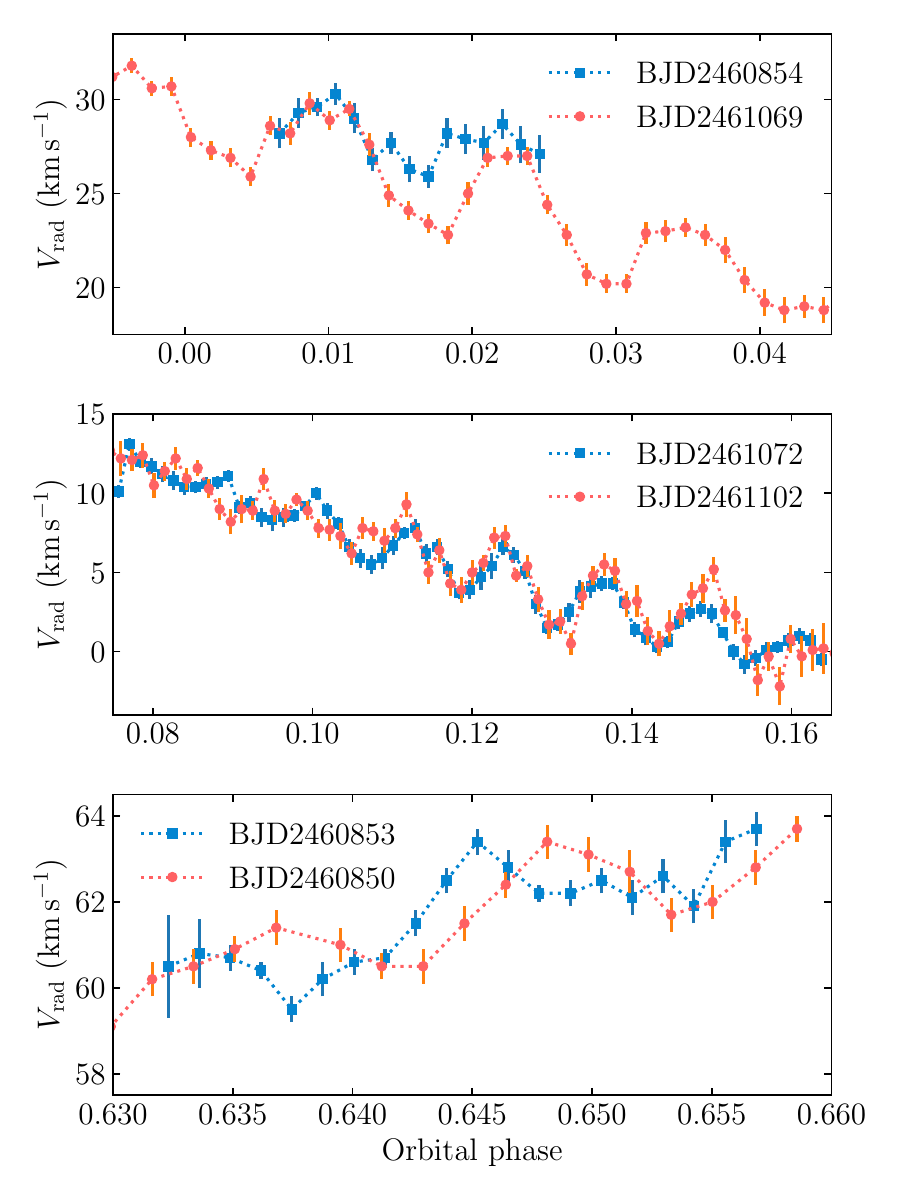}
      \caption[]{Radial velocity variations of the pulsating primary at different epochs.}
  \label{fig:RV_synchro}
\end{figure}

Extensive use of time-resolved spectroscopy extends the potential identification of gravitationally perturbed pulsations to higher spherical-harmonic orders inaccessible to photometry. By mapping the antinodes produced by non-radial pulsation and visible in our 2D-DFT analysis, we infer that azimuthal orders as high as $|m|=5$, possibly $\pm1$, are required to explain the recurrence of the antinodes at the same longitudes in different observing epochs. Furthermore, the character of the variability in radial velocities, shown, for example, in Fig.~\ref{fig:RV_synchro}, indicates synchronisation of stellar pulsations with the orbital motion. The agreement between the radial-velocity pulsations observed in different epochs and phased with the photometrically derived $P_\mathrm{orb}$ independently supports the high consistency of our analysis, including the detection of $\dot{P}=5.045$\,s\,yr$^{-1}$. Thus, we hypothesise that the discrepancy between the two orbital period values obtained in Sects.~\ref{sec:eclipmod} and \ref{sec:system} most likely originates from the multiplicity of \object{HD\,99458}, despite the fact that we detect no significant variation in the systemic velocity and did not detect a tertiary in speckle-interferometric observations.

In our analysis, \object{HD\,99458} shows none of the Fe, Si, and Ti anomalies reported by \citet{2019MNRAS.487.4230S}. In contrast to abundances enhanced by 1\,dex, we find most elements in \object{HD\,99458} to be mildly depleted. Such underabundant Population~I A-type stars are commonly classified as $\lambda$~Bootis stars \citep{2004IAUS..224..443P, 2017MNRAS.466..546M}. The abundance profile of these objects is characterised by depletion of Fe-peak elements of up to 2\,dex, but near-solar abundances of C, N, O, and S \citep[e.g.][]{2002A&A...381..959H}. In a large sample studied by \citet{2002MNRAS.336.1030P}, stars showing $\lambda$ Boo properties occupied the whole width of the main sequence, peaking at an age of $\approx1$\,Gyr. An asteroseismically derived age of $\sim0.9$\,Gyr and subsolar metallicity make \object{HD\,99458} a good $\lambda$ Boo candidate. The $\lambda$ Boo status of \object{HD\,99458} requires additional confirmation primarily due to depleted carbon and slightly deficient sulphur. Even so, the lack of strongly overabundant species typical of CP2 stars, together with the absence of a strong magnetic field, allows us to conclude that the primary component of \object{HD\,99458} is genuinely not Ap.

In summary, our study has established a new classification for \object{HD\,99458}, prompting a re-evaluation of the system as a whole. Beyond this specific result, our findings highlight how data quality and a comprehensive analytical approach influence the physical interpretation of observational data, underscoring the value of time-resolved spectroscopy.

\begin{acknowledgements}
The authors are very grateful to the referee for carefully reading our paper and for the thoughtful comments that helped us improve it.

The authors express their gratitude to V.~Kozlov (International Centre for Astronomical, Medical \& Ecological Research, Ukraine) and G.~Kapanadze (Georgian National Astrophysical Observatory, Georgia) for their assistance with data acquisition.

This work is supported by the Fundamental Fund of Thailand Science Research and Innovation (TSRI) through the National Astronomical Research Institute of Thailand (Public Organization) (FFB690078/0269).
This work has been supported by the DAAD (project No. 57442043) and the Erasmus+ programme of the European Union under grant number 2020-1-CZ01-KA203-078200.
O.K. acknowledges support from the Swedish Research Council (grant agreement no. 2023-0366) and the Swedish National Space Agency.
P.W. acknowledges financial support from the Polish National Science Centre (NCN) under grant No.\ 2023/50/A/ST9/00144. The computations were performed using resources provided by the Wroc\l{}aw Centre for Networking and Supercomputing (\url{http://wcss.pl}) under grant No.\ 265.
M.G. acknowledges support from the EU NextGenerationEU through the Recovery and Resilience Plan for Slovakia under project No. 09I03-03-V04-00262.
M.S. acknowledges the support by Inter-transfer grant no LTT-20015.
A.V.M. acknowledges partial financial support from the Russian Science Foundation (grant No. 25-12-00003) for the MSS data reduction.
J.\v{Z}. acknowledges the support of the PRODEX agreement No. 400012791.
This work is based on observations obtained at the Canada-France-Hawaii Telescope (CFHT), which is operated by the National Research Council of Canada, the Institut National des Sciences de l'Univers (INSU) of the Centre National de la Recherche Scientifique (CNRS) of France, and the University of Hawaii.
This study includes observations made with MRES across multiple programmes, including the most recent ID TNTC13\_011, at the Thai National Observatory, operated by the National Astronomical Research Institute of Thailand (Public Organisation).
This work has made use of data from the European Space Agency (ESA) mission Gaia (\url{https://www.cosmos.esa.int/gaia}), processed by the Gaia Data Processing and Analysis Consortium (DPAC, \url{https://www.cosmos.esa.int/web/gaia/dpac/consortium}). Funding for the DPAC has been provided by national institutions, in particular the institutions participating in the Gaia Multilateral Agreement.
\end{acknowledgements}

\section*{Data availability}
Reduced ESPaDOnS spectra are available at the CFHT archive maintained by the CADC at \url{https://www.cadc-ccda.hia-iha.nrc-cnrc.gc.ca/en/}, where they can be found via standard stellar designations. The remaining spectra, as well as any additional materials used in the current research, can be obtained from the authors upon request.

\bibliographystyle{aa}
\bibliography{hd99458_arxiv}

\begin{appendix}

  \section{Spectroscopy log and radial velocities}
  \hspace{10pt}

  \renewcommand{\arraystretch}{1.05}
  \tablecaption{Observational log with measured radial and rotational velocities of HD\,99458.}
  \label{tab:obslog}
  \tablefirsthead{\hline\hline
  	BJD $-$ & $T_\mathrm{exp}$  & SNR  & $V_\mathrm{r}$ & \vsini\ \\ 2\,400\,000       & (s)   &   &  (\kms) & (\kms) \\
  	\hline}
  \tablehead{\multicolumn{4}{l}{\textbf{\tablename\ \thetable.} continued} \\ \hline\hline
  	BJD $-$ & $T_\mathrm{exp}$  & SNR  & $V_\mathrm{r}$ & \vsini\ \\ 2\,400\,000       & (s)   &   &  (\kms) & (\kms) \\
  	\hline}
  \tabletail{\hline \\}
  \tablelasttail{\hline\hline
  \multicolumn{5}{p{0.9\linewidth}}{\small \textbf{Notes.} Columns list the mid-exposure barycentric Julian date, exposure time ($T_\mathrm{exp}$), maximum SNR, radial velocity (\rv) and rotational velocity (\vsini).} \\
  }
  \begin{center}
\begin{supertabular}{lcccc}
  	\multicolumn{5}{c}{HERMES}                                                       \\
  	58\,543.53811  & 600    & 100    &  $1.7 \pm 0.2$          & $46.8 \pm 0.4$          \\
  	58\,543.57144  & 300    & 70     &  $1.5 \pm0.2$ & $48.1 \pm0.3$ \\
  	58\,543.57549  & 300    & 70     &  $0.9 \pm0.2$ & $48.1 \pm0.3$ \\
  	58\,543.57955  & 300    & 70     &  $0.1 \pm0.2$ & $48.2 \pm0.3$ \\
  	58\,543.58360  & 300    & 70     &  $0.2 \pm0.2$ & $49.2 \pm0.3$ \\
  	58\,543.58765  & 300    & 70     &  $0.6 \pm0.3$ & $49.4 \pm0.4$ \\
  	58\,544.51535  & 900    & 135    &  $39.2\pm0.2$ & $48.0 \pm0.3$ \\
\end{supertabular}
  \end{center}

  \begin{table}[h!]
      \centering
      \caption{Radial velocities of \object{HD\,99458} measured from the spectra presented in \citet{2019MNRAS.487.4230S}.}
      \label{tab:oes}
      \begin{tabular}{lc | lc}
  \hline\hline
  BJD $-$   &  \rv  & BJD $-$    & \rv  \\
  2\,450\,000   &  (\kms)          & 2\,450\,000  & (\kms) \\
  \hline
   7\,828.44658  &  $28.3\pm0.4$            & 7\,853.51071  &  $66.7\pm0.4$  \\
   7\,828.46791  &  $30.5 \pm 0.5$  & 7\,874.35562  &  $4.1  \pm 0.6$  \\
   7\,828.49370  &  $32.5 \pm 0.6$  & 7\,854.37255  &  $32.9 \pm 0.5$  \\
   7\,829.42227  &  $65.4 \pm 0.3$  & 7\,874.37695  &  $6.1  \pm 0.9$  \\
   7\,829.44360  &  $63.4 \pm 0.4$  & 7\,884.34898  &  $31.3 \pm 0.5$  \\
   7\,839.43422  &  $35.0 \pm 0.7$  & 7\,884.35990  &  $30.6 \pm 0.4$  \\
\hline\hline
  \end{tabular}
  \end{table}

  \onecolumn
  \section{Frequencies in K2 and TESS data}

  \LTcapwidth=0.8\linewidth
  \begin{longtable}{l c c c c c c c c c c c}
  \caption{Master list of 122 pulsation frequencies identified in three photometric datasets:
    TESS sectors~45--46 (S45), TESS sector~72 (S72), and K2.}\label{tab:frequencies}
  \\
  \hline\hline
  ID & $f_\mathrm{best}$ & $\sigma_f$ & Src & $f_\mathrm{S45}$ & $A_\mathrm{S45}$ & $f_\mathrm{S72}$ & $A_\mathrm{S72}$ & $f_\mathrm{K2}$ & $A_\mathrm{K2}$ & $N$ & $A_\mathrm{max}$ \\
     & (\cpd)            & (\cpd)     &     & (\cpd)           & (mmag)           & (\cpd)           & (mmag)           & (\cpd)          & (mmag)       &     & (mmag)           \\
  \hline
  \endfirsthead
  \caption{continued} \\
  \hline
  ID & $f_\mathrm{best}$ & $\sigma_f$ & Src & $f_\mathrm{S45}$ & $A_\mathrm{S45}$ & $f_\mathrm{S72}$ & $A_\mathrm{S72}$ & $f_\mathrm{K2}$ & $A_\mathrm{K2}$ & $N$ & $A_\mathrm{max}$ \\
     & (\cpd)            & (\cpd)     &     & (\cpd)           & (mmag)           & (\cpd)           & (mmag)           & (\cpd)          & (mmag)       &     & (mmag)           \\
  \hline
  \endhead
  \hline
  \endfoot
  \hline\hline
  \multicolumn{12}{p{0.8\linewidth}}{\small \textbf{Notes.} Identifiers $f_\mathrm{n}$ are ordered by decreasing $A_\mathrm{max}$. The best frequency $f_\mathrm{best}$ and its uncertainty $\sigma_f$ were taken from the most precise detection (column Src). $N$ is the number of datasets in which the frequency is detected. Amplitudes are in mmag; em dashes indicate non-detection.} \\
  \endlastfoot
  $f_{1}$ & 31.52110 & 0.00020  & S45 & 31.52110 & 1.040 & 31.52150 & 2.810 & 31.52000 & 0.860 & 3 & 2.810 \\
  $f_{2}$ & 27.74160 & 0.00010  & S45 & 27.74160 & 1.800 & 27.74110 & 1.660 & 27.74120 & 1.000 & 3 & 1.800 \\
  $f_{3}$ & 29.72200 & 0.00020  & S72 & 29.72360 & 0.760 & 29.72200 & 1.630 & 29.72260 & 1.420 & 3 & 1.630 \\
  $f_{4}$ & 27.38670 & 0.00020  & S72 & 27.38300 & 1.550 & 27.38670 & 1.530 & 27.37780 & 0.800 & 3 & 1.550 \\
  $f_{5}$ & 28.11260 & 0.00020  & K2       & 28.11360 & 1.380 & 28.11250 & 1.460 & 28.11260 & 1.070 & 3 & 1.460 \\
  $f_{6}$ & 31.49280 & 0.00030  & S45 & 31.49280 & 0.680 & 31.49540 & 1.430 & 31.49340 & 0.630 & 3 & 1.430 \\
  $f_{7}$ & 27.68060 & 0.00020  & S45 & 27.68060 & 1.180 & 27.67880 & 0.990 & 27.68060 & 0.890 & 3 & 1.180 \\
  $f_{8}$ & 28.87760 & 0.00020  & S45 & 28.87760 & 1.090 & 28.87630 & 1.000 & 28.87670 & 1.070 & 3 & 1.090 \\
\end{longtable}

\newpage
\section{2D-DFT diagrams}\label{ap:dft}
\begin{figure}[h!]
  \centering
  \includegraphics[width=0.9\textwidth]{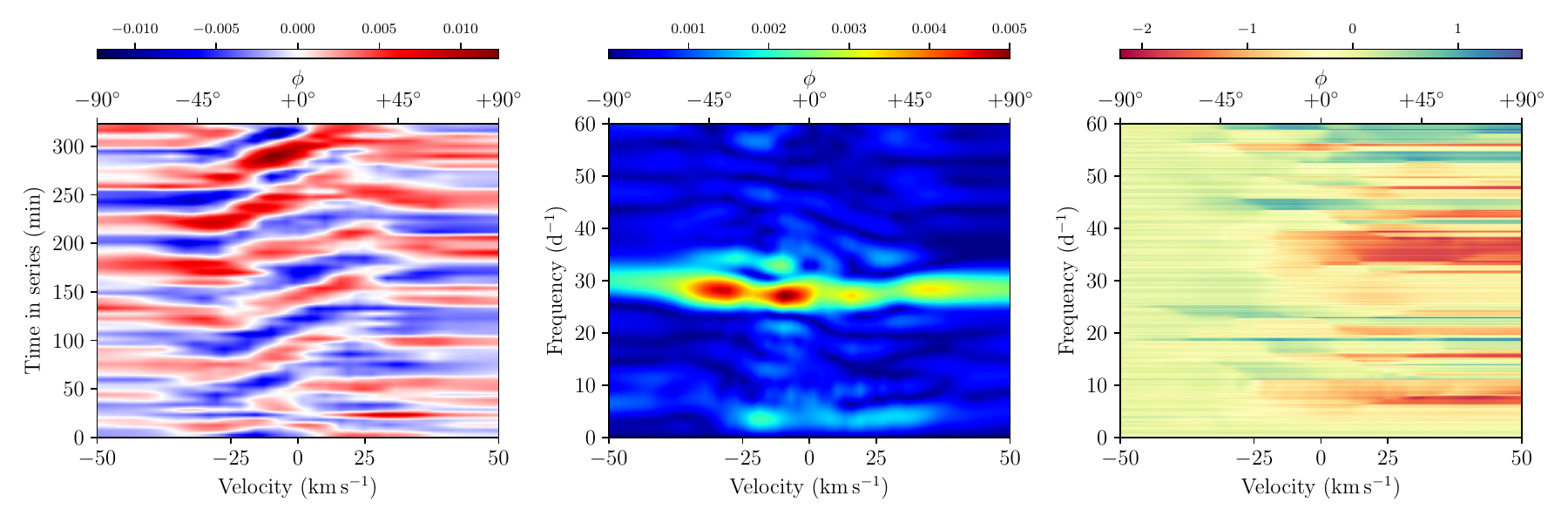}
  \includegraphics[width=0.9\textwidth]{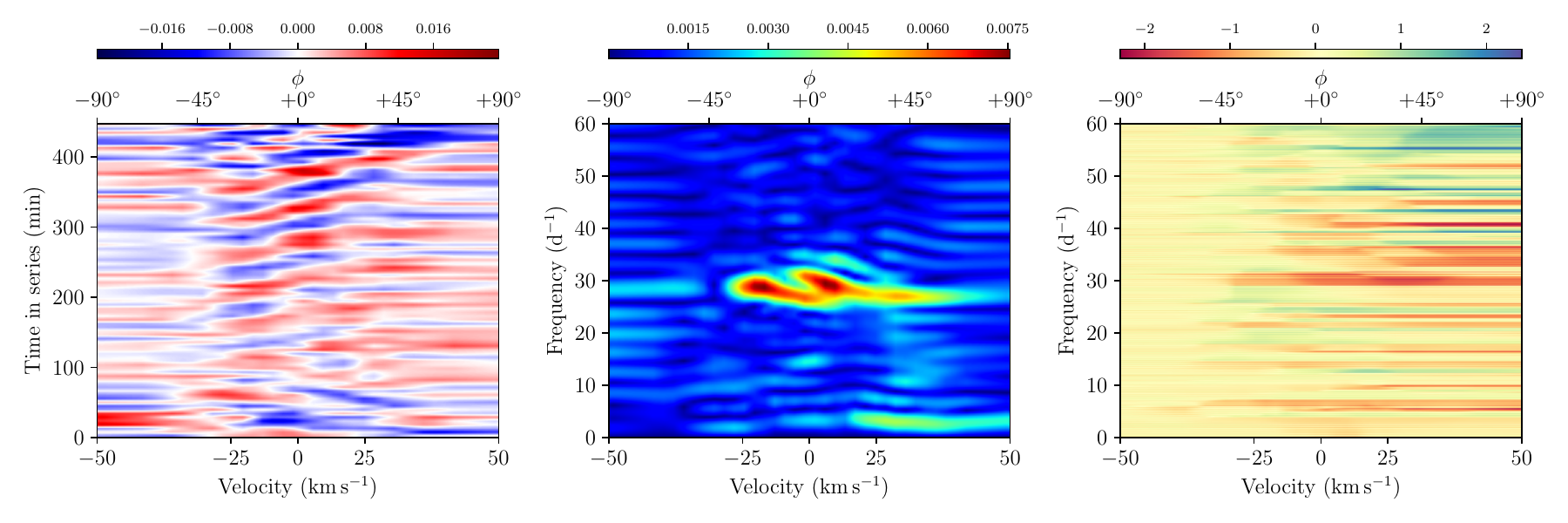}
  \includegraphics[width=0.9\textwidth]{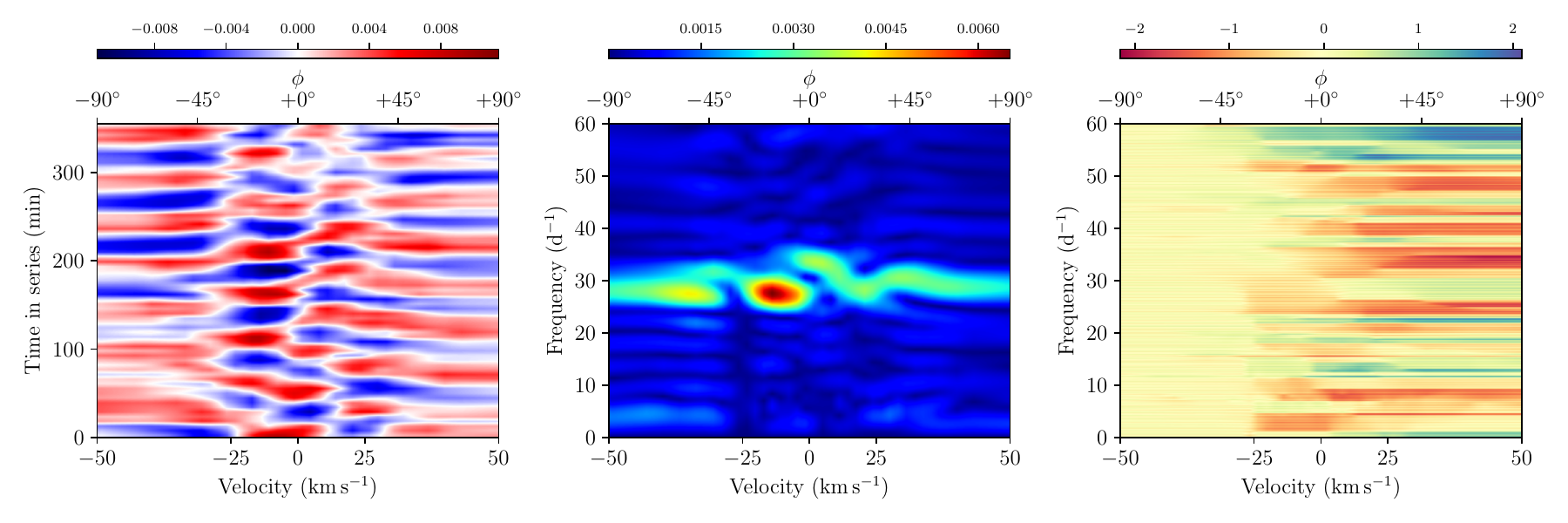}
  \includegraphics[width=0.9\textwidth]{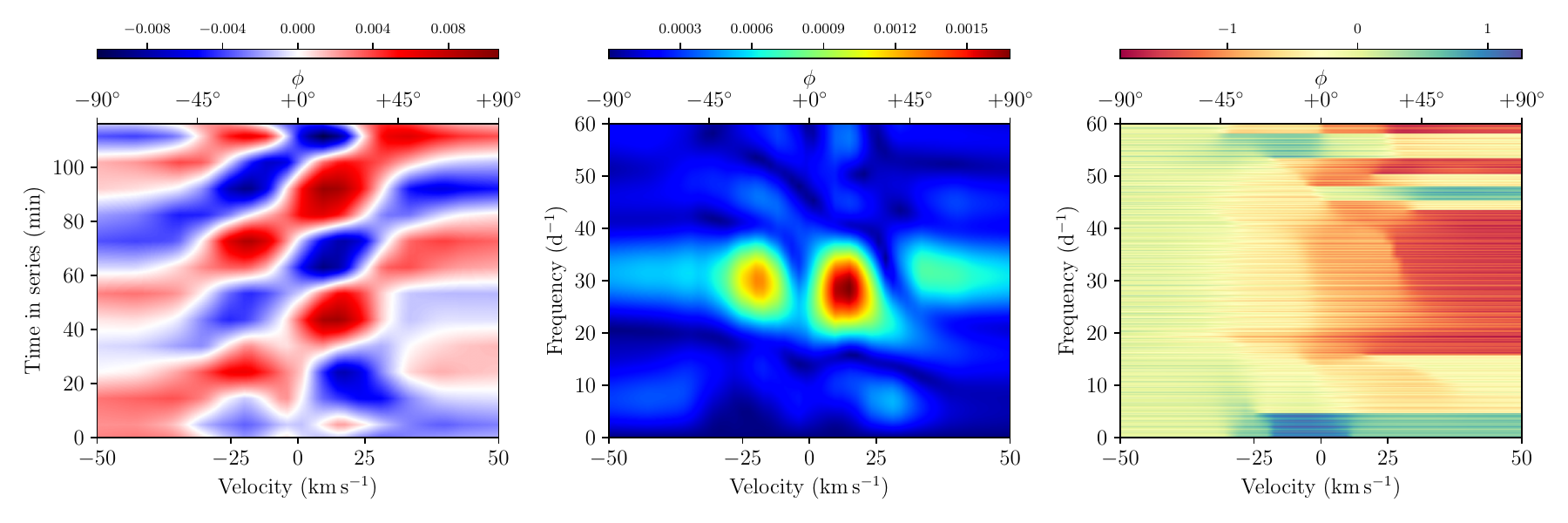}
\caption[]{Top to bottom: 2D-DFT summary of the LPV analysis of datasets DS1--DS4. The corresponding orbital phases are $\varphi_\mathrm{orb}=0.964$--0.048, 0.056--0.170, 0.072--0.163, and 0.185--0.214, respectively. The left panels show LPV in residual spectra, the middle panels demonstrate amplitude variability in the 2D Fourier transform of LPV, and the right panels show the corresponding phase.}
         \label{fig:2DDFT_grp1}
\end{figure}

\begin{figure}[h!]
\centering

\includegraphics[width=0.9\textwidth]{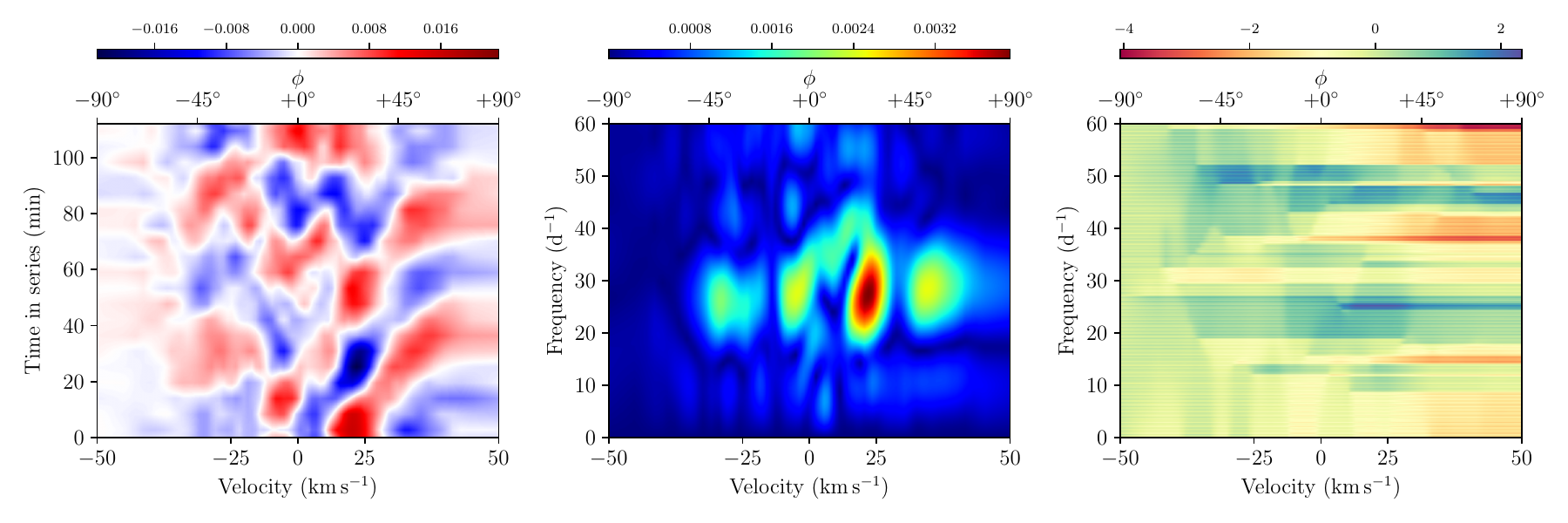}
\includegraphics[width=0.9\textwidth]{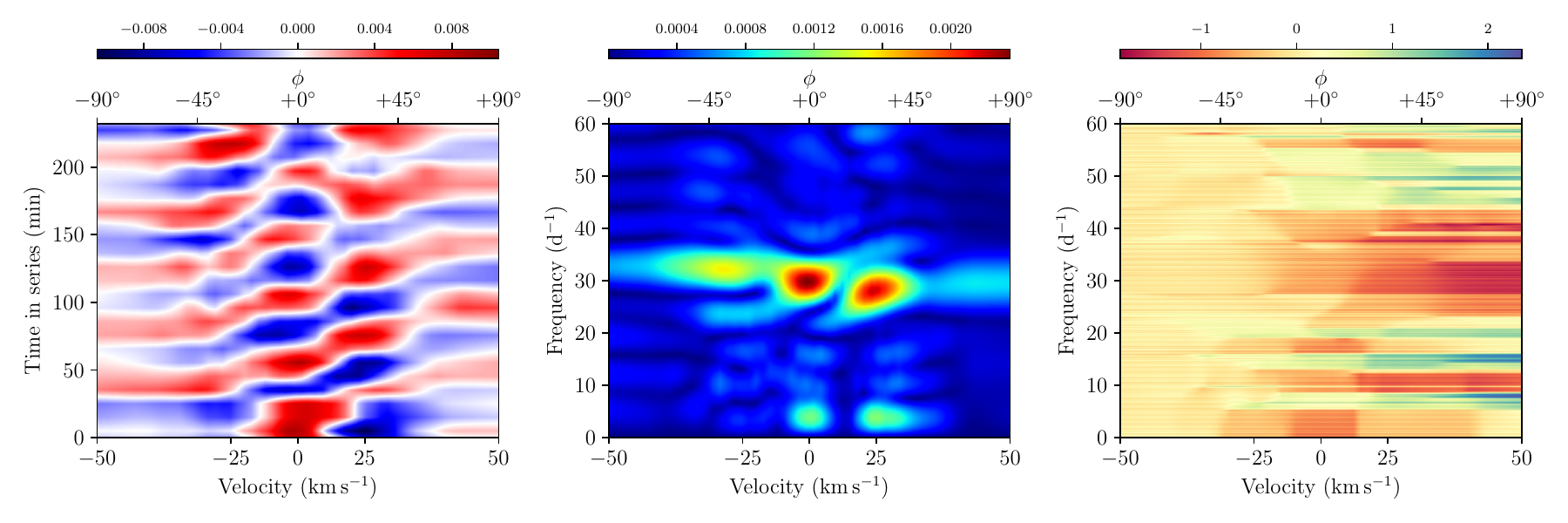}
\includegraphics[width=0.9\textwidth]{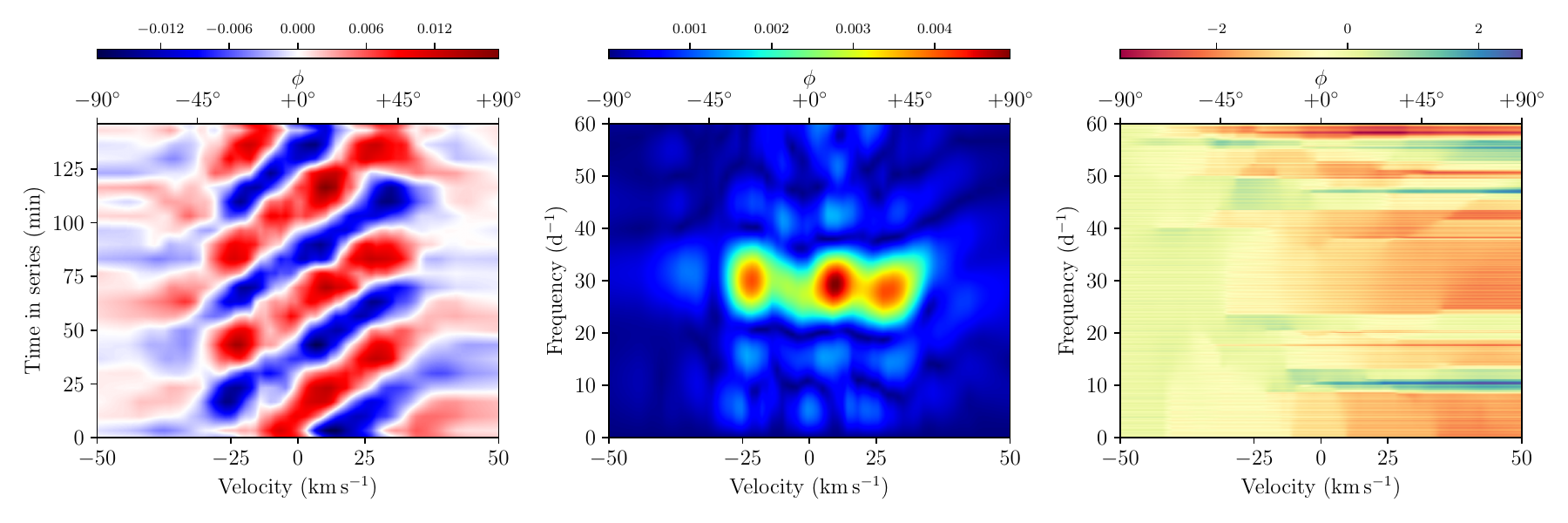}
\includegraphics[width=0.9\textwidth]{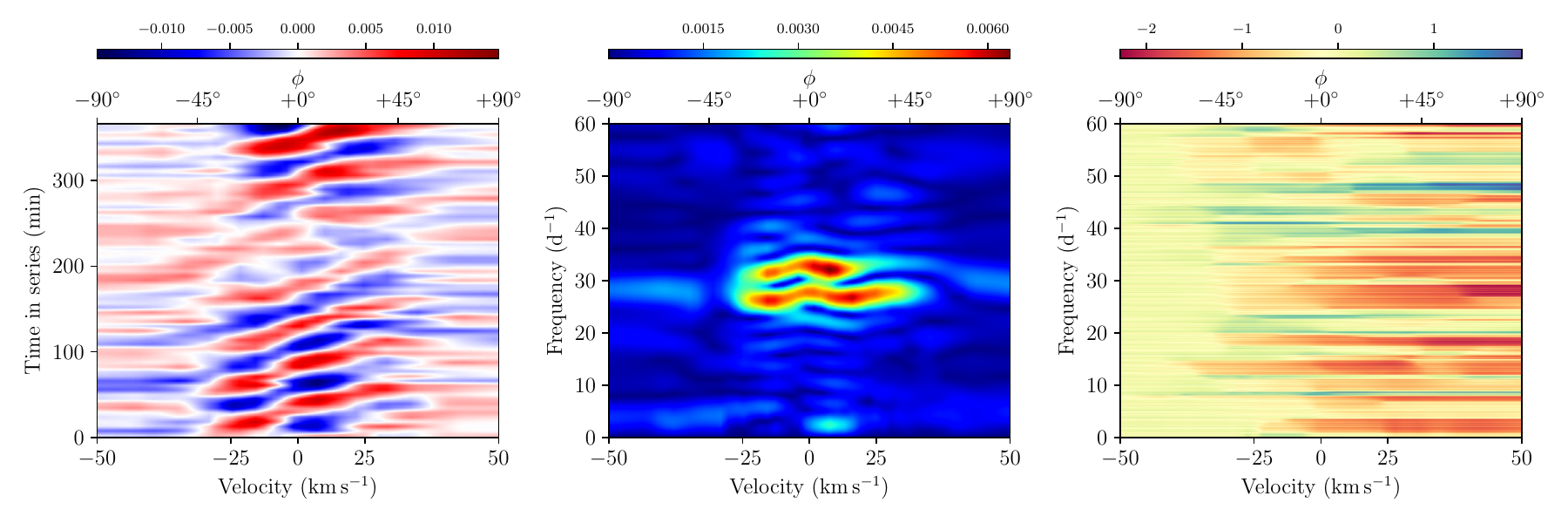}
\caption[]{Top to bottom: 2D-DFT summary of the LPV analysis of datasets DS5, DS7--DS9. The corresponding orbital phases are $\varphi_\mathrm{orb}=0.357$--0.386, 0.527--0.586, 0.621--0.658 and 0.705--0.799, respectively. The left panels show LPV in residual spectra, the middle panels demonstrate amplitude variability in the 2D Fourier transform of LPV, and the right panels show the corresponding phase.}
\label{fig:2DDFT_grp2}
\end{figure}

\end{appendix}
\end{document}